\documentclass{aa}  

\usepackage{graphicx}
\usepackage{txfonts}
\bibpunct{(}{)}{;}{a}{}{,} % to follow the A&A style
\begin{document}

   \title{Comparing UV/EUV line parameters and magnetic field in a quiescent prominence with tornadoes} %Catchy title, yeah!

\titlerunning{UV/EUV line parameters \emph{vs.} magnetic field}

 %  \subtitle{}

   \author{
          %\inst{1}
	 P. J. Levens \inst{1}
	  \and
	  N. Labrosse\inst{1}
	  \and
	  B. Schmieder\inst{2}
	    \and
	   A. L\'{o}pez Ariste \inst{3,4}
	   \and
	  L. Fletcher \inst{1}
          }

   \institute{SUPA School of Physics and Astronomy, University of Glasgow, Glasgow, G12 8QQ, UK
   \email{p.levens.1@research.gla.ac.uk}
	 \and
	LESIA, Observatoire de Paris, PSL Research University, CNRS, Sorbonne Universit\'es, UPMC Univ. Paris 06, Univ. Paris Diderot, Sorbonne Paris Cit\'e, 5 place Jules Janssen, F-92195 Meudon, France
		\and
	IRAP - CNRS UMR 5277. 14, Av. E. Belin. 31400 Toulouse. France
		\and
	Universit\'{e} de Toulouse, UPS-OMP, Institut de Recherche en Astrophysique et Plan\'{e}tologie, Toulouse, France	
              %Goddard Space Flight Centre, 8800 Greenbelt Rd., Greenbelt, MD 20771, USA
              }
         %    University of Alexandria, Department of Geography, ...\\
         %    \email{c.ptolemy@hipparch.uheaven.space}
         %    \thanks{The university of heaven temporarily does not
         %            accept e-mails}
       %  CISL/HAO, National Center for Atmospheric Research, P.O. Box 3000,Boulder, CO 80307-3000, USA

   \date{Received ...; accepted ...}

% \abstract{}{}{}{}{} 
% 5 {} token are mandatory
 
  \abstract
% context heading (optional)
{Understanding the close relationship between the plasma and the magnetic field is important to describe and explain the observed complex dynamics of solar prominences.}
% aims heading (mandatory)
{We determine if a close relationship between plasma and magnetic field parameters measured in a well-observed solar prominence with high spatial resolution can be found.}
% methods heading (mandatory)
{We select a prominence observed on 15 July 2014 from space (\textit{IRIS}, \textit{Hinode}, \textit{SDO}) and from the ground (THEMIS). We perform a robust co-alignment of the data sets using a 2D cross-correlation technique. We derive the magnetic field parameters from spectropolarimetric measurements of the \ion{He}{i} D$_3$ line taken by THEMIS. Line ratios and line-of-sight velocities from the \ion{Mg}{ii} h and k lines observed by \textit{IRIS} are compared with magnetic field strength, inclination and azimuth. Electron densities are calculated using \textit{Hinode}/EIS \ion{Fe}{xii} line ratios and also compared with THEMIS and \textit{IRIS} data.}
% results heading (mandatory)
{We find \ion{Mg}{ii} k/h ratios of {  around 1.4 everywhere, similar to values found previously in prominences}. We also find that the magnetic field is strongest {  ($\sim$ 30 G)} and predominantly horizontal in the tornado-like legs of the prominence. The k$_3$ Doppler shift is found to be between $\pm$ 10 km s$^{-1}$ {  everywhere}. Electron densities at a temperature of $1.5 \times 10^6$~K are found to be {  around 10$^9$ cm$^{-3}$}. No significant correlations are found between the magnetic field parameters, and any of the other plasma parameters inferred from EUV spectroscopy, which may be explained by the large differences in the {  temperatures} of the lines used in this study.}
% conclusions heading (optional), leave it empty if necessary
{This is the first time that a detailed statistical study of plasma and magnetic field parameters has been carried out at high spatial resolution in a prominence. Our results provide important constraints on future models of the plasma and magnetic field in these structures.}

   \maketitle
%
%________________________________________________________________

\section{Introduction}

Solar ``tornadoes'' in prominences have recently been the focus of several papers debating their true nature. 
With the launch of the \textit{Solar Dynamics Observatory} and its high resolution Atmospheric Imaging Assembly \citep[\textit{SDO}, AIA;][]{Lemen2012} imager, several authors \citep{Su12,Li2012,panesar13,Wedemeyer2013} noted tornado-like structures in the solar atmosphere. 
These were seen as dark columns of apparently-rotating material, absorbing background coronal emission. 
Some authors \citep[e.g.][]{Panasenco14} argued that the observed motions can be interpreted as oscillations in the plane of the sky. 

More recent observations with spectroscopic instruments have allowed measurements of the line-of-sight velocities and several plasma diagnostics in these structures. 
\citet{Su2014} and \citet{Levens2015} used the Extreme-ultraviolet Imaging Spectrometer \citep[EIS;][]{Culhane07} on the \textit{Hinode} satellite \citep{Kosugi2007} to measure Doppler velocities in a tornado, finding an anti-symmetric pattern along the axis of the column at plasma temperatures above 10$^6$ K. 
This pattern has also been seen in H$\alpha$ \citep{Wedemeyer2013} and \ion{He}{i} 10830~\AA\ \citep{Orozco2014}, {  however it is unclear whether these observations are really showing rotation}. 
Other observations have cast doubt on the presence of long-term rotational patterns -- \cite{2017A&A...597A.109S} show that Doppler patterns suggestive of rotational motions over a short period of time disappear on longer time scales. 
These authors detected changing patterns of Doppler shifts, {  indicating that we are seeing oscillations with periods on the order of 40 to 60 minutes.}%with 4-6 minutes oscillation periods on top of longer periods of oscillations.

%{\bf \citet{Schmieder2017} studied a prominence using the \textit{Interface Region Imaging Spectrograph} \citep[\textit{IRIS};][]{DePontieu2014}, demonstrating the instruments capabilities for studying prominence dynamics. 
%In that prominence they observed fast moving `blobs' of plasma, and were able to reconstruct the 3D shape of the prominence using \textit{IRIS} spectra and slit-jaw images. 
{\citet{Schmieder2017} used the high spectral and spatial resolution of the \textit{Interface Region Imaging Spectrograph} satellite \citep[\textit{IRIS};][]{DePontieu2014} to reconstruct the 3D trajectories of plasma `blobs' in a helical prominence.
Despite the plasma tracing highly curved paths in the plane of the sky, it was found that the actual trajectories of the plasma was along horizontal paths, indicating that the apparent plasma motion can differ significantly from the real motion.}

Some work has also now been done on measuring the magnetic field in tornado-like prominences \citep{Levens2016,Levens2016b,Martinez2016}. 
\citet{Levens2016,Levens2016b} found that the field in these tornadoes is largely horizontal, parallel to the limb, with field strengths of between 10 and 50 G. 
This does not support the twisted models suggested in \citet{Su2014} or as modelled by \citet{Luna2015}, rather suggesting the plasma is in a horizontal field with dips over parasitic polarities on the solar surface \citep{Aulanier1998}. 
\citet{Martinez2016} find a mix of possible field orientations, which they interpret as evidence for a twisted field. 

The link, however, between the magnetic field and the plasma parameters that are measurable remains unclear. 
In this work we aim to study the magnetic field parameters, as measured by the T\'elescope H\'eliographique pour l'Etude du Magn\'etisme et des Instabilit\'es Solaires (THEMIS) instrument in the Canary Islands, in comparison with plasma parameters from the \ion{Mg}{ii} h and k lines, measured by \textit{IRIS}.
We study the magnetic field strength and orientation, and look for correlations with characteristics derived from \ion{Mg}{ii} line profiles (velocities, optical thickness, intensity ratios, amongst others). 

We introduce the observations and the instruments used in Section \ref{sec:observations}, as well as outlining the diagnostic parameters that can be derived from these observations. 
Section \ref{sec:coalignment} details the methods for co-aligning the various data sets that are used. 
In Sect.~\ref{sec:correlation} we present our results and correlation plots between parameters available from \textit{IRIS} and THEMIS. 
Section \ref{sec:eis_correlation} contains results and correlations between the electron density, measured by \textit{Hinode}/EIS and magnetic field parameters from THEMIS. 
We also look for correlations between EIS and \textit{IRIS}, which is shown in Sect.~\ref{sec:iris_eis}.
Section \ref{sec:conclusion} includes our concluding remarks on this work.
%Section \ref{sec:discussion} is our discussions on these results, and Section \ref{sec:conclusion} houses our concluding remarks on this work.%is our discussions on these results, and Section \ref{sec:conclusion} houses our concluding remarks on this work.

%__________________________________________________________________

\section{Observations and diagnostic parameters available}
\label{sec:observations}

\begin{figure*}
\begin{center}
\includegraphics[width=0.45\hsize,clip=true,trim=0.9cm 0 2.5cm 0]{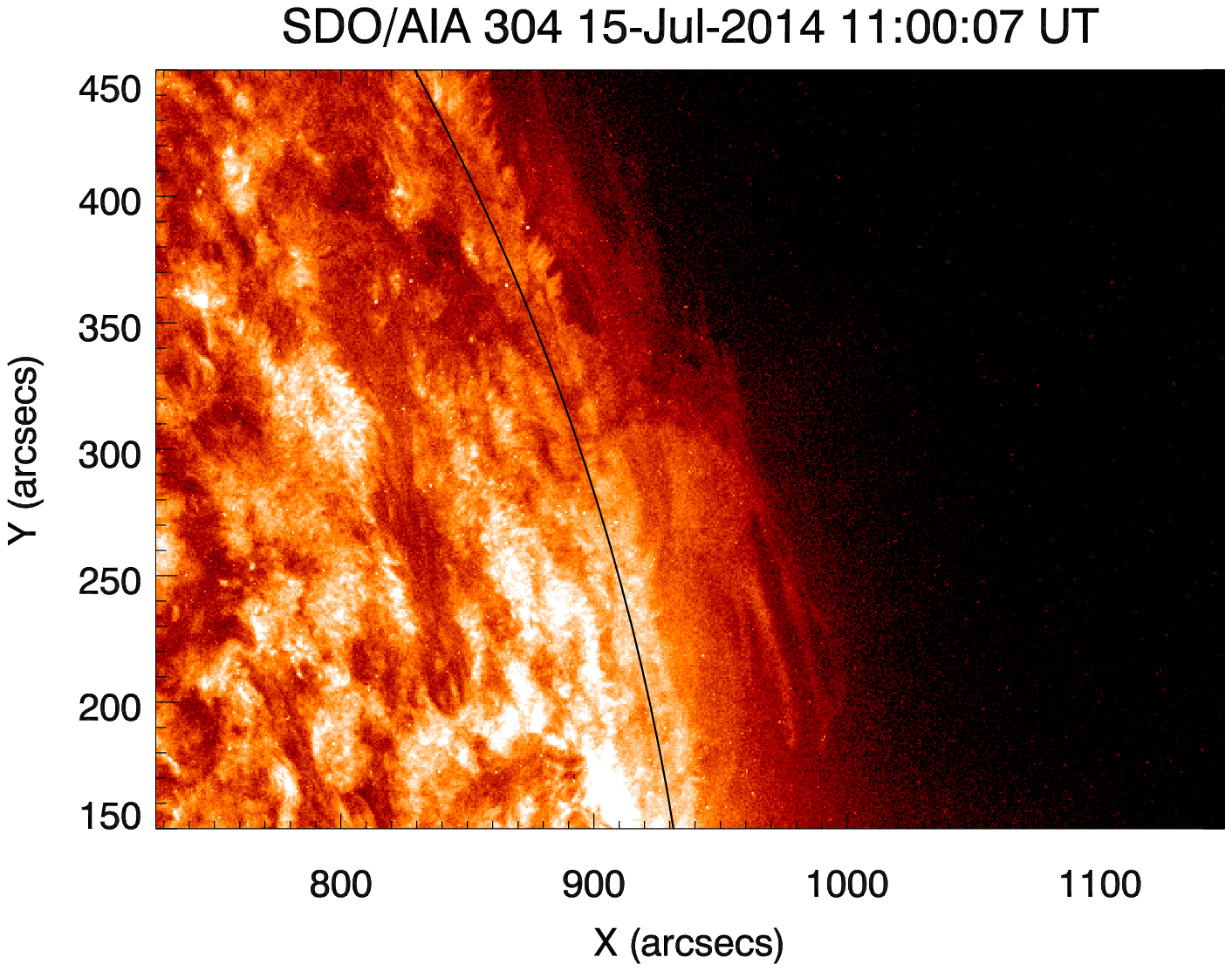}
\includegraphics[width=0.45\hsize,clip=true,trim=0.9cm 0 1.8cm 0]{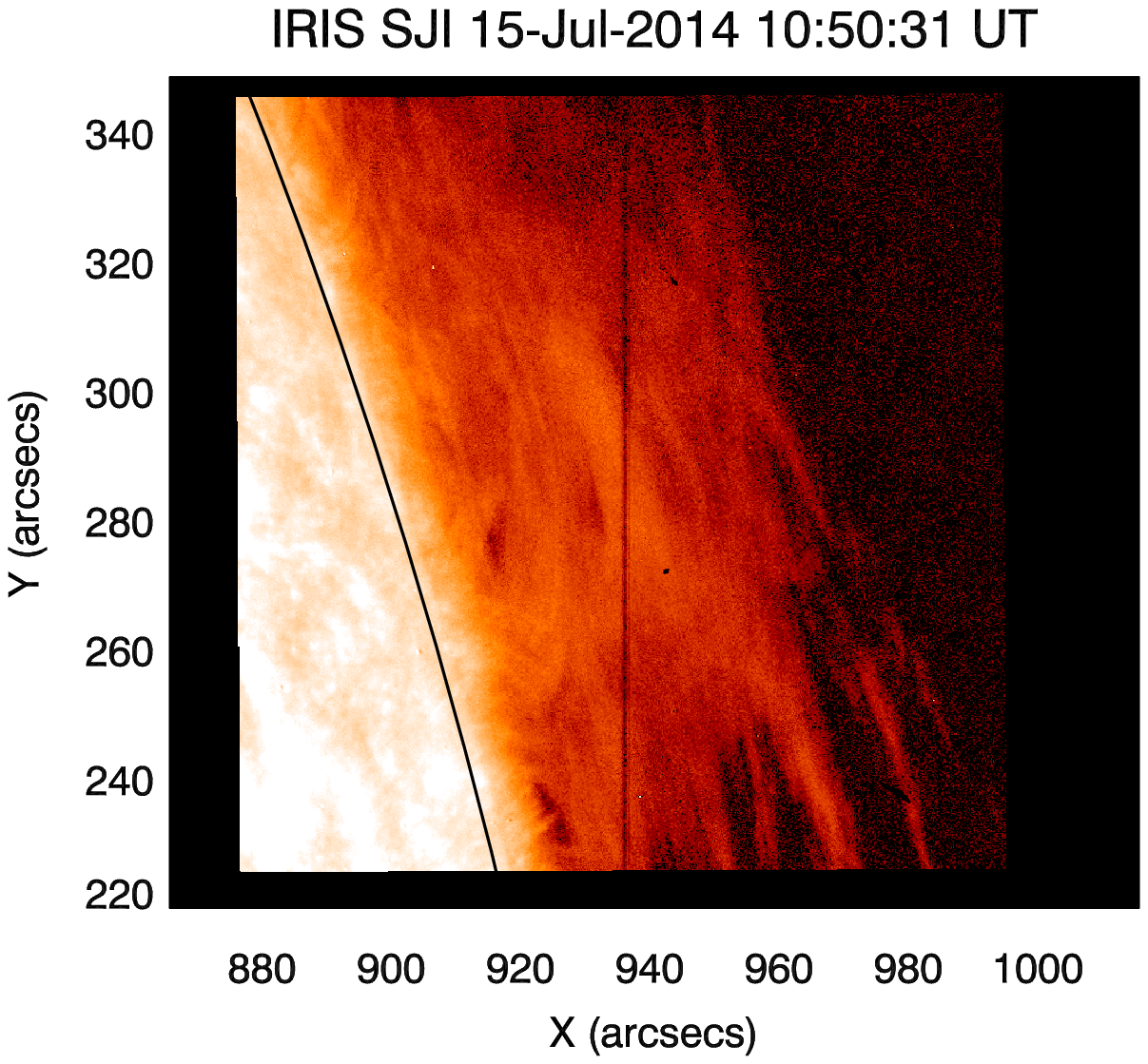}
\\
\includegraphics[width=0.45\hsize,clip=true,trim=0.9cm 0 1.8cm 0]{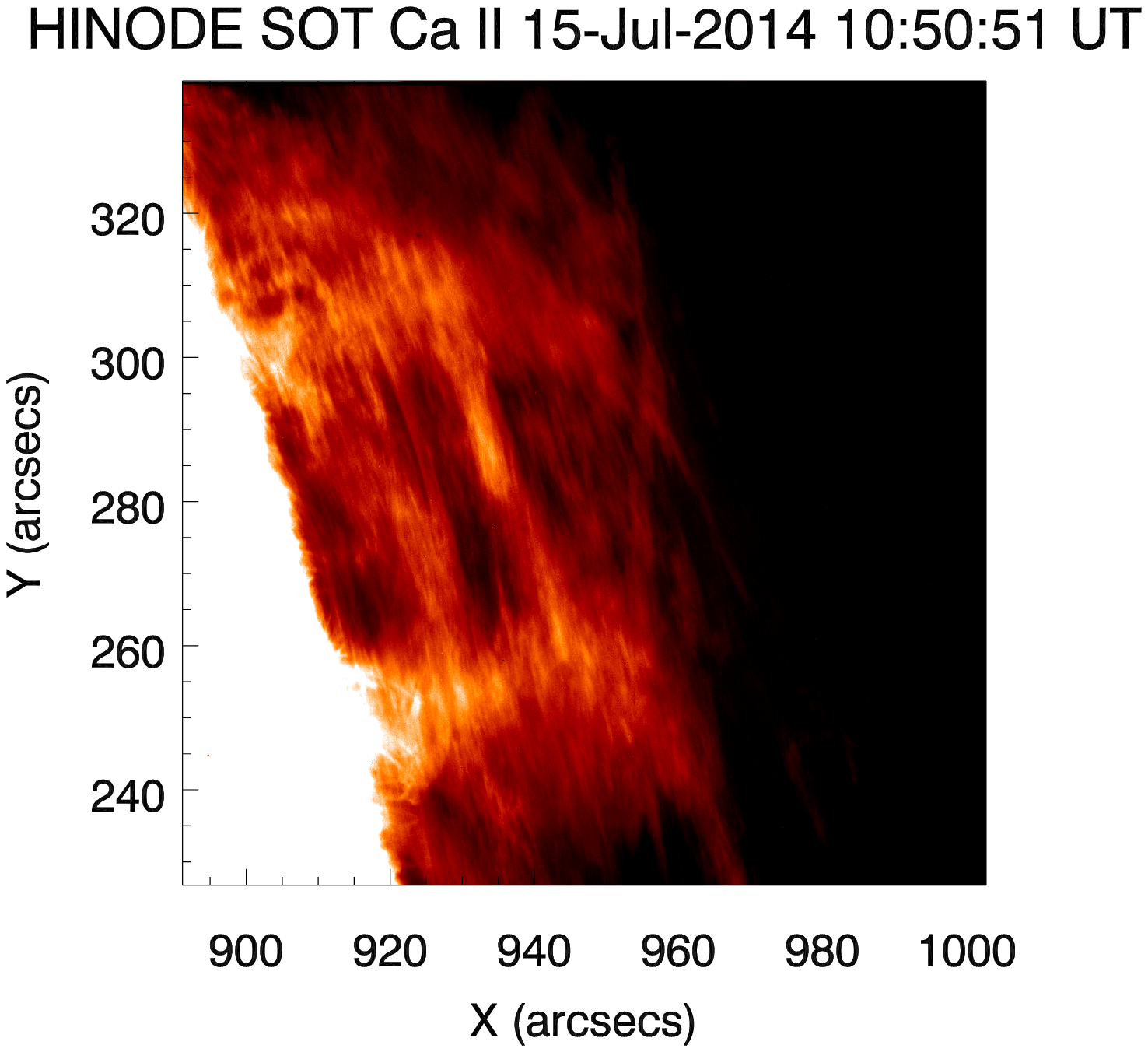}
\includegraphics[width=0.45\hsize,clip=true,trim=0.9cm 0 1.8cm 0]{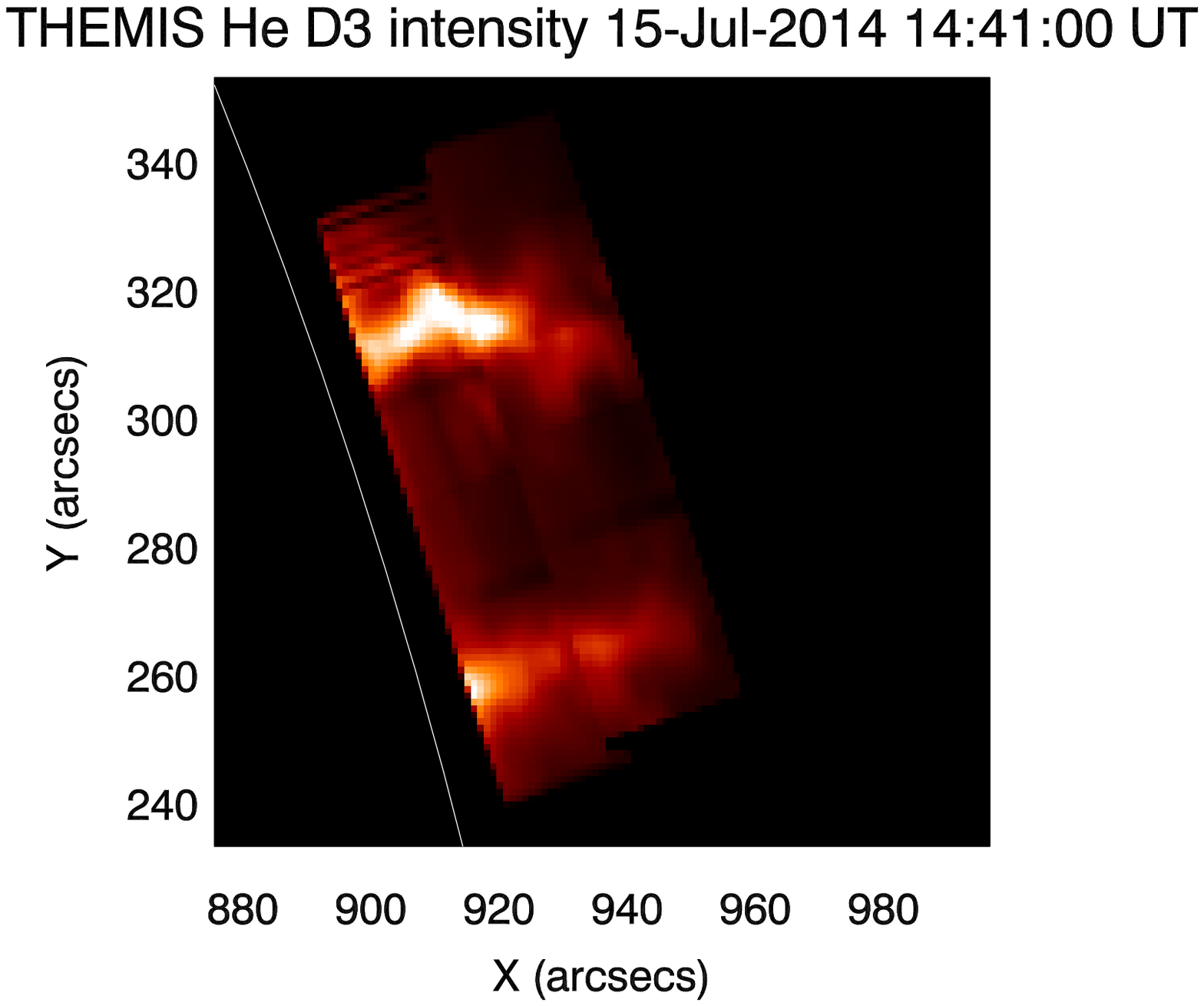}
\caption{Observations by several instruments on 15 July 2014. \textit{Clockwise from upper left:} AIA observation in 304~\AA. \textit{IRIS} SJI map using the \ion{Mg}{ii} filter. THEMIS map, made using the two rasters obtained on that day in the \ion{He}{i} D$_3$ line (rotated and aligned with the SOT image). \textit{Hinode} SOT map in \ion{Ca}{ii}.}
\label{fig:aia}
\end{center}
\end{figure*}

\begin{figure}
\begin{center}
\includegraphics[width=\hsize,clip=true,trim=0.9cm 0 2.5cm 0]{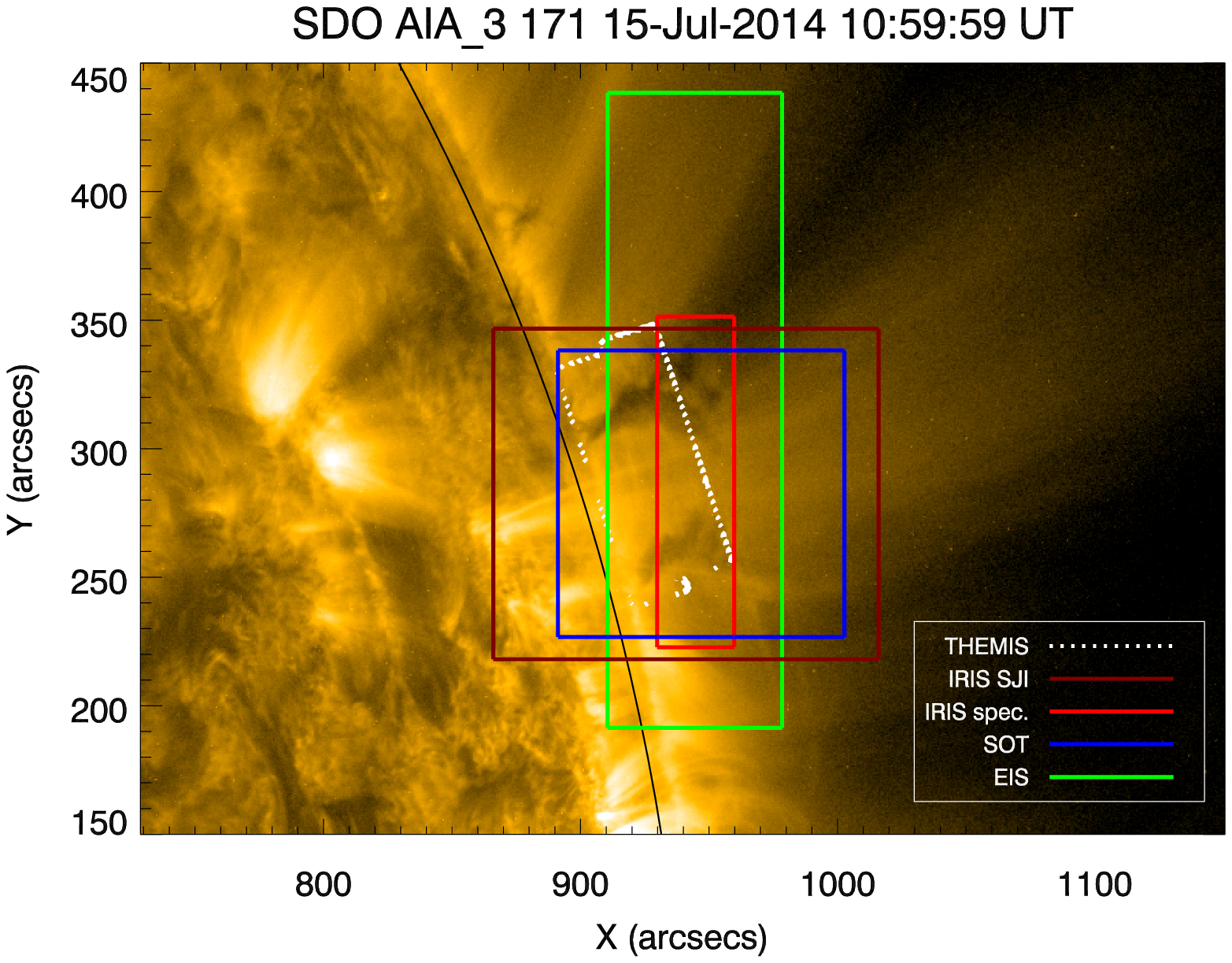}
\caption{AIA observation from 15 July 2014 in 171~\AA. The coloured boxes indicate the fields of view of the other instruments used in this study.}
\label{fig:aiaboxed}
\end{center}
\end{figure}

The prominence observed on 15 July 2014 above the western solar limb was seen brightly in emission by AIA in the 304~\AA\ passband (Fig.~\ref{fig:aia}), whereas in the coronal filters (e.g. 171~\AA, Fig.~\ref{fig:aiaboxed}) we see two columns of dark material, which are absorbing the background emission. 
The emission seen in {  the AIA image of} Fig.~\ref{fig:aia} is dominated by the \ion{He}{ii} doublet at 303.78~\AA, which are optically thick emission lines, formed mostly by resonant scattering of emission from the solar surface below the prominence \citep{Labrosse2001}. 
The 171~\AA\ images from AIA, however, are dominated by emission from \ion{Fe}{ix}, formed at around 1 MK. 
The dark columns are comprised mostly of cooler material, hydrogen and helium, that absorbs the emission coming from the corona behind.
These columns have been observed to display {  oscillatory} behaviour when viewed over time \citep{Su2012,Su2014,Levens2016}, {  which has been interpreted as tornado-like rotation,} so here we refer to them as ``tornadoes''.

Plasma parameters for this prominence were explored in \citet[][herein Paper I]{Levens2016}, but here we aim to take a deeper look at the diagnostics available. 
Full details of the observations and more about the dynamics of the prominence are covered in Paper I.

%{\bf In a recent piece of work Schmieder et al. (2017, \textit{in review}) looked at a prominence in the same region, but two days after the observation considered here. 
%This 17 July prominence appeared quite differently to the one on 15 July, despite apparently being part of the same prominence complex. 
%On 17 July the prominence showed looping threads above a main body, which are in fact horizontal structures that appear looped due to the projection effect. 
%This is quite different dynamically to the prominence on 15 July, as there was not the same absorption features in the AIA movies on the 17 July.}

%{\citet{Schmieder2017} studied a prominence observed two days later, also at the north-west limb. 
%The two prominences belong to two different magnetic systems and they display very different behaviour. 
%On July 17 their prominence appeared much more active than the July 15 prominence, exhibiting fast moving `blobs' of plasma observed by \textit{IRIS}. 
%They reconstructed the 3D shape of the prominence observed over two and a half hours using \textit{IRIS} spectra and slit-jaw images.}

\subsection{\textit{IRIS}}

\subsubsection{Observations}
The \textit{IRIS} observations from the 15 July 2014 consisted of  16-step coarse rasters between 08:00 UT to 11:00 UT. 
Spatial {scale} is 0.167\arcsec\ in y, {with an actual spatial resolution of around 0.35\arcsec, and} a step size of 2\arcsec\ in x. 
Exposure time was 5.4 s per slit position, giving a raster cadence of 86.4 s.
These rasters contained both near-ultraviolet (NUV, 2783--2834~\AA) and far-ultraviolet (FUV, 1332--1348~\AA\ and 1390--1406~\AA) lines. 
In this paper we focus on the NUV channel, namely the \ion{Mg}{ii} h and k lines at 2803.5~\AA\ and 2796.35~\AA\ respectively. 
{Raw data is calibrated to level-2 \citep{DePontieu2014}, where dark current subtraction, geometrical correction, and flat field correction have each been accoundted for.}

{Slit-jaw images (SJI) were taken in the broadband filters centred on 2976~\AA\ (\ion{Mg}{ii}) and 1330~\AA\ (\ion{C}{ii}). 
The cadence for SJI was 11 s, and the FOV covered 119\arcsec\ $\times$ 119\arcsec.}

\subsubsection{Plasma diagnostics}
\label{ssec:iris_diagnostics}
\begin{figure}
\begin{center}
\includegraphics[width=0.9\hsize]{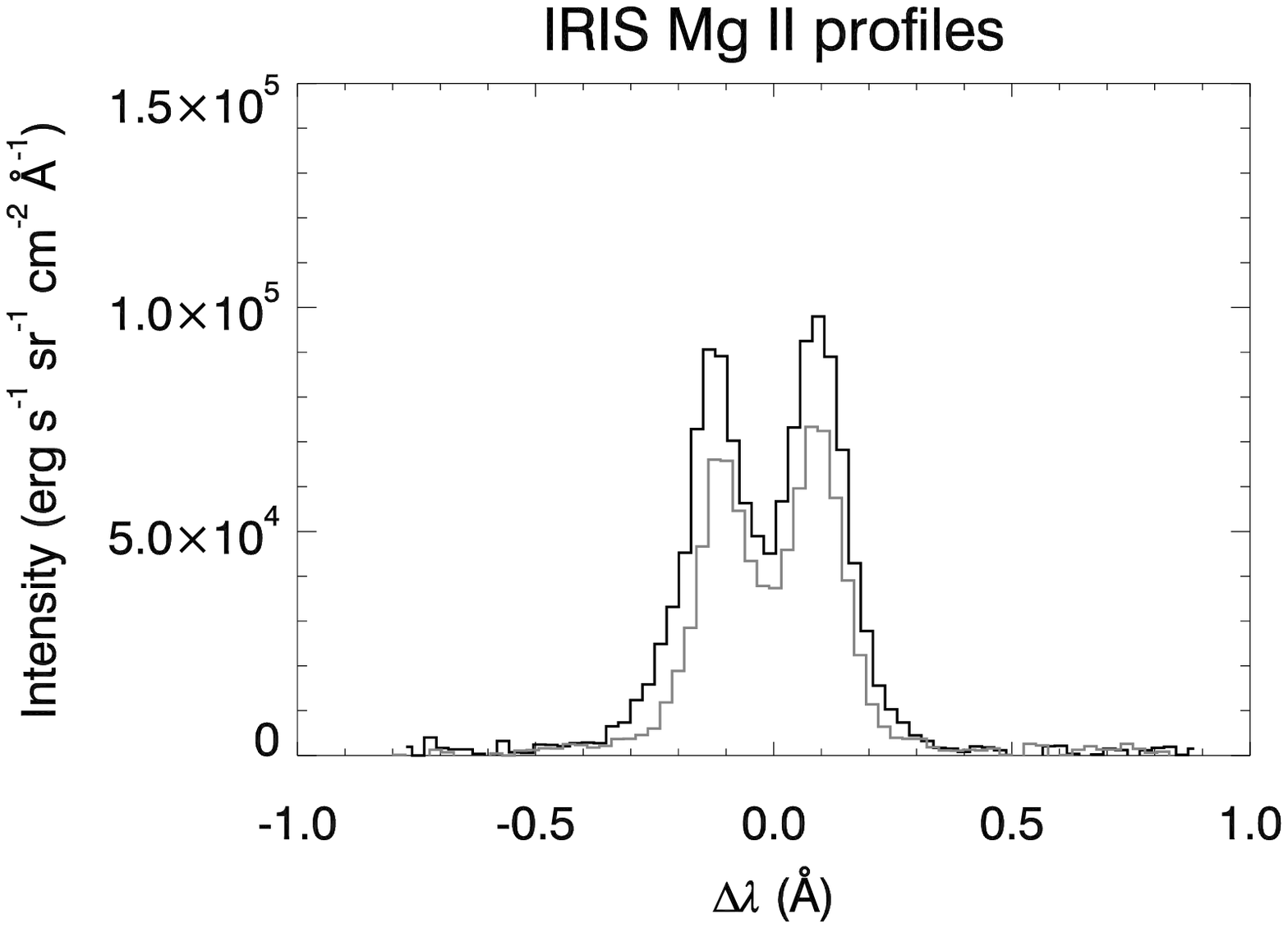}
\includegraphics[width=0.9\hsize]{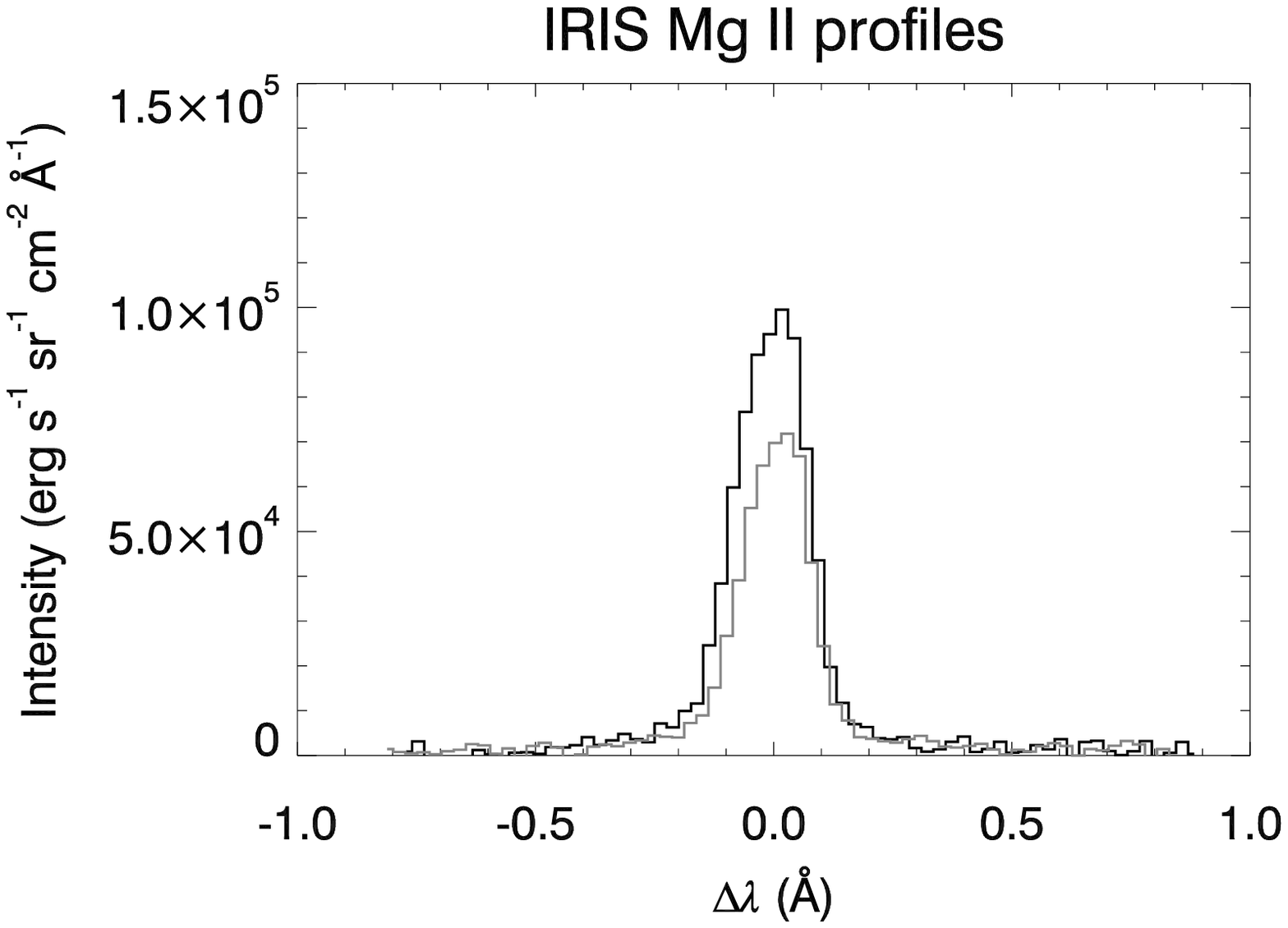}
\includegraphics[width=0.9\hsize]{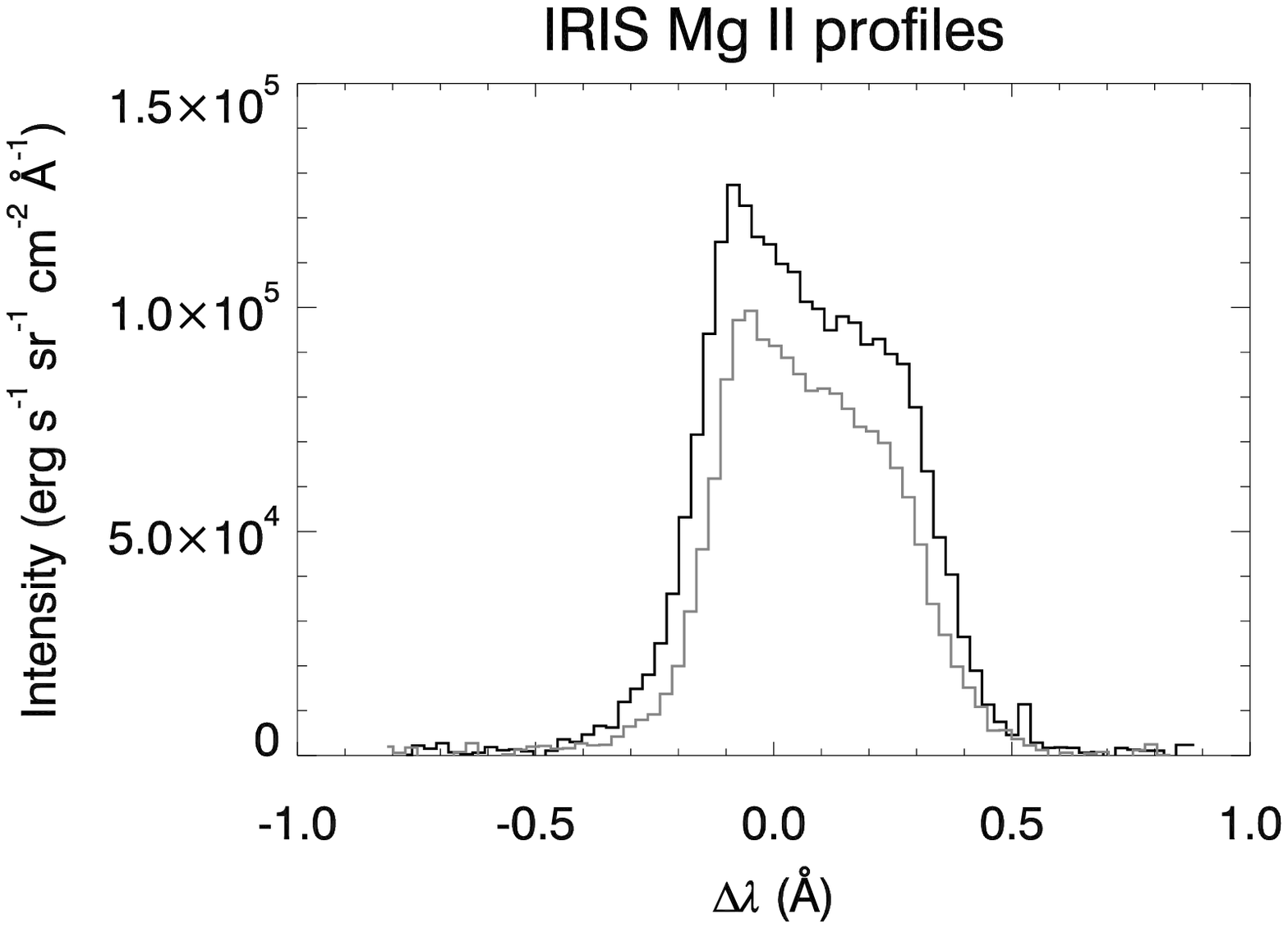}
\caption{Example \ion{Mg}{ii} k and h line profiles from the prominence of 15 July 2014. Each is taken from the raster that started at 10:21 UT. Shown are both k (black lines) and h (grey lines) profiles. \textit{Top:} Reversed profiles. \textit{Middle:} Single-peaked profiles. \textit{Bottom:} Complex profiles.}
\label{fig:profiles}
\end{center}
\end{figure}

The  \ion{Mg}{ii} h and k lines are optically thick in prominences, evident from observed centrally reversed profiles (Paper 1). However they are not always reversed \citep{Vial1982,Schmieder2014,Vial2016}, often showing non-reversed, single-peaked profiles. 
This behaviour can be explained by different physical conditions in different regions of the prominence \citep{Heinzel14,Heinzel2015}. 
In the prominence observed on 15 July 2014 we find a mixture of reversed and non-reversed profiles, see Fig.~\ref{fig:profiles} and Paper I.

As is usually done, we call an \ion{Mg}{ii} line profile \emph{reversed} when there are two {  distinct} peaks in the line, known as the k$_2$ or h$_2$ peaks for the k and h lines respectively, surrounding a region of lower intensity (the line core, known as k$_3$ and h$_3$ respectively). 
Typical reversed profiles can be seen in Fig.~\ref{fig:profiles} (top). 
By \emph{single-peaked} profiles, we refer to line profiles showing only one peak which is close to the line centroid (as would be determined, for instance, by an approximate Gaussian fit). 
Typical examples of single-peaked profiles are shown in the middle panel of Fig.~\ref{fig:profiles}. 
A large fraction of observed profiles do not fit in any of these two categories however, and we therefore call them \emph{complex} profiles (Fig.~\ref{fig:profiles} bottom). 
Complex profiles may have no clear central reversal, or a broad, flat line core, or may have just one peak but away from the line centre (at a distance $>0.04$~\AA), or show several intensity peaks. 

The existence of such complex profiles has {  two} possible explanations. 
The first is as in \citet{Schmieder2014}, where {  numerous threads with different} velocity components along the line of sight create multiple-component profiles. 
{  In this scenario, single-peaked Gaussian profiles blend together when observed by \textit{IRIS}.}
%As was discussed in Paper I, our prominence shows relatively low velocities in the \ion{Mg}{ii} lines (less than $\sim$10 km s$^{-1}$), so we would {\bf perhaps} not expect large Doppler shifted components. 
%However, the analysis in Paper I focussed on the overall line profile, using single Gaussian fits to characterise the centroid position, but this approach cannot account for the {\bf possibility of} multiple components consisting of  narrow Gaussian profiles. 
{  \citet{Schmieder2014} found, using multiple narrow Gaussians, that there were observable flows of up to 80 km s$^{-1}$ in a quiescent prominence.} 
%We therefore cannot rule out this possibility for the complex profiles. 
A second possibility is that of a reversed profile with one of the peaks missing. 
{  This could be due to seeing multiple optically thick threads with different velocity components along the line of sight. 
This model was explored by \citet{Gunar2008} to explain asymmetrical Lyman line profiles and by \citet{Labrosse2016} for helium lines in prominences. 
It has been shown that a combination of a number of reversed profiles with different line-of-sight velocities can create an emergent reversed line profile with one peak missing \citep{Gunar2008}. }
%Another mechanism for removing one of the peaks is by assuming that the reversal minimum position is shifted relative to the two maxima, therefore the emission in one of the peaks is absorbed. }
This could occur due to the optical thickness difference across the \ion{Mg}{ii} h and k line profiles, and the differences in where the component parts of those profiles  are formed in the prominence, under the assumption of a multi-thread model. % \citep{Labrosse2016}. 
%We consider a situation where, for the k line shown in Figure \ref{fig:profiles} (bottom), the line wings and the k$_2$ peaks are formed under the optically thin regime, meaning an integration along the line of sight. 
%The k$_3$ reversal is then formed at the surface closest to us. 
%As seen in Paper I and \citet{Heinzel14} the reversal levels seen in this prominence reveal that the line core is very optically thick. 
The complex profiles could then be explained by `stationary' line wings and k$_2$ peaks resulting from an integration along the line of sight of several threads, and a shifted k$_3$ core due to the motion of the frontmost thread. 
This, then, absorbs emission in one of the k$_2$ regions, reducing the emission from that peak. 
%{\bf For this prominence we prefer the final solution, with optically thick plasma in the front part of the prominence absorbing one of the peaks, however the other options have their own merits. 
{  In this data set there do not appear to be highly Doppler shifted profiles, as was discussed in Paper I, so we do not believe that these profiles are a blend of multiple line-of-sight components. 
Also discussed in Paper I was the high optical thickness of the \ion{Mg}{ii} resonance lines in this prominence, apparent from comparing the levels of reversal seen to the models of \citet{Heinzel14}. 
This points towards the second scenario as the most likely explanation for the `complex' profiles seen here. 
A further investigation into these profiles may be useful, as they appear to be common in prominences. }
{  Regardless of the formation mechanism}, we need to characterise these profiles in order to use them in this analysis. 
%We here consider the second scenario, as the majority of the prominence appears to be optically thick, meaning we should not see a number of velocity components along the line of sight. 

To make an estimate of the level of reversal in the \textit{complex} profiles we make the assumption that the centroid of the line, calculated from the moments of the distribution, is the k$_3$ location, and the maximum intensity of the profile corresponds to the k$_2$ peak. 
We use an algorithm presented in Waller et al. (2017, \textit{in prep.}) to automatically measure the \ion{Mg}{ii} line profile characteristics.
Complex profiles are found automatically by comparing the centroid of the distribution to the position of the peak of the distribution. 
For a truly single-peaked profile, these two values should not be more than a few m\AA\ apart. % \textbf{CHECK VALUE}. 
However, for a complex profile they are more separated. 
We set a limit of 0.04~\AA\ on the distance between the positions of the line centroid and the peak for a profile to be classed as complex.

\subsection{THEMIS}

\subsubsection{Observations}

THEMIS ran two raster studies using the MulTi Raies (MTR) spectrograph \citep{LARS00} on 15 July 2014. 
%{\bf These rasters took 30 minutes to complete}
The first raster started at 14:41 UT {and} the second at 16:55 UT, {each taking 30 minutes to complete}, putting these rasters after the observations made by \textit{IRIS} and \textit{Hinode}. 
However, the conditions of the prominence and tornadoes did not change significantly in the interval between the sets of observations. 
Therefore, we can assume that parameters obtained in the morning are not dissimilar to those obtained in the afternoon by THEMIS, but we must keep in mind  that there is this temporal discrepancy between the space-based and ground-based observations.
In both rasters the slit of the spectrograph was orientated parallel to the limb, and each consisted of 30 slit positions separated by 2\arcsec. 
{Original spatial pixel size is 0.23\arcsec, but }
data used here has been binned to have square pixels of 1\arcsec $\times$ 1\arcsec. 
{Exposure time is 1.5 s per Stokes parameter, with 6 Stokes parameters per cycle, and 10 repeats of each cycle at each slit position to increase SNR. }
Paper I contains {further} details of the study. 

The THEMIS MTR instrument is a spectropolarimeter, giving observations of the four Stokes parameters (\textit{I}, \textit{Q}, \textit{U}, and \textit{V}) using the \ion{He}{i} D$_3$ line. 
The data was handled using the DeepStokes procedure \citep{LAARMSDG09}, then the Stokes profiles were {  treated} using the code of \citet{LAC02} and \citet{Casini2003} which is based on Principal Component Analysis. 
After this treatment, the resulting observed profiles are compared to those in a model database of over 90000 profiles, generated for \ion{He}{i} accounting for both the Hanle and Zeeman effects \citep{LAC02}. 
The most similar profile in the database to the observed profile is taken to be the solution, which gives us information about the magnetic field strength and orientation at each pixel. 
Again, more details are found in Paper I and references therein.

\subsubsection{Plasma Diagnostics}

\begin{figure}
\begin{center}
\includegraphics[width=\hsize,clip=true,trim=0.9cm 0 1.0cm 0]{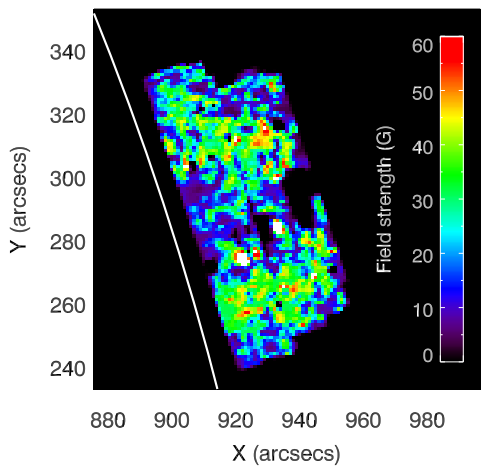}
\caption{Magnetic field map for the prominence on 15 July 2014, calculated using the \ion{He}{i} D$_3$ line from THEMIS. Image is made using both rasters from that day, one starting at 14:41 UT and the second starting at 16:55 UT. Solar limb position is shown in white.}
\label{fig:themis_bfield}
\end{center}
\end{figure}

The main parameters that THEMIS provides are the magnetic field strength and the orientation of the field, namely the field inclination and azimuth. 
We also have the intensity image in \ion{He}{i} D$_3$ {(Fig.~\ref{fig:aia}, \textit{lower right})}, which gives us information about the spatial structure of the prominence in that wavelength. 
{Figure~\ref{fig:themis_bfield} shows the magnetic field map for the tornadoes on 15 July 2014. }

\subsection{\textit{Hinode}}

\subsubsection{Observations}

\textit{Hinode} was observing co-temporally with \textit{IRIS}, with both the Solar Optical Telescope \citep[SOT;][]{Suematsu2008,Tsuneta2008} and EIS aquiring data. 

During this study, SOT observed using the \ion{Ca}{ii} H filter centred at 3968.5~\AA, with images taken with a 30 s cadence and a pixel size of 0.109\arcsec. 
SOT observed from 10:21 UT until 11:06 UT, covering the last part of the \textit{IRIS} observing time. 
{The SOT FOV was 112\arcsec\ $\times$ 112\arcsec\ and is shown as a blue box in Fig.~\ref{fig:aiaboxed}. 
Raw SOT data is calibrated to level-1 using the standard SSW routine (\texttt{fg\_prep}).}

EIS ran a 35 step raster using the 2\arcsec\ slit, covering a 70\arcsec\ $\times$ 248\arcsec\ {area (green box, Fig.~\ref{fig:aiaboxed})}. 
The study used was \texttt{madj\_qs}, which has an exposure time of 50 s, meaning that a full raster is made in around 30 minutes. 
Only one raster was achieved during the observing time, starting at 10:34 UT and ending at 11:06 UT. 
\texttt{madj\_qs} observes eleven wavelength channels in both the long-wavelength CCD and short-wavelength CCD of EIS, allowing access to spectral lines formed at a range of plasma temperatures from around $8 \times 10^5$ K to more than $2 \times 10^6$ K. 
{EIS data is acquired at level-0, and hot/warm pixels, dark currents and cosmic-ray hits are removed using the SSW routine \texttt{eis\_prep}. 
This routine also perfoms absolute calibration of the data, leaving it ready for analysis at level-1. }

{In order to isolate spectral lines from EIS, Gaussian functions are fitted. 
This allows us to derive plasma parameters at each pixel for each line, creating a series of maps using each Gaussian fit. }

\subsubsection{Plasma Diagnostics}

SOT brings us intensity images for a spectral window around the \ion{Ca}{ii} H line. 
These are useful as they provide another optically-thick intensity image that can be used for co-aligning data sets. 

EIS provides spectra for a large number of EUV spectral lines from  ionised elements. 
The plasma emitting these lines is mostly optically thin and at coronal temperatures, however some cooler, optically thick lines are also observed, such as \ion{He}{ii} 256~\AA. 
Optically thin lines {are} fitted with Gaussian profiles. 
This approximation allows us to derive  plasma properties from the Gaussian parameters, such as line-of-sight velocities and line FWHMs, as well as electron densities when density-sensitive pairs of lines are observed together. 
This kind of analysis was done on a solar tornado {  observed by} EIS \citep{Su2014,Levens2015}, where velocities of less than 5 km s$^{-1}$ were found in the tornado, and a split Doppler pattern across the tornado led the authors to conclude that the tornado is rotating around a central axis. 
\citet{Levens2015} also presented density diagnostics, finding electron densities, $n_e$, of around $10^9$ cm$^{-3}$ noting that there appears to be lower electron density in the tornado than in the surrounding corona at a temperature of $1.5 \times 10^6$ K. 
The analysis done previously on tornadoes with EIS has mostly focused on coronal temperature plasma, as the cooler lines observed by EIS are often difficult to interpret. 
For example, the \ion{He}{ii} line is heavily blended with  lines formed at a higher temperature, making it difficult to distinguish the cooler component. 
%Here we approximate this blend as being made up of two Gaussian components - one for the cool plasma and one for the hot plasma. 
%Although not perfect, this is a good enough approximation to get intensity images of the \ion{He}{ii} emission which can be used for co-aligning the EIS data with other data sets. 
%The other Gaussian component cannot be used for any analysis, as there are still  unresolved lines within it, each formed at different plasma temperatures. 

%__________________________________________________________________

\section{Co-alignment of data}
\label{sec:coalignment}

The aim of this work is to do a statistical analysis on a pixel by pixel basis, so we require that the data sets from different instruments are well aligned. 
%For the data on 15 July 2014, we noted that the pointing from \textit{Hinode} was not correct. 
%The pointing data from the FITS header for both SOT and EIS were noticably wrong -- Apparent limb positions were far from the visible solar limb in SOT images. 
%This discrepency was found to be around 45\arcsec\ in the y direction and around 10\arcsec\ in the x direction. 
%We need a more accurate method for aligning the data in order to perform a robust analysis on it. 
Here we use a 2D cross-correlation method making use of a Mean Absolute Difference (MAD) algorithm in the \texttt{get\_correl\_offsets} SSW routine. 
To {cross-correlate data sets with different spatial resolutions, it is necessary to} reduce the spatial resolutions to that of the lowest resolution image by binning the data. 
{This re-binning is done individually for each pair of images to be co-aligned. }
The correlations used are described in detail below.

In this analysis we concentrate on one {image} from each of the data sets. 
Due to the discrepancies in observing time between space-based and ground-based observatories on this day, we unfortunately cannot have well temporally-aligned data. 
However, {from AIA images/movies,} the prominence and tornadoes do not change much over the course of the day. 
{Differential rotation does not have any influence on a prominence over the course of a few hours, so we do not expect it to affect the prominence observed here.}

\subsection{\textit{IRIS} with \textit{SDO}/AIA}

For the co-alignment of the data sets we take the AIA 304~\AA\ image from 11:00 UT as the base image (Fig.~\ref{fig:aia} top left panel), {to line up with the end of the observing time of \textit{IRIS} and \textit{Hinode}}. 
All other data sets will eventually be aligned with this image. 
The prominence appears as an extended structure in both AIA 304~\AA\ images and \textit{IRIS} \ion{Mg}{ii} images, with horizontal fine structure. 
The similarity is due to the fact that the (dominant) \ion{He}{ii}  emission line seen in the AIA 304~\AA\ passband and the \ion{Mg}{ii} h and k lines are both extremely optically thick, with $\tau_{Mg} \sim \tau_{He}$ (Paper I). %have similar optical thicknesses, giving the prominence a similar appearance in all lines formed under that regime. 

Figure \ref{fig:aia} (top right panel) shows the \textit{IRIS} SJI from 10:51 UT. 
\textit{IRIS} passed through the South Atlantic Anomaly at the end of its observing time, so the data after this time was unusable.
We use data from the end of the observing time for the space-based telescopes so as to minimise the time between these data and that from THEMIS, which observed in the afternoon. 

%Passing both the AIA image and SJI to the cross-correlation routine brings the two data sets in line. 
{  We cross-correlate the AIA and SJI data, finding} that an offset of $x$ = 0.81\arcsec\ and $y$ = -1.65\arcsec\ needs to be applied to the \textit{IRIS} SJI to bring it in line with the AIA image. 
%This was found to be the average offset over the time that \textit{IRIS} observed, and was applied to all time steps. 
%It turns out that only a small offset is required to align AIA and \textit{IRIS}, but we still perform the correlation in order to have all data sets co-aligned in a consistent manner. 

\subsection{\textit{Hinode}/SOT with \textit{IRIS}}

The \ion{Ca}{ii} H line observed by SOT is an optically thick emission line. 
However, it has been seen that the H and K lines of calcium are not as optically thick as the h and k lines of magnesium. 
This is evident from Fig.~\ref{fig:aia} (\textit{IRIS} and SOT images), where we see more of the column-like structure in the prominence in the SOT image, but not in the \textit{IRIS} image. 
We note, however, that the SOT images largely still show the horizontal structure common to both \textit{IRIS} and AIA 304~\AA\ images, so we can use these images for the co-alignment of the data set from SOT. 
{  The \textit{IRIS} SJI data has high spatial resolution so it is used to co-align the SOT data. }
%Due to the \textit{IRIS} SJI data being of higher spatial resolution than that of AIA, it was the \textit{IRIS} SJI that was used to perform the alignment of SOT images. 
%This results in the least amount of spatial binning of the SOT data possible. 

%\begin{figure}
%\begin{center}
%\caption{}
%\label{fig:iris}
%\end{center}
%\end{figure}
%
%\begin{figure}
%\begin{center}
%\caption{}
%\label{fig:sot}
%\end{center}
%\end{figure}

The cross-correlation routine was run on the SOT images and {  the resulting} offset of $x$ = 11.22\arcsec\ and $y$ = 45.64\arcsec\ was applied to the SOT images. %\textbf{NL: This is Huge... Peter, are you sure it's in arcsec and not in pixels?}
%As with the \textit{IRIS} offsets, this is the average SOT offset found over the entire time that \textit{IRIS} and \textit{Hinode} were observing co-temporally. %after a manual correction of around 45\arcsec\ in y and 10\arcsec\ in x was applied, as mentioned previously. 
%After these two changes, the SOT images became well aligned with \textit{IRIS}, which itself is aligned with AIA. 
%We therefore also have SOT and AIA images well aligned.

\subsection{THEMIS with \textit{Hinode}/SOT}

Aligning THEMIS with the other instruments is more of a challenge than for other data sets, due to  the way THEMIS observes. 
THEMIS observations of prominences are generally done with the slit orientated parallel to the solar limb at a certain position angle (PA), measured anti-clockwise from solar north, and the first slit position is somewhere near the limb. 
This means that we do not have a traditional `pointing' for the THEMIS maps, and we must make sure that the map is orientated properly before performing any further alignment. 
To begin with, the THEMIS data is orientated with the slit parallel to the {  solar limb, and the slit along the x-axis of the resulting raster.} %, meaning that the solar limb is also parallel to the x-axis.
On the 15 July 2014 THEMIS made two such images, one at the bottom of the prominence and one near the top. 
These have been spliced together, as the offset between them in y (height above limb) was known. 
They are then taken to be one image for the day in this analysis. 
To orientate these images we rotate the THEMIS map by 360$^\circ$ minus the PA at which the prominence was observed, which on the 15 July 2014 was PA = 288$^\circ$, giving an image roll of 72$^\circ$. 
%{\bf A rough co-alignment is carried out by eye, and then the images are passed to the cross-correlation routine to perform the alignment in detail. }
%Due to the fact that there is no pointing information in the THEMIS header, our original map is by default centred on (0\arcsec, 0\arcsec) in solar coordinates. 
%We must therefore move it to be centred on an appropriate position, near to the position of the prominence, in order for the correlation routine to properly handle the offset -- if it tries to cross-correlate two images that are too far apart to begin with then it can quickly reach the iteration limit and fail to find an appropriate offset for the images. 
%\begin{figure}
%\begin{center}
%\caption{}
%\label{fig:themis}
%\end{center}
%\end{figure}
The rotated THEMIS image can be seen in Fig.~\ref{fig:aia} (bottom right panel). 
The centre of this rotated map was shifted manually, {by eye,} to be centred on (936\arcsec, 292\arcsec), {deemed to be approximately the centre of the THEMIS FOV}.

Notably the \ion{He}{i} D$_3$ line observed by THEMIS is optically thin, meaning that the bright tornado columns that we see are the integration of all points along the line-of-sight, with the majority of the emission coming from the prominence legs themselves. 
%This is slightly problematic, as 
{  Since} all the other observations showing the prominence in emission are in optically thick lines, {  it is necessary to find common features that can be used for co-alignment}. 
As was noted previously and in Paper I, the prominence legs are visible in the \ion{Ca}{ii} images from SOT. 
%{\bf Since all other observations which show the prominence in emission are of optically thick lines, we use the prominence legs fo}
We are therefore able to apply a threshold on an SOT image so that only the tornadoes are visible in emission, also blocking out the solar limb to replicate the appearance of the THEMIS image. 
In doing so, we are able to use \texttt{get\_correl\_offsets} to align the THEMIS data with the rest of the data sets. 
The offset for the THEMIS image is found to be $x$ = 0.91\arcsec, $y$ = 0.61\arcsec, meaning that the centre of the THEMIS map is (936.91\arcsec, 292.61\arcsec).

\subsection{\textit{Hinode}/EIS with \textit{SDO}/AIA}
\label{ssec:eisaia}

\begin{figure*}
\begin{center}
\includegraphics[width=0.30\hsize,clip=true,trim=6cm 0 7cm 2.1cm]{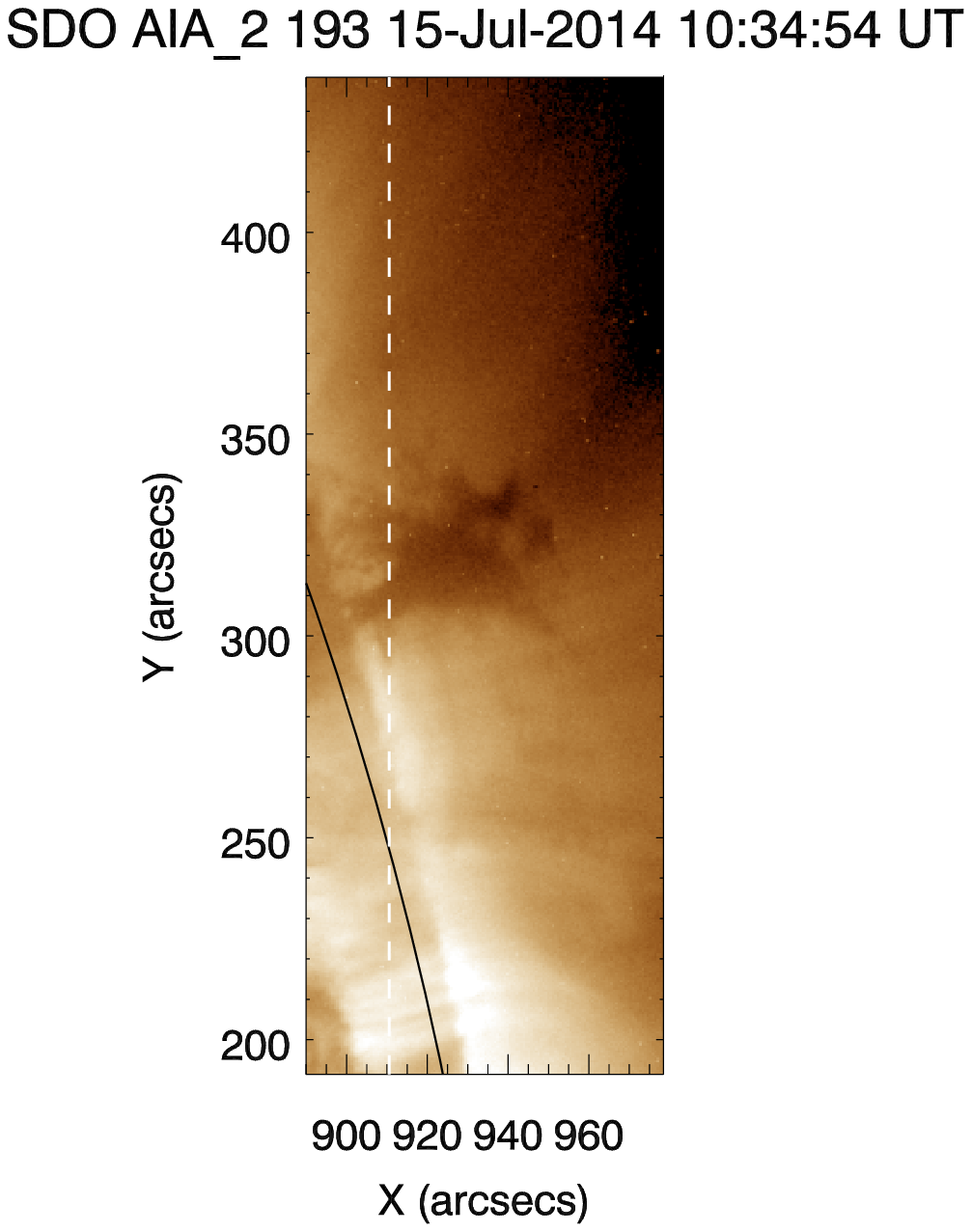}
\includegraphics[width=0.30\hsize,clip=true,trim=6cm 0 7cm 2.1cm]{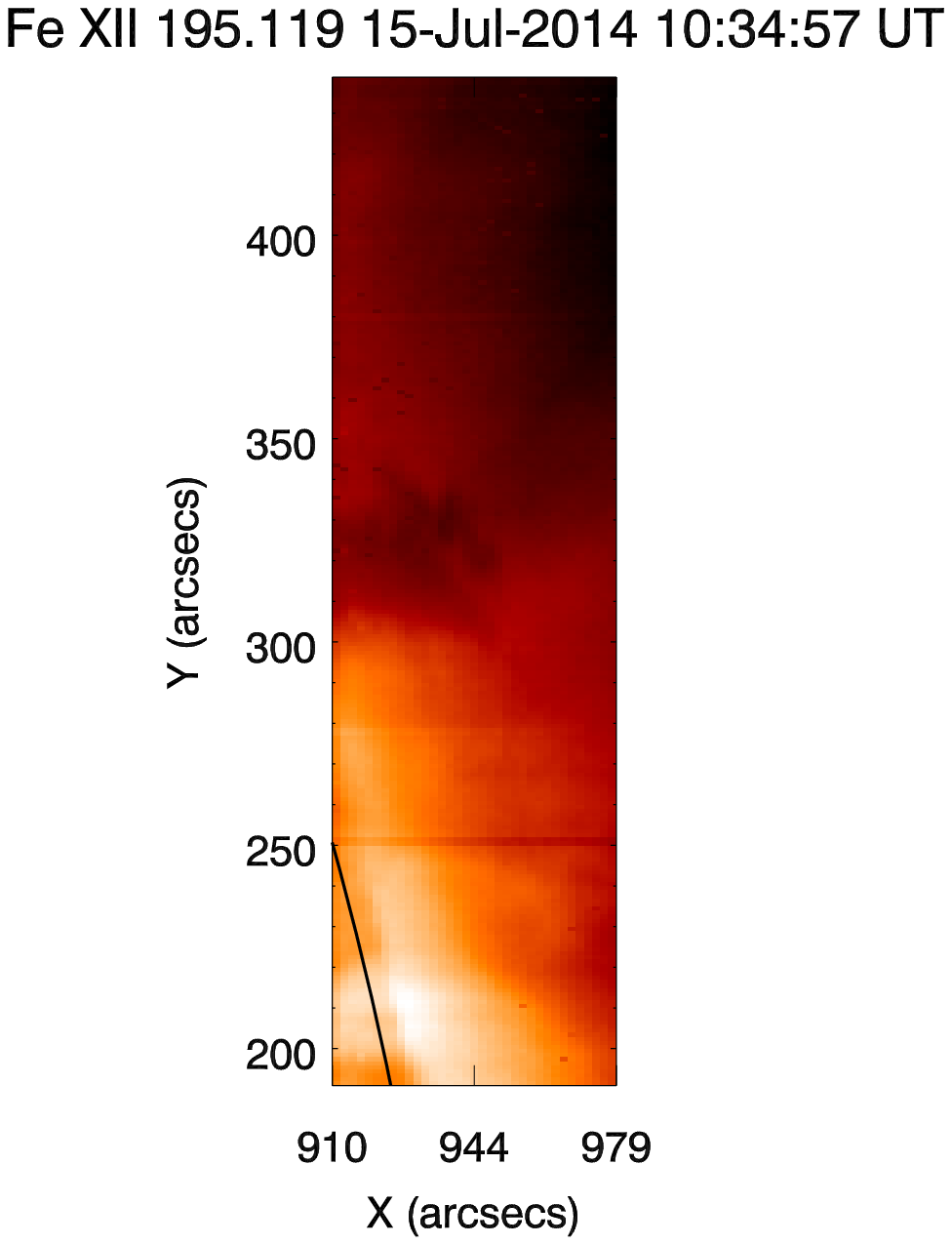}
\includegraphics[width=0.30\hsize,clip=true,trim=6cm 0 7cm 2.1cm]{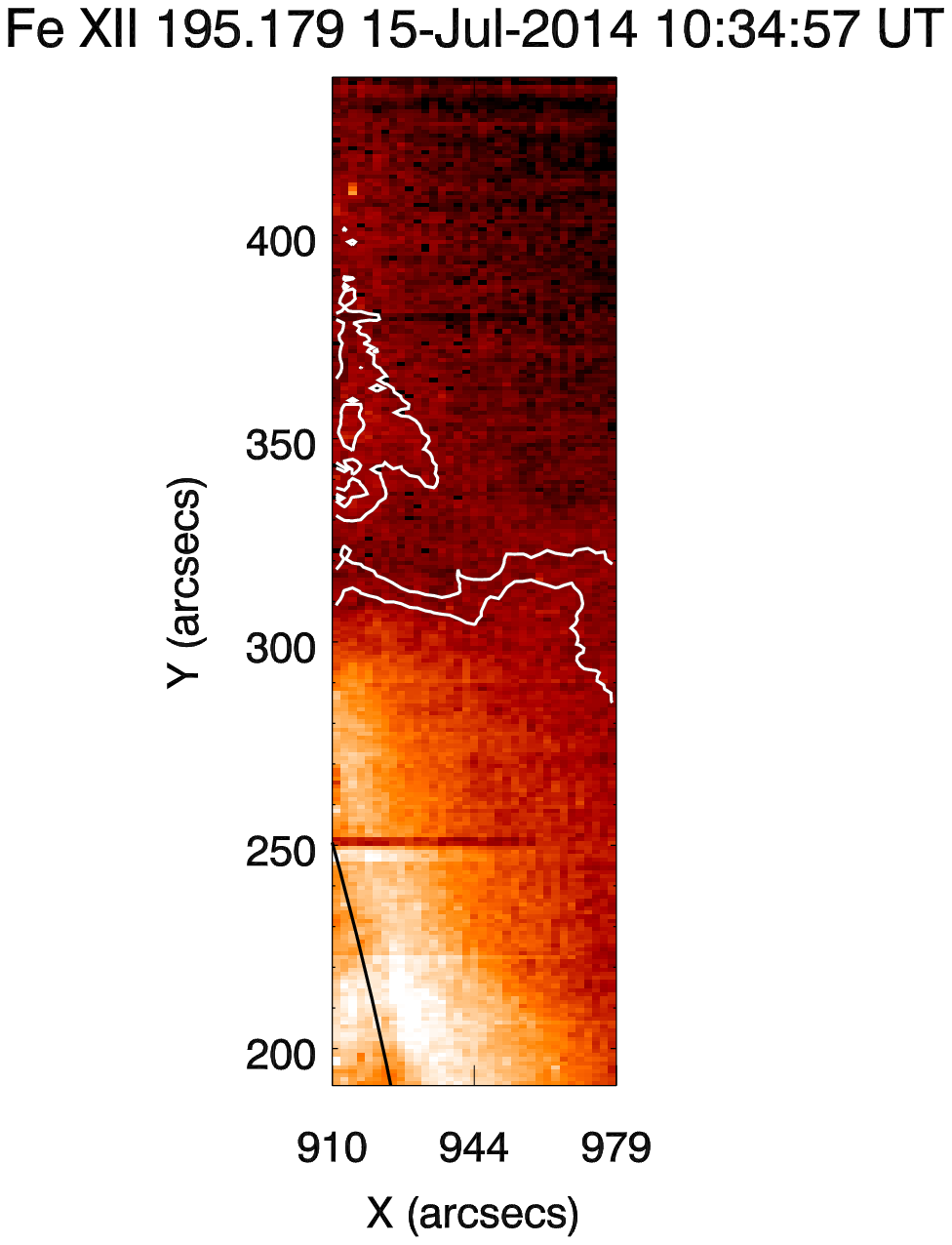}
\caption{Maps from the 15 July 2014 for \ion{Fe}{xii} lines, showing views of the prominence by AIA and EIS. \textit{Left:} AIA 193 \AA\ filter. {  Vertical} dashed line shows the edge of the EIS FOV. \textit{Middle:} EIS 195.119~\AA. \textit{Right:} EIS 195.179~\AA\ with {22\% and 25\%} contours of 195.119 \AA\ (this line pair is used for density diagnostics in Sect.~\ref{sec:eis_correlation}). The observations were made at 10:34 UT. The dark horizontal features in the EIS {maps} are artifacts from the EIS detectors}%, which are  \textit{Left panel:} 195.119~\AA. \textit{Right panel:} 195.179~\AA\ with contours of 195.119\AA.}
\label{fig:eis_int}
\end{center}
\end{figure*}

There is a known offset between the two EIS CCDs \citep{Young2009,Graham2015}, which is accounted for using the \texttt{eis\_ccd\_offset} routine. 
This routine ensures that the data at each wavelength is self-consistently spatially aligned. 
After each of the EIS {maps} at different wavelengths have been co-aligned, they must be aligned with the rest of the data sets. 
%We use AIA 193~\AA\ and EIS \ion{Fe}{xii} 195~\AA\ for this alignment -- The {\bf dominant emission seen in these images} is formed by the same ion (\ion{Fe}{xii}) under the same conditions. 
{  Since the emission in AIA images at 193 \AA\ is dominated by \ion{Fe}{xii}, we use them alongside EIS \ion{Fe}{xii} 195 \AA\ data for this co-alignment.}
Simple inspection of Fig.~\ref{fig:eis_int} shows that {  EIS and AIA} images are very similar and have a good signal-to-noise ratio. 
%However, as has been noted previously \citep{2011A&A...531A..69L,Young2007}, the \ion{He}{ii} line observed by EIS is heavily blended with lines formed at higher temperatures, so we cannot directly use the data for the co-alignment. 
%We employ a double-Gaussian fit to the blended 195~\AA\ spectral line observed by EIS, which allows us to separate the two \ion{Fe}{xii} components. %\textbf{NL: It is really necessary for the co-alignment of EIS and AIA? This statement seems a bit out of place.} %cooler component on the blue side of the blend from the number of hotter lines on the red side. 
%Although not entirely accurate in terms of wavelength position, this approximation is sufficient to isolate enough of the \ion{He}{ii} component to provide an image that can be used for alignment. 
We find that an offset of $x$ = 4.08\arcsec\ and $y$ = 18.24\arcsec\ needs to be applied to the EIS {maps} in order to bring them in line with the rest of the data.

\vspace{1cm}

\begin{figure}
\begin{center}
\includegraphics[width=\hsize]{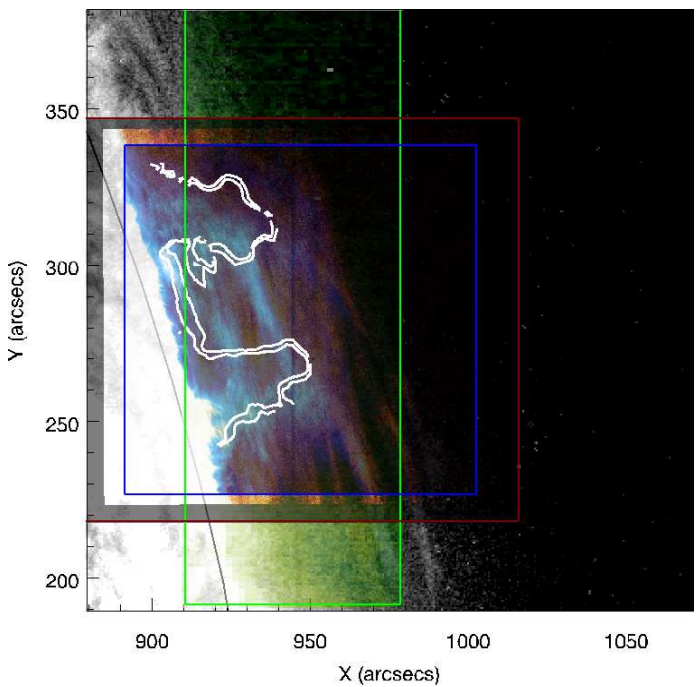}
\caption{Composite image of the prominence observed on 15 July 2014 showing, after co-alignment: the EIS raster in \ion{He}{ii} 256~\AA\ (green), the \textit{IRIS} SJI in \ion{Mg}{ii} (red), and an SOT \ion{Ca}{ii} image (blue). The white contours show the THEMIS D$_3$ intensity image. The background image is an AIA 304~\AA\ image (greyscale)}
\label{fig:coaligned}
\end{center}
\end{figure}

The result of our co-alignment is shown in Fig.~\ref{fig:coaligned}. 
This figure reveals how several structures observed with different instruments are related.
{  The uncertainty on the spatial co-alignment of all data sets is estimated to be 2\arcsec, the spatial resolution of the lowest resolution data used.} 

Now that we have successfully co-aligned the data from THEMIS, AIA, EIS, SOT, and \textit{IRIS}, we search for correlations on a pixel-by-pixel basis between the magnetic field and plasma parameters inferred from these observations. 
We first focus on THEMIS and \textit{IRIS} data.

%__________________________________________________________________

\section{Correlation between THEMIS and IRIS data}
\label{sec:correlation}

We use \ion{Mg}{ii} line ratios, and select some \ion{Mg}{ii} line parameters from Table~4 of \citet{Pereira2013}, to study the properties of the plasma as seen by \textit{IRIS}. 
To compare the plasma properties with the parameters of the magnetic field inferred from the THEMIS measurements, we first need to find the area that corresponds to the overlap between the \textit{IRIS} raster and the THEMIS rasters. 
A mask of the area of overlap between \textit{IRIS} and THEMIS {  is created, and with this mask we can find the appropriate pixels in both rasters to produce correlation plots on a pixel-by-pixel basis.} 
However, as the spatial resolution of \textit{IRIS} is higher {in y} than that of THEMIS used here, we first average the \textit{IRIS} data over 1\arcsec\ {in y} to match the resolution of the THEMIS data. 
{In the x-direction \textit{IRIS} has a 2\arcsec\ step, meaning that for each re-binned \textit{IRIS} pixel there are two THEMIS magnetic field pixels. 
These are considered separately and are both compared to the same \textit{IRIS} pixel.}

\begin{figure}
\begin{center}
\includegraphics[width=0.9\hsize,trim=3cm 0 2cm 0,clip=true]{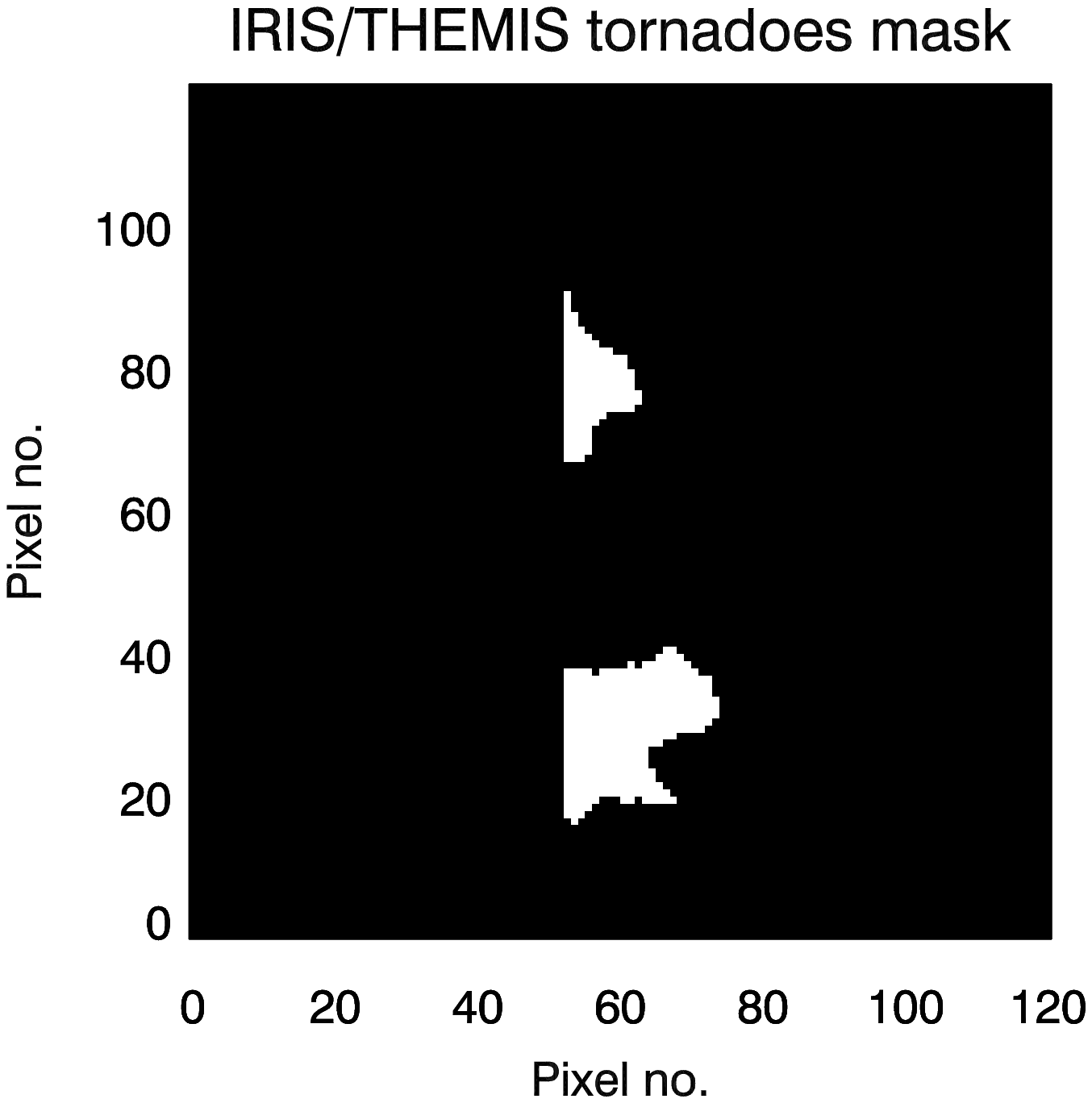}
\includegraphics[width=0.9\hsize,trim=3cm 0 2cm 0,clip=true]{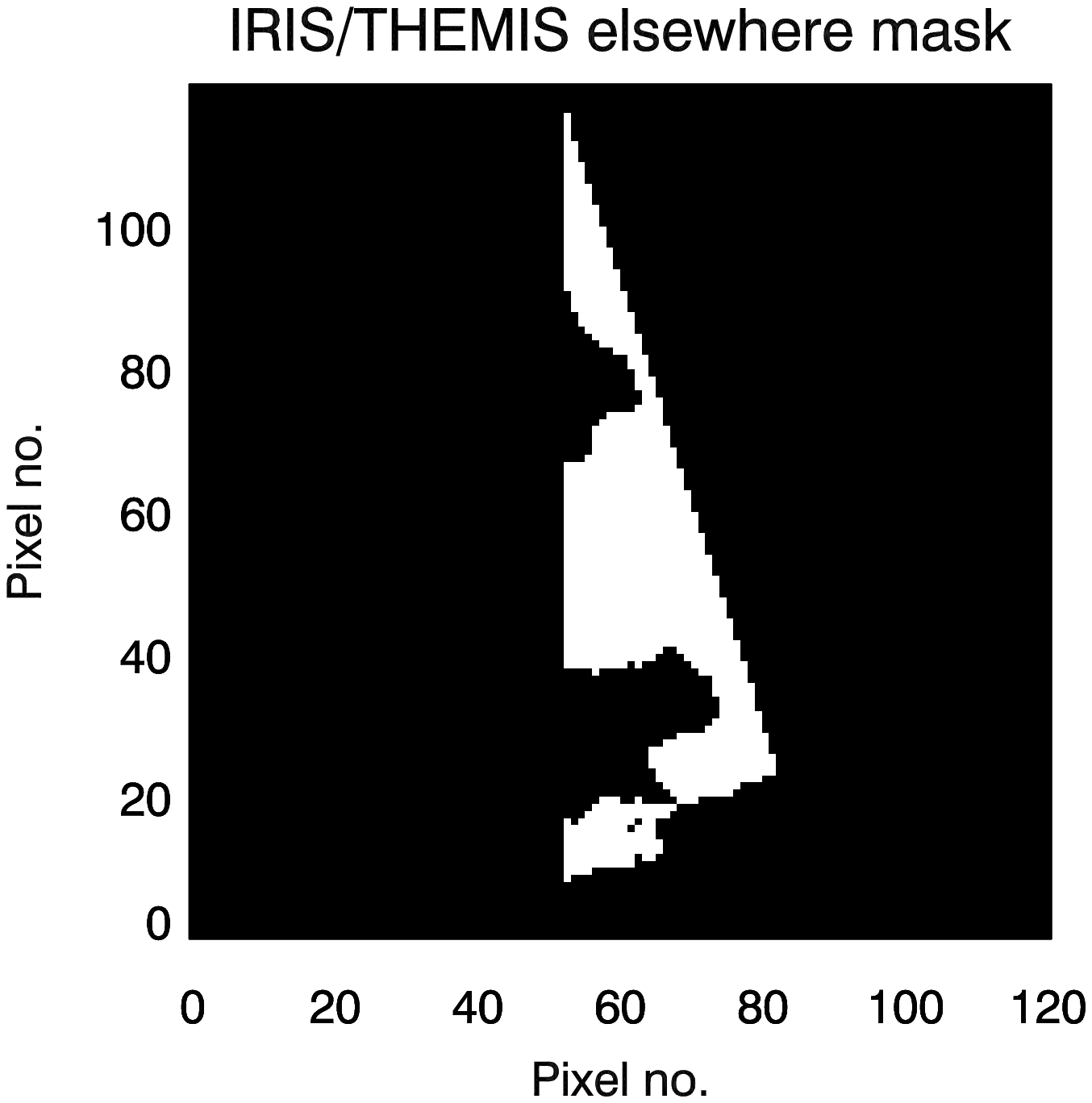}
\caption{Overlapped area between \textit{IRIS} and THEMIS maps for data sets on 15 July 2014. White area shows overlap. \textit{Top:} Areas including the two tornadoes. \textit{Bottom:} Area including the rest of the prominence.}
\label{fig:overlap}
\end{center}
\end{figure}

Using the {  intensity map of the} THEMIS raster, we can identify the locations of the tornadoes. 
This then allows us to create more masks for the pixels in the tornadoes and those outside them, in the rest of the prominence. 
The entirety of the overlapped area covered by the THEMIS raster is filled with prominence in the \textit{IRIS} raster. 
These masks are shown in Fig.~\ref{fig:overlap}. 

%\begin{figure}
%\begin{center}
%\includegraphics[width=0.9\hsize,trim=3cm 0 2cm 0,clip=true]{IRIS_THEMIS_overlap_tornadoes.eps}
%\includegraphics[width=0.9\hsize,trim=3cm 0 2cm 0,clip=true]{IRIS_THEMIS_overlap_not_tornadoes.eps}
%\caption{Masks of regions covering the two tornadoes (\textit{top}) and the rest of the prominence (\textit{bottom}) for the overlap between THEMIS and \textit{IRIS} rasters on the 15 July 2014.}
%\label{fig:masks}
%\end{center}
%\end{figure}

There is only a partial overlap between the two rasters, and therefore we only see the top of the northern tornado as seen by THEMIS and just over half of the southern one in the \textit{IRIS} raster. 
%We can, however, see parts of both tornadoes, but it is worth noting that we cannot see the `base' of either column. 

\subsection{\ion{Mg}{ii} k$_2$/k$_3$ ratio}

By measuring the level of reversal of the k line (equivalently for h), {  which is defined as being the \ion{Mg}{ii} k$_2$/k$_3$ intensity ratio}, we can compare these observations to models of the \ion{Mg}{ii} lines in prominences. 
\citet{Heinzel14} gives us a grid of models with which to compare observations, allowing us to narrow down values for physical parameters such as gas pressure and temperature of the prominence. 
This analysis  was done in Paper I for parts of this prominence, where it was found that the {  k$_2$/k$_3$ ratio} was between 1 (non-reversed) and 2.8. 
{  \ion{Mg}{ii} k line profiles with a strong central reversal (k$_2$/k$_3$ > 2) are reminiscent of those found in the chromosphere \citep{Leenaarts2013,Pereira2013}, and indicate that in these pixels the prominence plasma has a large optical thickness, probably due to high pressures \citep{Heinzel14}.}

\begin{figure*}
\begin{center}
\includegraphics[width=0.45\hsize,clip=true,trim= 2.7cm 0 0 0]{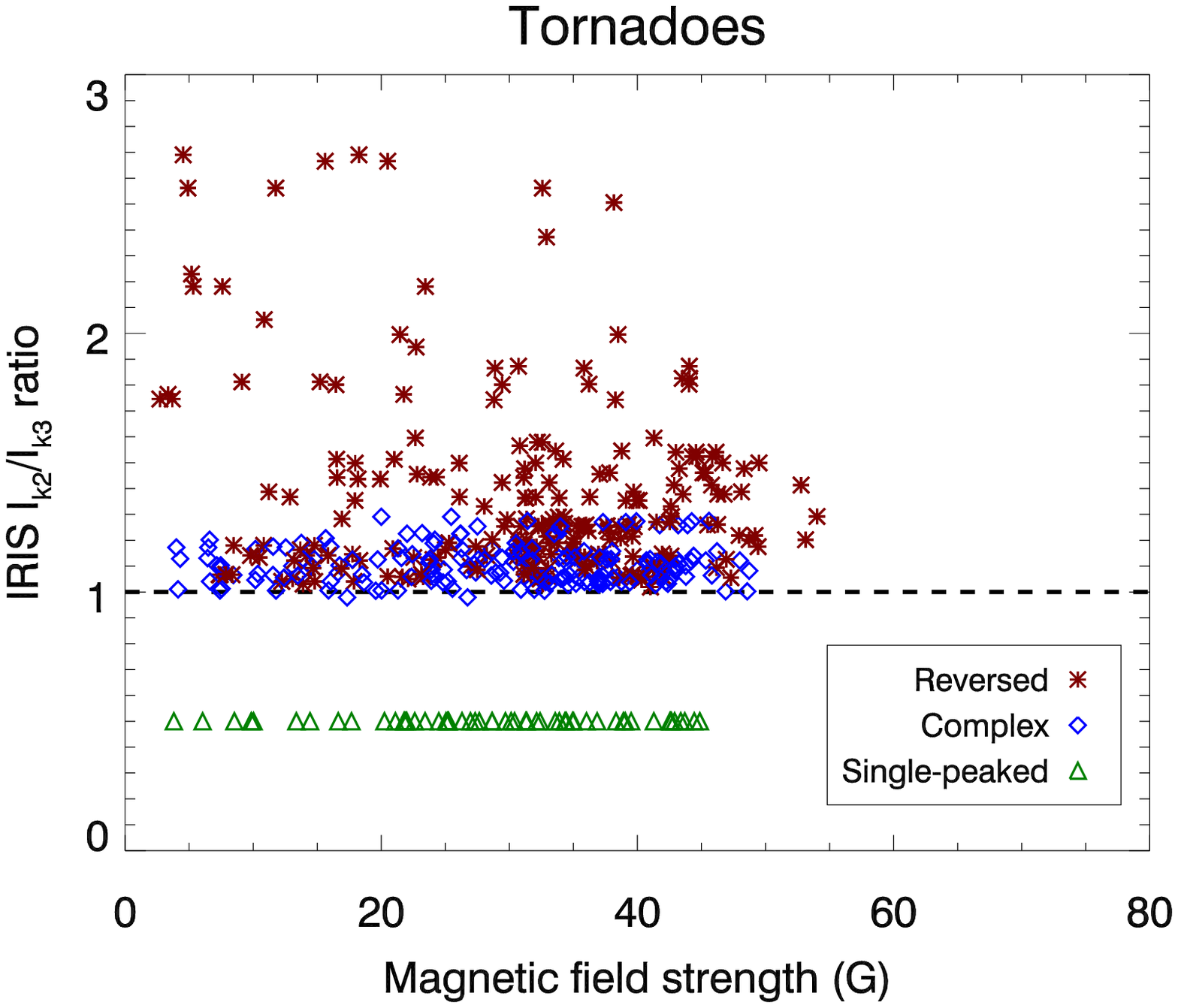}
\includegraphics[width=0.45\hsize,clip=true,trim= 2.7cm 0 0 0]{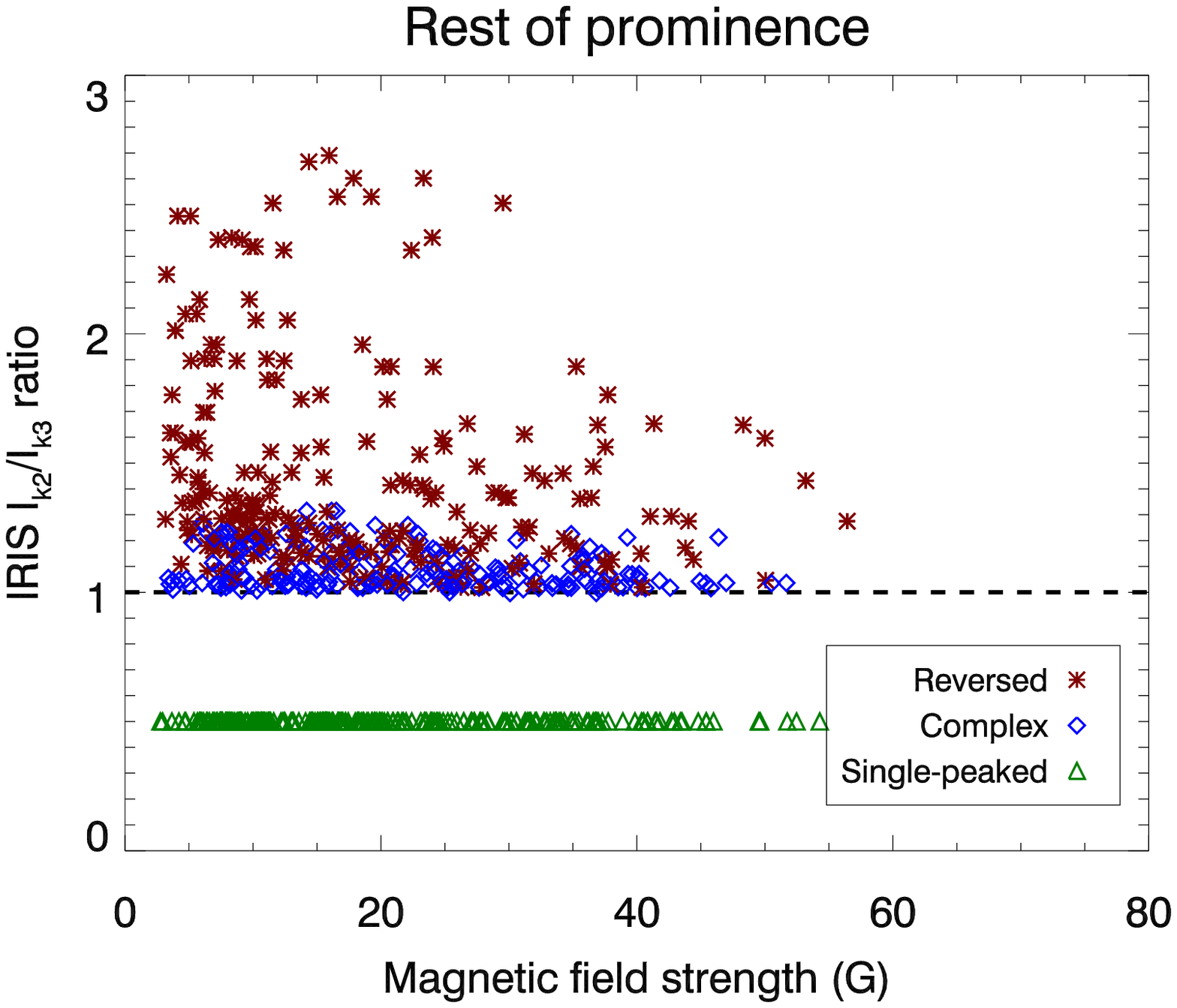}
\\
\includegraphics[width=0.45\hsize,clip=true,trim= 2.7cm 0 0 0]{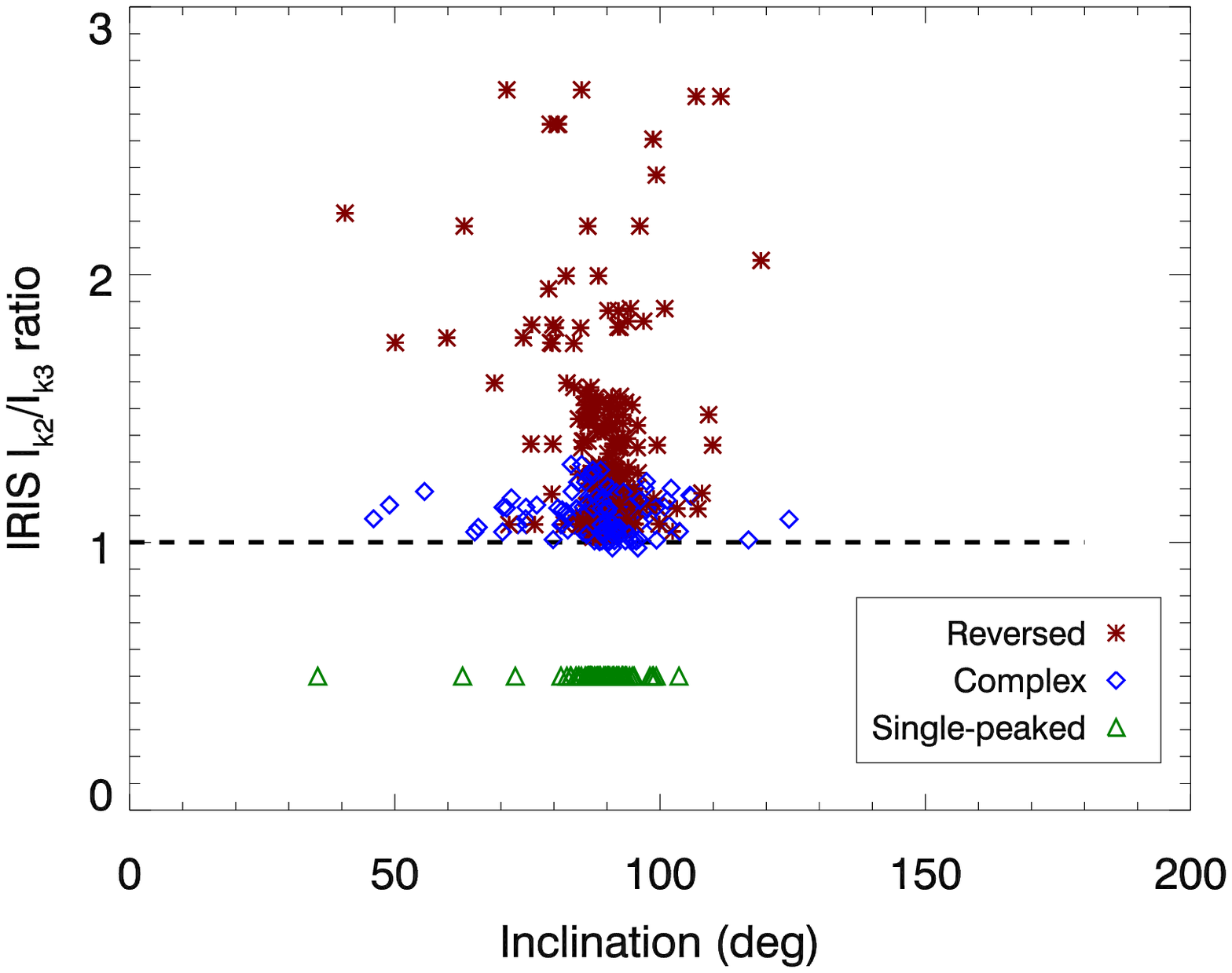}
\includegraphics[width=0.45\hsize,clip=true,trim= 2.7cm 0 0 0]{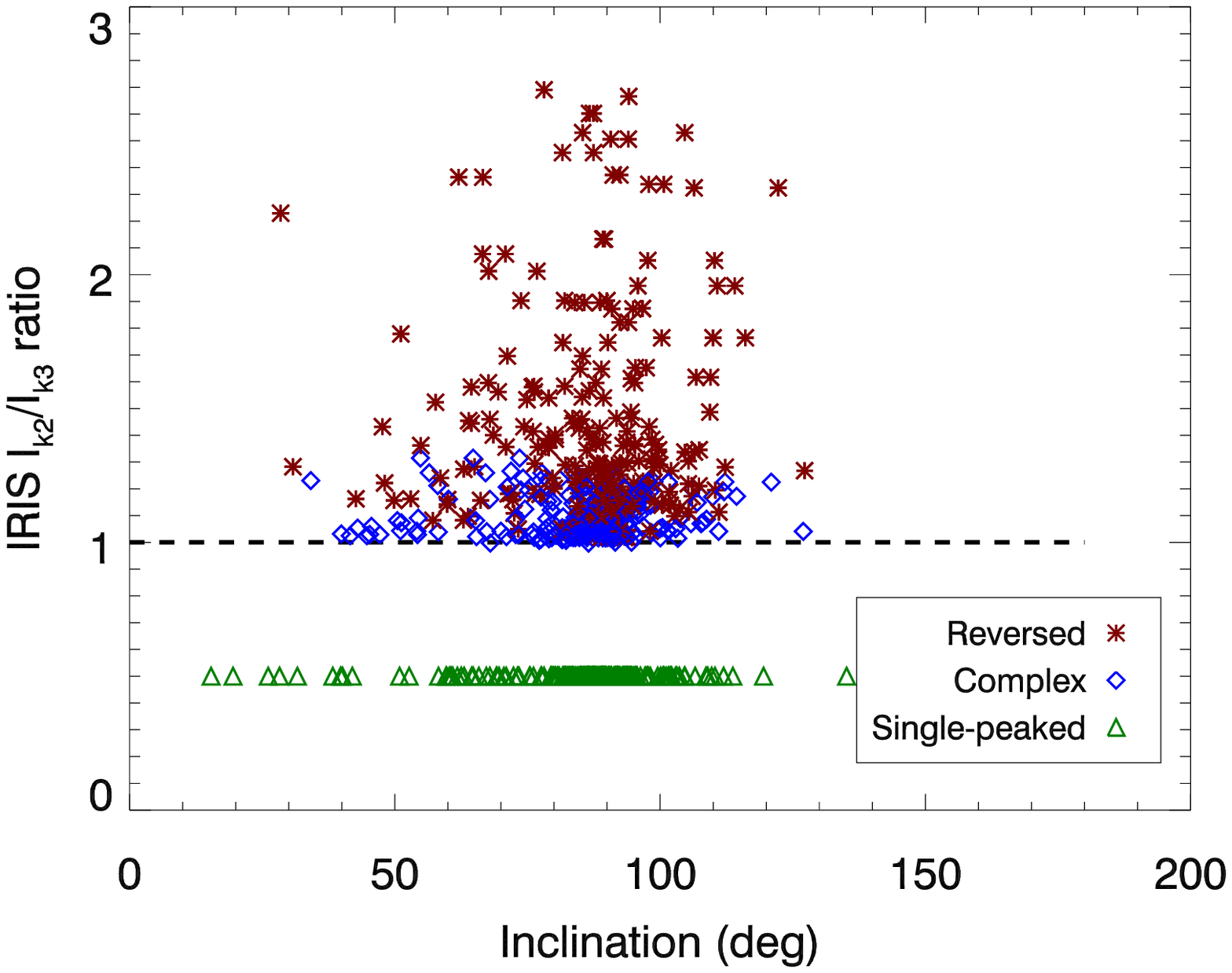}
\\
\includegraphics[width=0.45\hsize,clip=true,trim= 2.7cm 0 0 0]{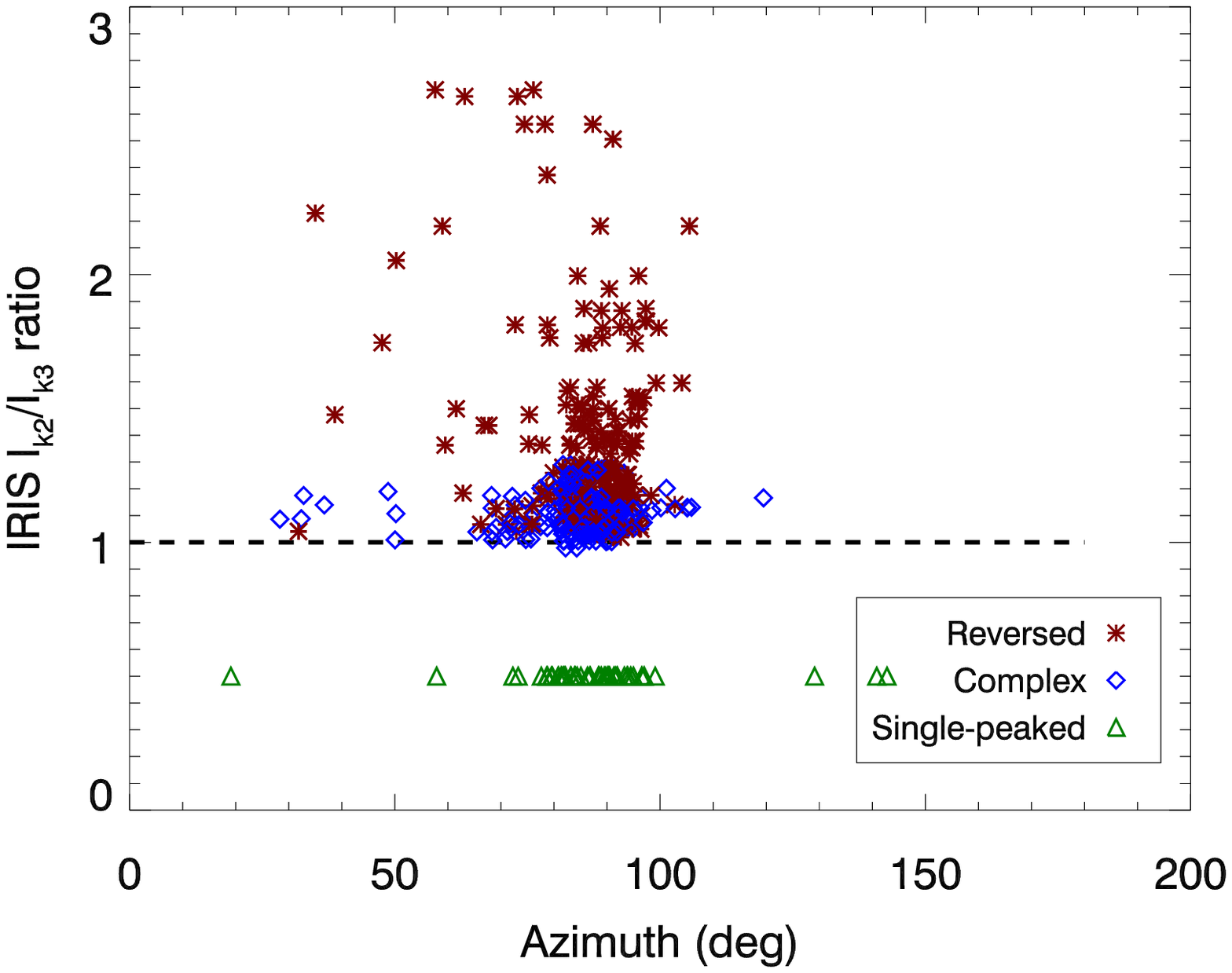}
\includegraphics[width=0.45\hsize,clip=true,trim= 2.7cm 0 0 0]{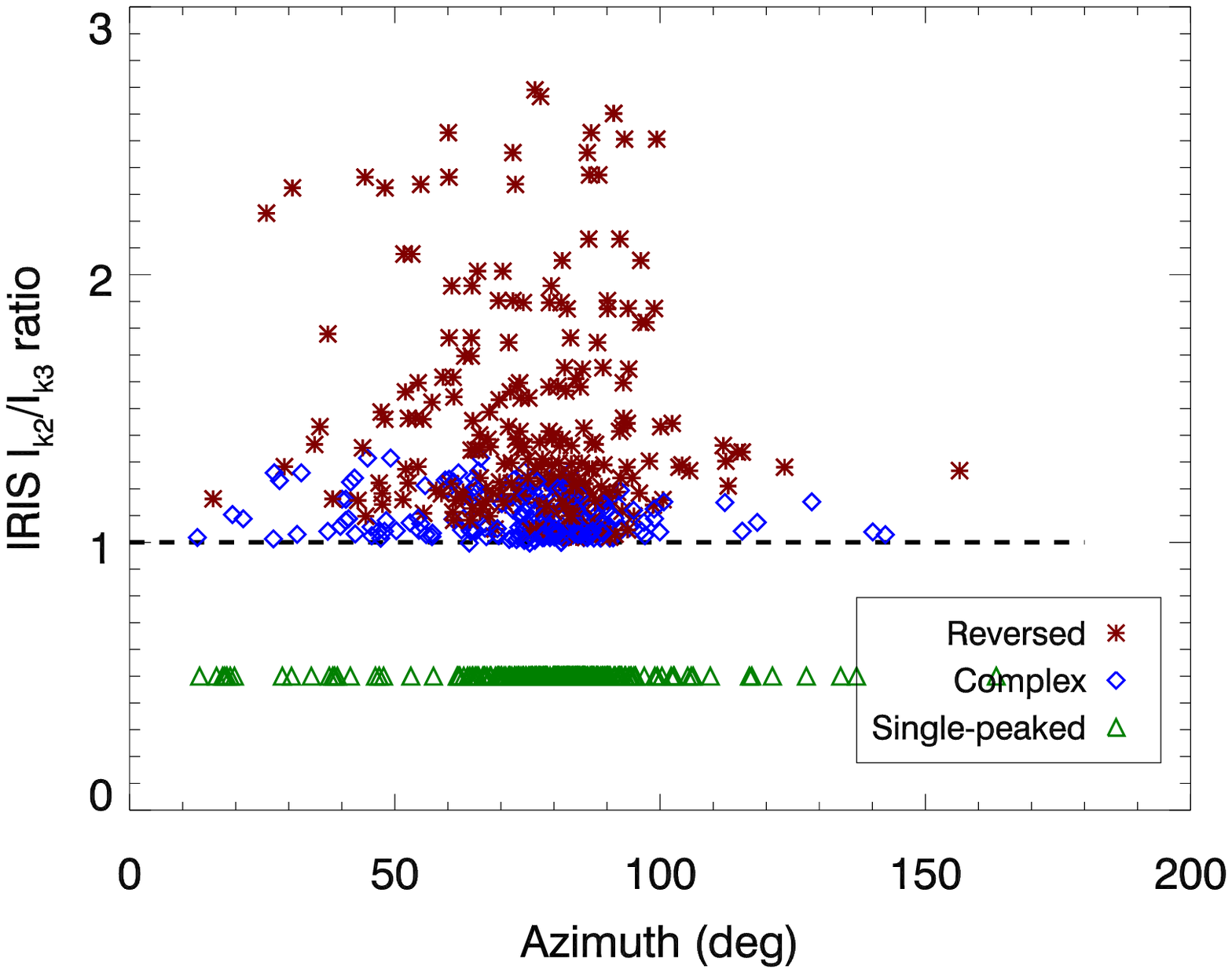}
\caption{Plots showing the k$_2$/k$_3$ ratio of \ion{Mg}{ii}  against magnetic field parameters. Left column is points in the tornadoes, right column is points everywhere else in the prominence. Red asterisks are points where \ion{Mg}{ii} k profiles are reversed, green triangles are where \ion{Mg}{ii} k profiles are single peaked (manually placed at a ratio of 0.5 to distinguish them). Blue diamonds correspond to complex profiles, described in Sect.~\ref{ssec:iris_diagnostics}. The black dashed line provides a cutoff for reversed profiles, where a ratio of 1 corresponds to a single-peaked profile. {  Inclination is with respect to the local vertical, and azimuth is with respect to the line of sight.}}
\label{fig:k2k3}
\end{center}
\end{figure*}

\begin{figure}
\begin{center}
\includegraphics[width=0.9\hsize,trim=1cm 0 0.7cm 1cm,clip=true]{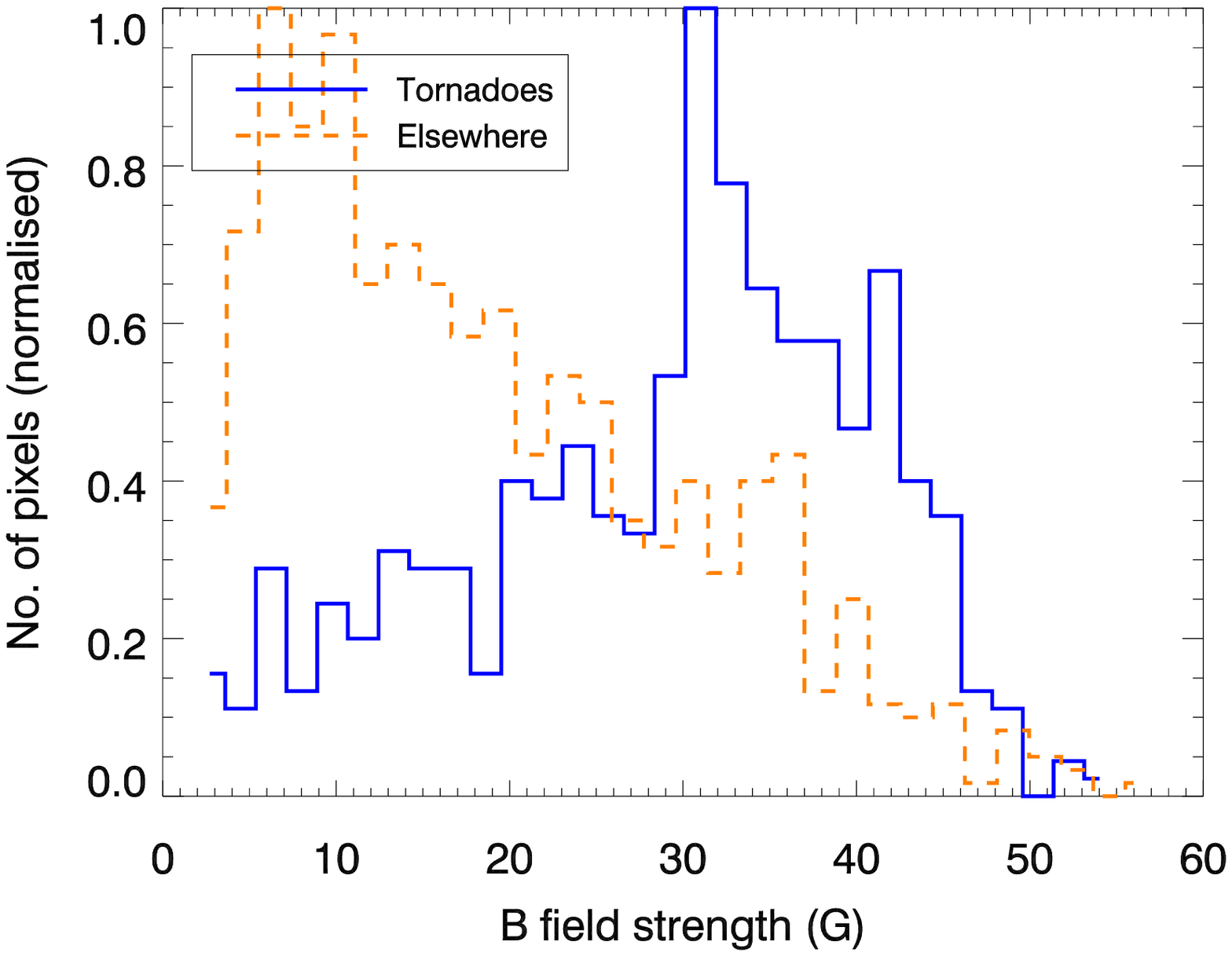}
\includegraphics[width=0.9\hsize,trim=1cm 0 0.7cm 1cm,clip=true]{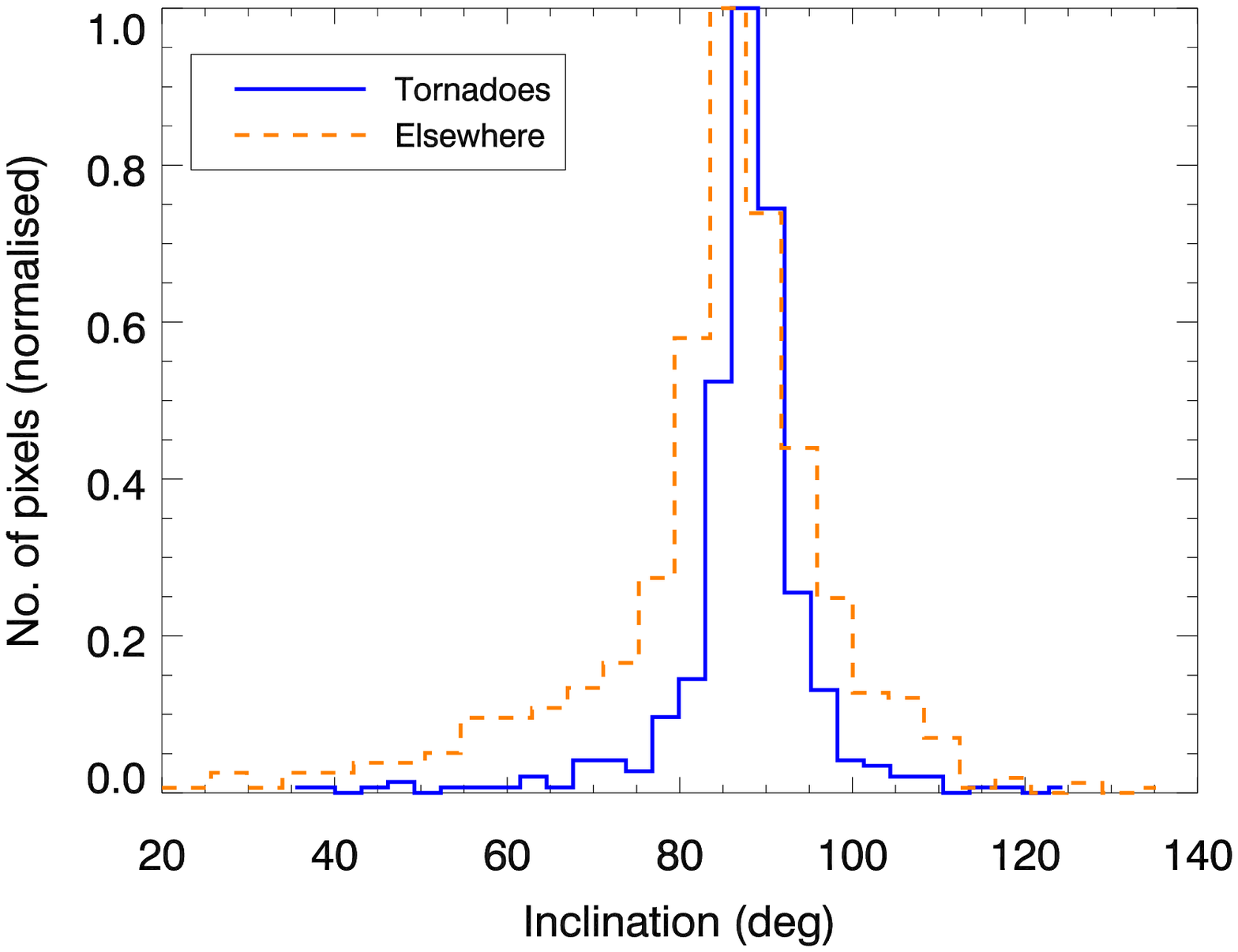}
\includegraphics[width=0.9\hsize,trim=1cm 0 0.7cm 1cm,clip=true]{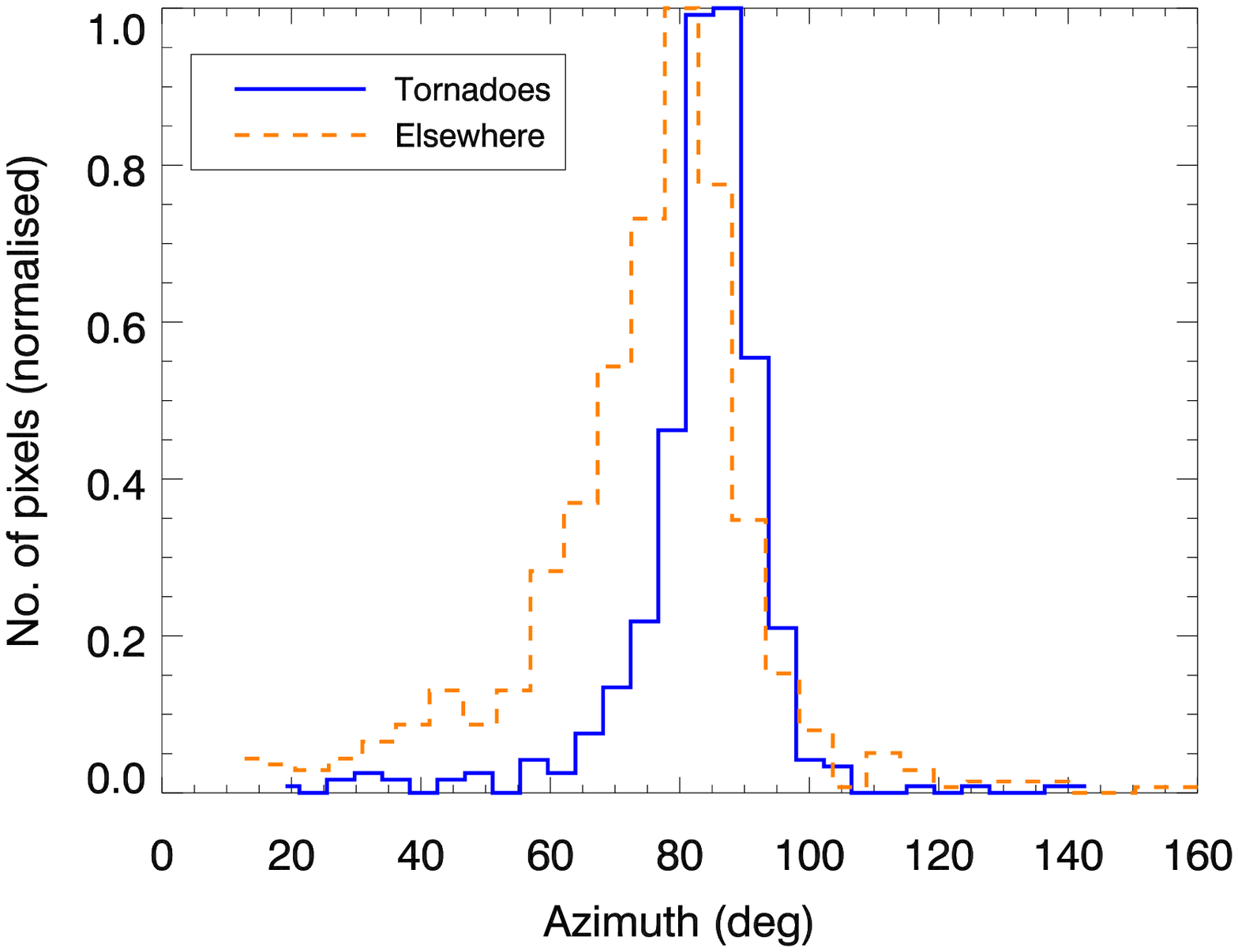}
\caption{Histograms of magnetic field parameters from THEMIS showing the distribution of values in both the tornadoes case ($\sim$ 500 pixels) and the rest-of-prominence case ($\sim$ 1100 pixels). \textit{Top:} Magnetic field strength. \textit{Middle:} Inclination. \textit{Bottom:} Azimuth.}
\label{fig:themis_hist}
\end{center}
\end{figure}

Is the level of reversal related spatially to the magnetic field parameters? 
To investigate this question we have plotted  the ratio k$_2$/k$_3$  in Fig.~\ref{fig:k2k3} against the magnetic field strength,  inclination and azimuth. 
We present data separately for pixels defined as being `tornado' pixels, and `rest-of-prominence' pixels, using the masks described above. 
{  The value for the k$_2$/k$_3$ ratio for non-reversed profiles has been arbitrarily set to 0.5 in Fig.~\ref{fig:k2k3} to distinguish them. 
A black dashed line shows the cutoff value for reversed profiles. 

The top panels of Fig.~\ref{fig:k2k3} show the k$_2$/k$_3$ ratio vs. magnetic field strength in the tornadoes and the rest of the prominence. 
From these plots it appears that the field strength is generally higher in the tornadoes than the rest of the prominence, but there are no clear correlations between the field strength and the k$_2$/k$_3$ ratio. }
%These plots can tell us a number of things. 
%Although no clear correlation is immediately obvious, there are some points worth noting. 
From the top panel of Fig.~\ref{fig:themis_hist} it is clear that there is a higher field strength in the tornadoes than outside of them. %, and there are relatively a much lower number of unreversed profiles in the tornadoes. 
In fact, points in the rest of the prominence have a mean field strength of 20~G, whereas in the tornadoes the majority of points have a value of around 30~G. 
%This is mostly due to the fact that in \ion{He}{i} D$_3$ there is little to no signal outside of the bright tornado columns, meaning it is difficult for the inversion code cannot return a meaningful value at these locations.
%We also note that the most reversed profiles generally correspond to low to medium magnetic field strengths, lower than around 45 G, whereas in places we find field strengths of above 80 G. 

For the field inclination we see most of the points are clustered around an inclination of 90$^\circ$ (i.e. horizontal) in Fig.~\ref{fig:k2k3} (middle panels). 
This is consistent with previous studies on this data \citep{Levens2016,Levens2016b}, where it was found that the field is horizontal with respect to the limb.
We see a few outliers in the tornadoes, but not a significant number. 
{In the rest of the prominence there is a larger spread of values, but the field is still predominantly horizontal, as seen in Fig.~\ref{fig:themis_hist} (middle panel)}
%In the rest of the prominence, we can see a `tail' of lower reversal levels (less than $\sim$1.6) stretching to lower inclinations. 
%However, as Fig.~\ref{fig:themis_hist} (middle panel) shows, this tail is actually a relatively small number of points, and most pixels have an inclination of 90$^\circ$.

A similar structure is seen in the k$_2$/k$_3$ ratio against azimuth (Fig.~\ref{fig:k2k3}, bottom panels), where most of the points are clustered around a mean value of around {85$^\circ$ in the tornadoes and 77$^\circ$ elsewhere} (Fig.~\ref{fig:themis_hist}, bottom panel). 
There is  no clear correlation that  can be seen. %, but again we have a `tail' stretching towards lower {  azimuths in the rest-of-prominence case.}
%We question what could cause this tail, and why we do not see it stretching to higher inclinations as well. 
%We note here that the `rest-of-prominence' case is the part in the THEMIS data that has the lowest signal (as can be seen in Fig.~\ref{fig:aia}, bottom right panel), so may be affected by noise. 
%If we are seeing the effect of {  an} ambiguity {  that has not been accounted for} in these points we would expect to see a tail towards higher inclinations and azimuths as well, but that is not what we see. 
%We conclude that these points are probably due to a low SNR in those parts of the raster, carrying a large uncertainty. %\textbf{IS IT WHAT WE THINK? What about actual uncertainties on the plots?}
%{We deduce that these points are mostly artifacts of the map rotation procedure, which necessarily re-bins the data to maintain square pixels and the same spatial resolution. 
%Points at the edge of the THEMIS field of view are therefore averaged with some points outside of the FOV where there is no data, resulting in a reduction of their data value by up to $\sim$ half. 
%However, as it is a small number of points we do not believe that the affected points change the overall profile past introducing this tail.}

{  We note that some of the points for the `complex' profiles have k$_2$/k$_3$ ratios lower than 1 in Fig.~\ref{fig:k2k3}. 
This is due to the fact that the peaks given by the peak-finder algorithm (Waller et al. 2017, \textit{in prep.}) are averaged over a number of pixels, but this averaging is not done for the intensity at the centroid of the distribution. 
Therefore in certain `complex' profiles where the peak is very sharp and narrow (i.e. one spectral pixel wide) the averaged peak intensity can be lower than the centroid intensity. 
As it is not a significant number of points and they are all `complex' profiles, we conclude that they are anomalies due to the automated handling of the profiles. }

\begin{figure}
\begin{center}
\includegraphics[width=0.9\hsize,trim=1cm 0 0.7cm 1cm,clip=true]{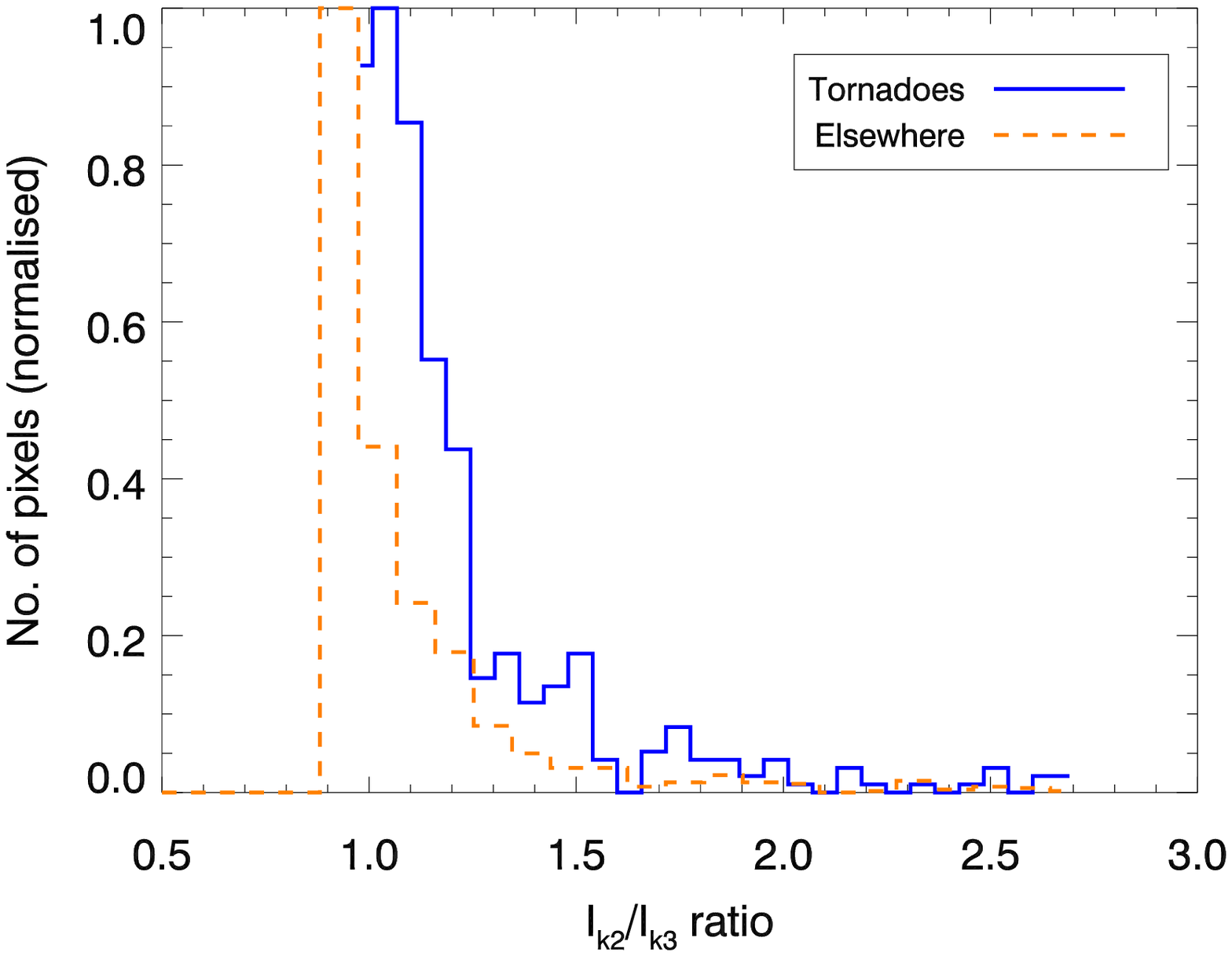}
\includegraphics[width=0.9\hsize,trim=1cm 0 0.7cm 1cm,clip=true]{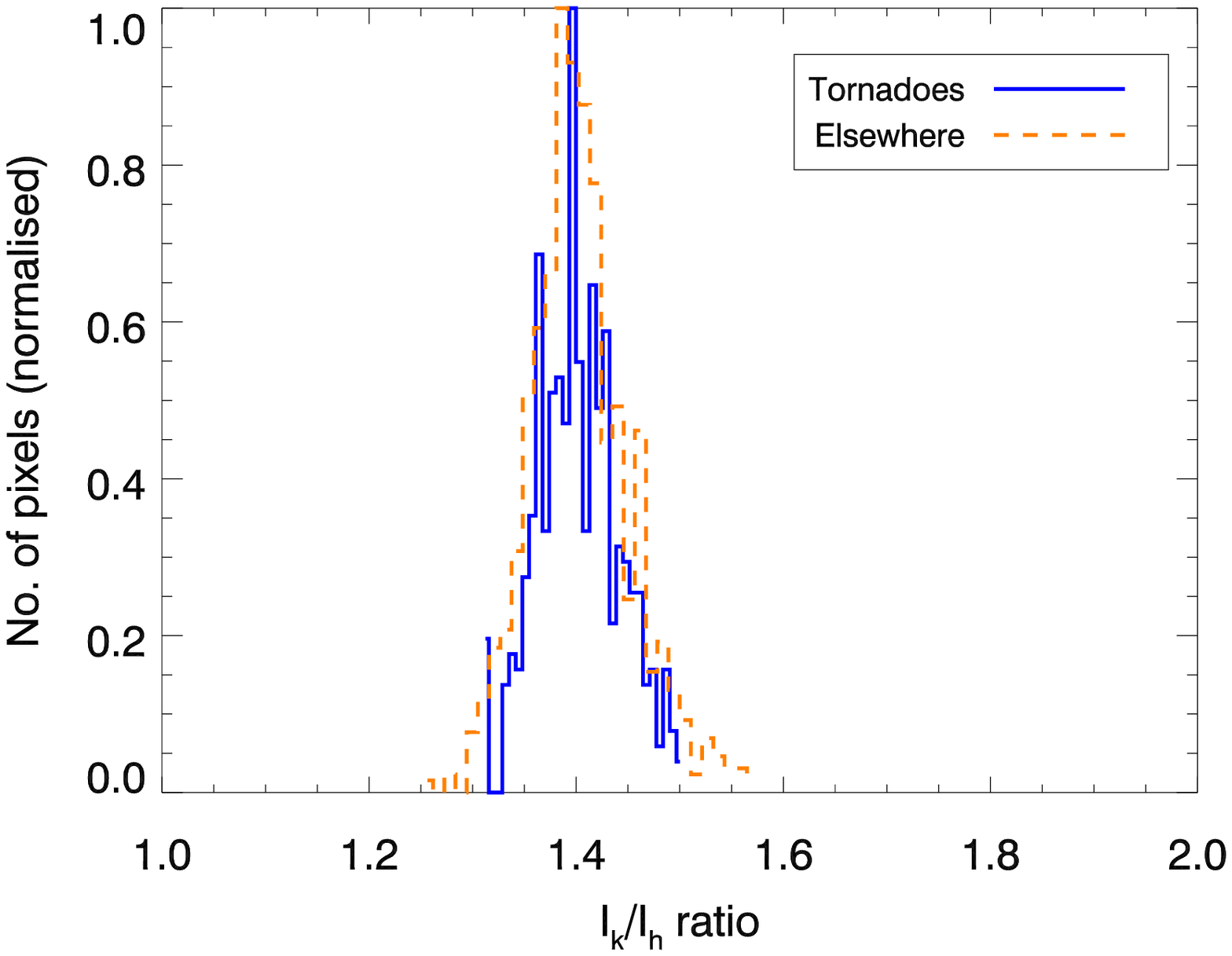}
\includegraphics[width=0.9\hsize,trim=1cm 0 0.7cm 1cm,clip=true]{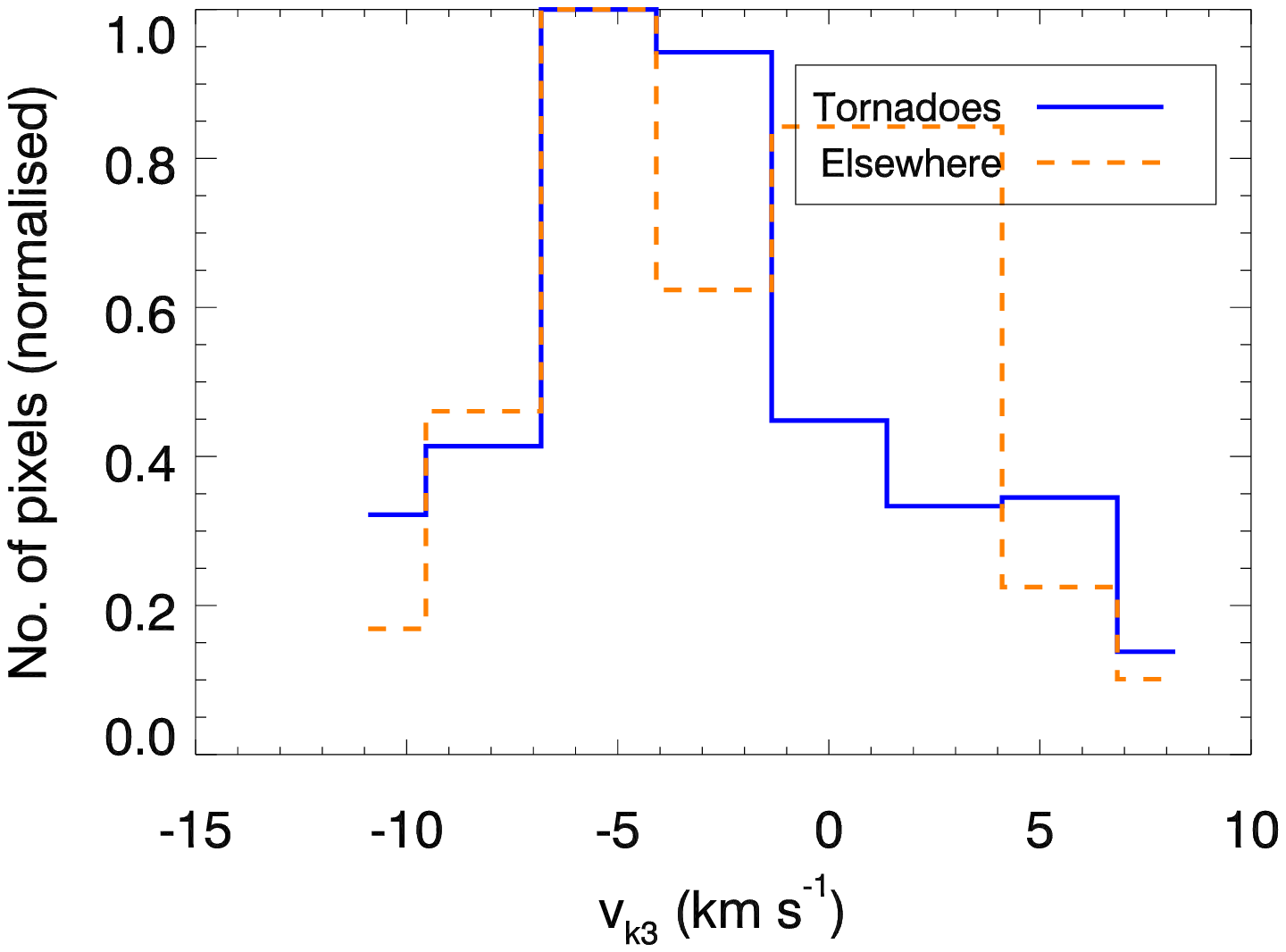}
\caption{Histograms of parameters derived from the \ion{Mg}{ii} lines showing the distribution of values in both the tornadoes case ($\sim$ 500 pixels) and the rest-of-prominence case ($\sim$ 1100 pixels). \textit{Top:} k$_2$/k$_3$ ratio (level of reversal). \textit{Middle:} k/h ratio. \textit{Bottom:} k$_3$ Doppler shift.}
\label{fig:iris_hist}
\end{center}
\end{figure}

%Figure \ref{fig:iris_hist} shows histograms of the spread of values found in the three \ion{Mg}{ii} parameters considered here. 
The top panel of Fig.~\ref{fig:iris_hist} shows the distributions of the k$_2$/k$_3$ ratio in the tornadoes (solid blue) and elsewhere in the prominence (dashed orange). 
We take moments of the distribution, which tells us that the mean reversal level in the tornadoes is 1.23, whereas in the rest of the prominence it is lower, with a mean value of 1.14. 
{  Standard deviation is 0.3 in both cases. }
There are also relatively more points extending to higher reversal levels in the tornadoes than elsewhere. 

\subsection{\ion{Mg}{ii} I$_k$/I$_h$ ratio}

The relative intensities of the h and k lines of \ion{Mg}{ii}, referred to here as the k/h ratio, can tell us something about the formation of these lines.  
Under normal chromospheric conditions, where the \ion{Mg}{ii} ions are collisionally excited, we would expect a k/h ratio of 2 \citep{Leenaarts2013} and the plasma to be optically thin. 
%However, if the ratio of these oscillator strengths is equal to 4 then we have a situation where radiation is playing an important role in the atomic excitation \citep{Harra2014}. 
Previous prominence observations have shown typical k/h ratios of around 1.5 \citep{Schmieder2014,Liu2015,Vial2016,Levens2016}, and NLTE models of \ion{Mg}{ii} have found similar expected values \citep{Heinzel14}. 
Values lower than 2, as have been observed, suggest the presence of scattering  in the emitting region, i.e. a departure from the optically thin case.

\begin{figure*}
\begin{center}
\includegraphics[width=0.45\hsize,clip=true,trim= 2.3cm 0 0 0]{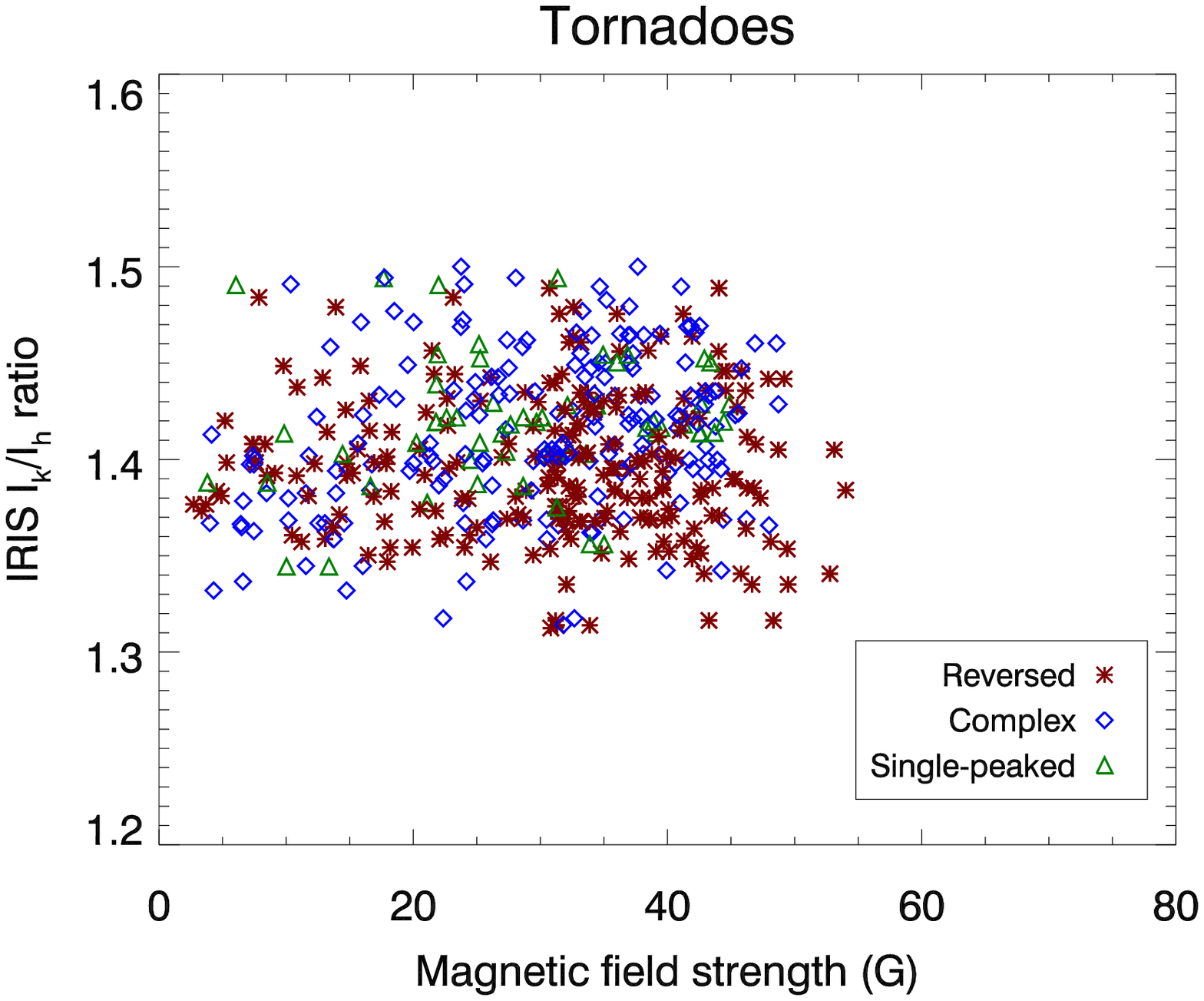}
\includegraphics[width=0.45\hsize,clip=true,trim= 2.3cm 0 0 0]{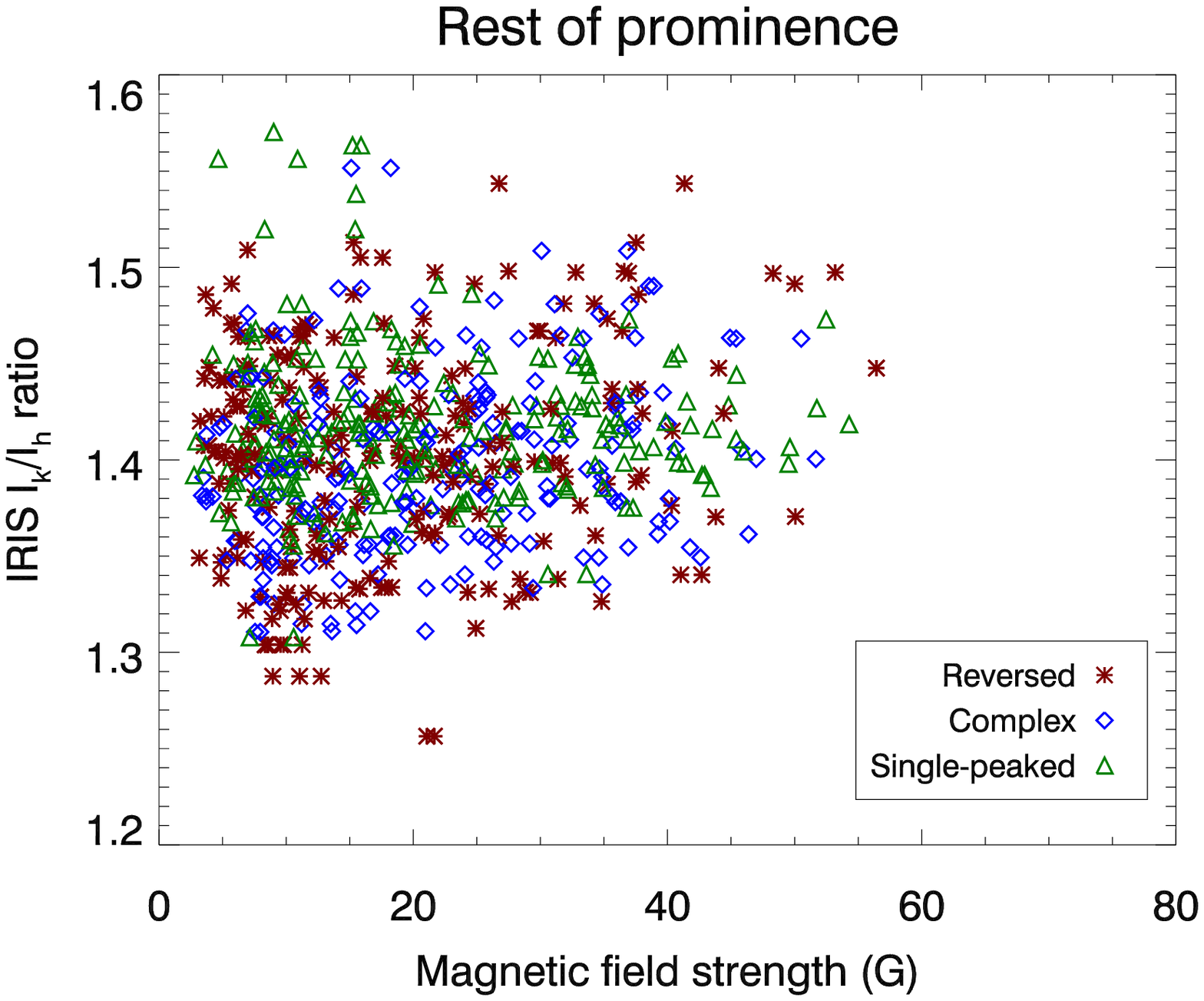}
\\
\includegraphics[width=0.45\hsize,clip=true,trim= 2.3cm 0 0 0]{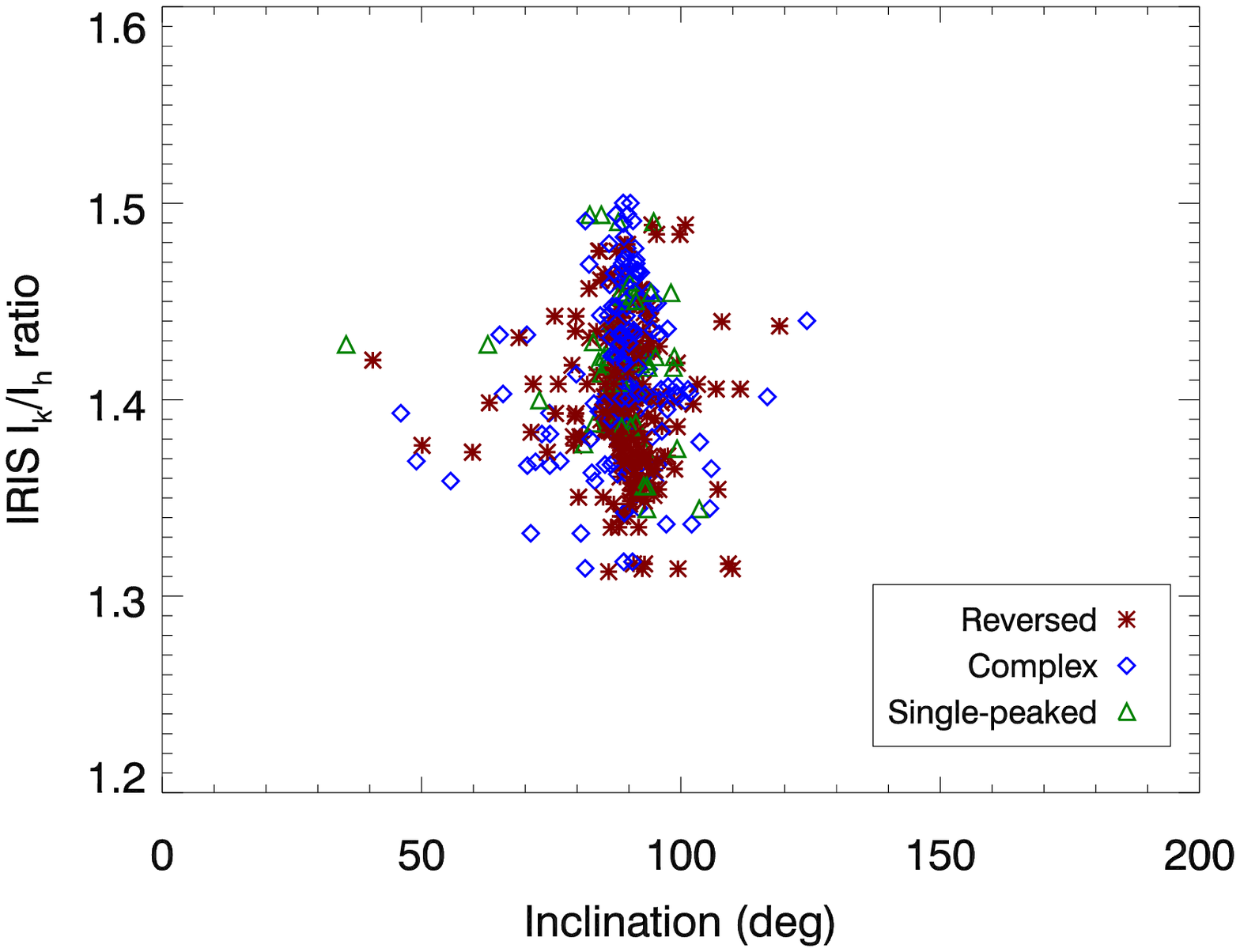}
\includegraphics[width=0.45\hsize,clip=true,trim= 2.3cm 0 0 0]{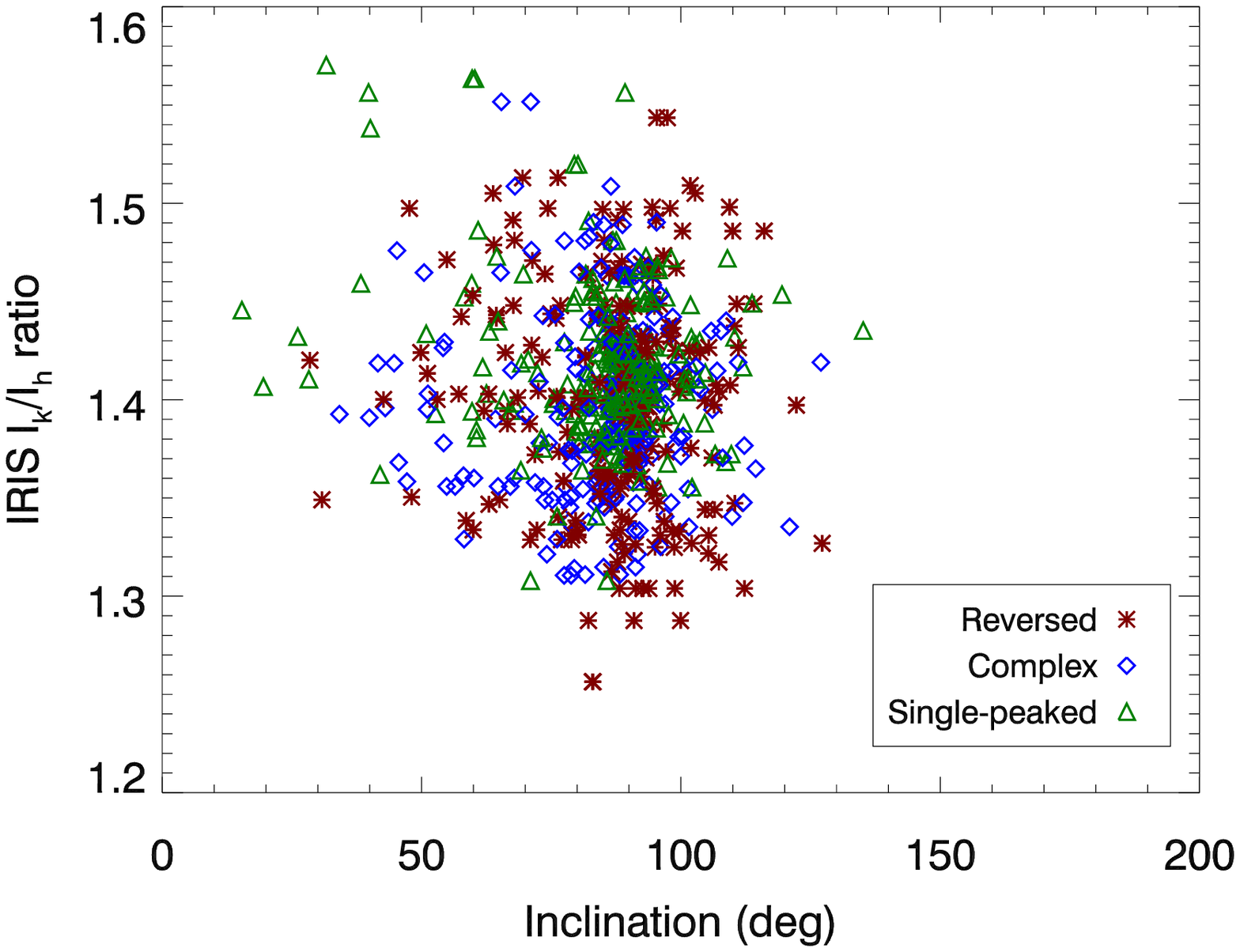}
\\
\includegraphics[width=0.45\hsize,clip=true,trim= 2.3cm 0 0 0]{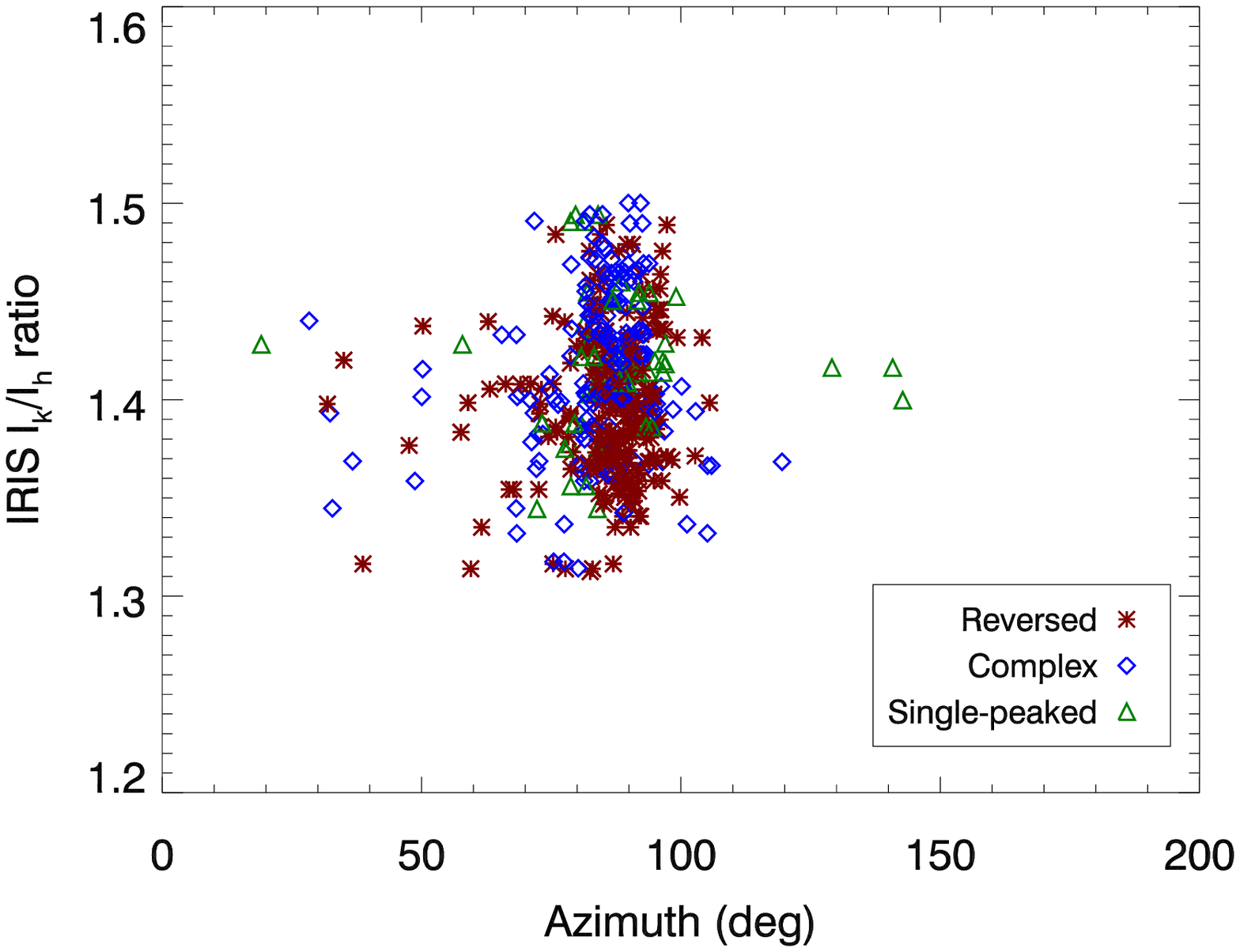}
\includegraphics[width=0.45\hsize,clip=true,trim= 2.3cm 0 0 0]{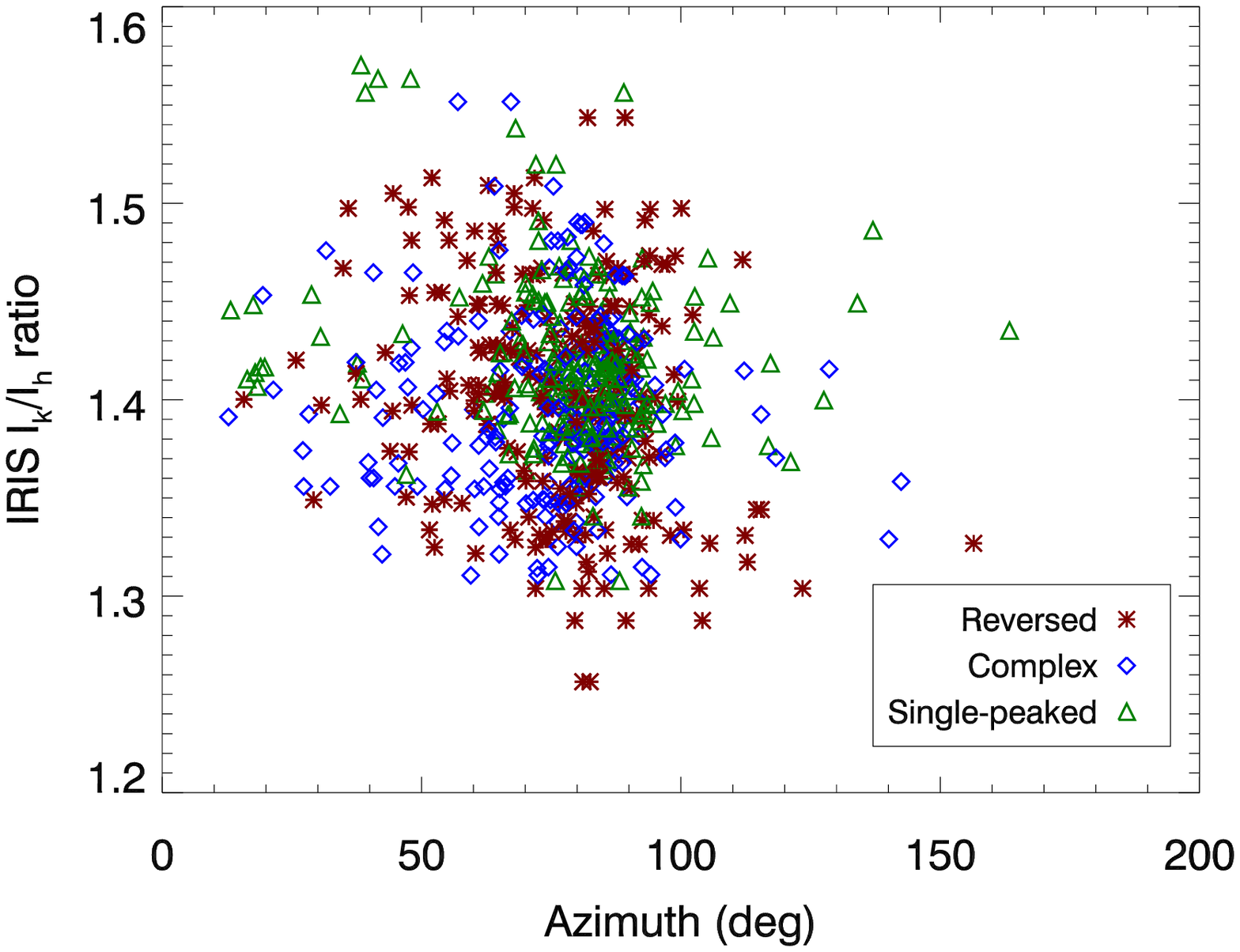}
\caption{Plots showing the \ion{Mg}{ii} k/h ratio  against  magnetic field parameters. Left column is points in the tornadoes, right column is points everywhere else in the prominence. Red asterisks are points where \ion{Mg}{ii} k line profiles are reversed, green triangles are where \ion{Mg}{ii} k profiles are single peaked. Blue diamonds correspond to complex profiles, described in Sect.~\ref{ssec:iris_diagnostics}. {  Inclination is with respect to the local vertical, and azimuth is with respect to the line of sight.}}
\label{fig:khrat}
\end{center}
\end{figure*}

Figure \ref{fig:khrat} shows the k/h ratio against the magnetic field parameters, with symbols and colours having the same meaning as in Fig.~\ref{fig:k2k3}. 
%The symbols and colours have the same meanings as in Figure \ref{fig:k2k3}. 

In the left-column panels (in the tornadoes) we see again that the data is more clustered towards higher field strengths than in the right-hand column, where there are many more points at lower field strength values. 
We find a similar behaviour to that seen in Fig.~\ref{fig:k2k3} in terms of inclination and azimuth. 
We find k/h ratios similar to those reported previously of between 1 and 2. %, however we have ratios stretching to much higher values too -- as high as 3.5 in some places. 
%This may suggest that radiative excitation is playing a more important role here (perhaps due to relatively lower densities) than in other prominence observations with \textit{IRIS}, especially in the tornadoes where most of the points are clustered at values of k/h above 2 (Fig.~\ref{fig:iris_hist}). 
We take moments of the k/h ratio distributions (Fig.~\ref{fig:iris_hist}, middle panel) to find that in both the tornadoes and the rest of the prominence the ratio has a mean value of 1.41, {  with standard deviation of 0.05}. %, and in the rest of the prominence it is lower, at 2.16. 

We also note that there are many more reversed and complex profiles in the tornado case than single peaked profiles. 
In the rest of the prominence we see relatively more single peaked profiles.

\subsection{\ion{Mg}{ii} k$_3$ Doppler shift}

The k$_3$ feature at the \ion{Mg}{ii} k line centre corresponds to the most optically thick part of that line. 
When we observe features of a prominence at wavelengths near line centre, we are really only looking at emission from the front-most layer(s) of material. 
Emission from material behind that is almost all re-absorbed or scattered out of the line of sight at these wavelengths. 
By then measuring the deviation of this k$_3$ feature of reversed profiles from rest wavelength, we can theoretically measure the velocity of the parts of the prominence emitting in \ion{Mg}{ii} that are closest to us, the observer. 
%With an increase in magnetic field strength, and hence magnetic pressure, we may expect also an increase in the velocity of the plasma in the magnetic field. \textbf{Would we? -- NL: Not sure. Gas pressure should be invoked too.}
Does an increase in magnetic field strength, and hence magnetic pressure, corresponds to higher plasma velocities? This is what could be expected in a low-$\beta$ plasma -- observed motions in a low-$\beta$ plasma will be caused by the magnetic pressure. 

\begin{figure*}
\begin{center}
\includegraphics[width=0.45\hsize,clip=true,trim= 2.1cm 0 0.6cm 0]{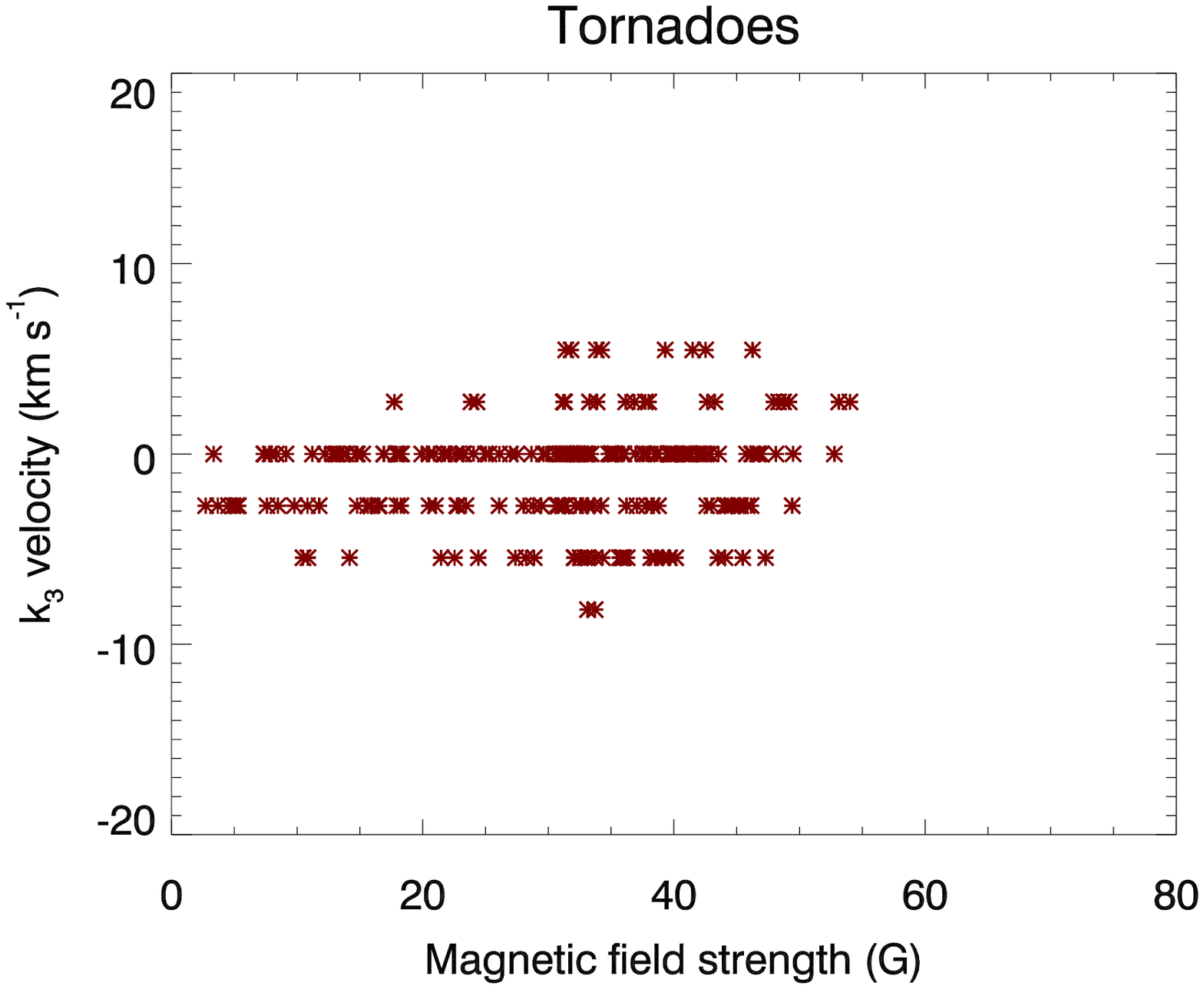}
\includegraphics[width=0.45\hsize,clip=true,trim= 2.1cm 0 0.6cm 0]{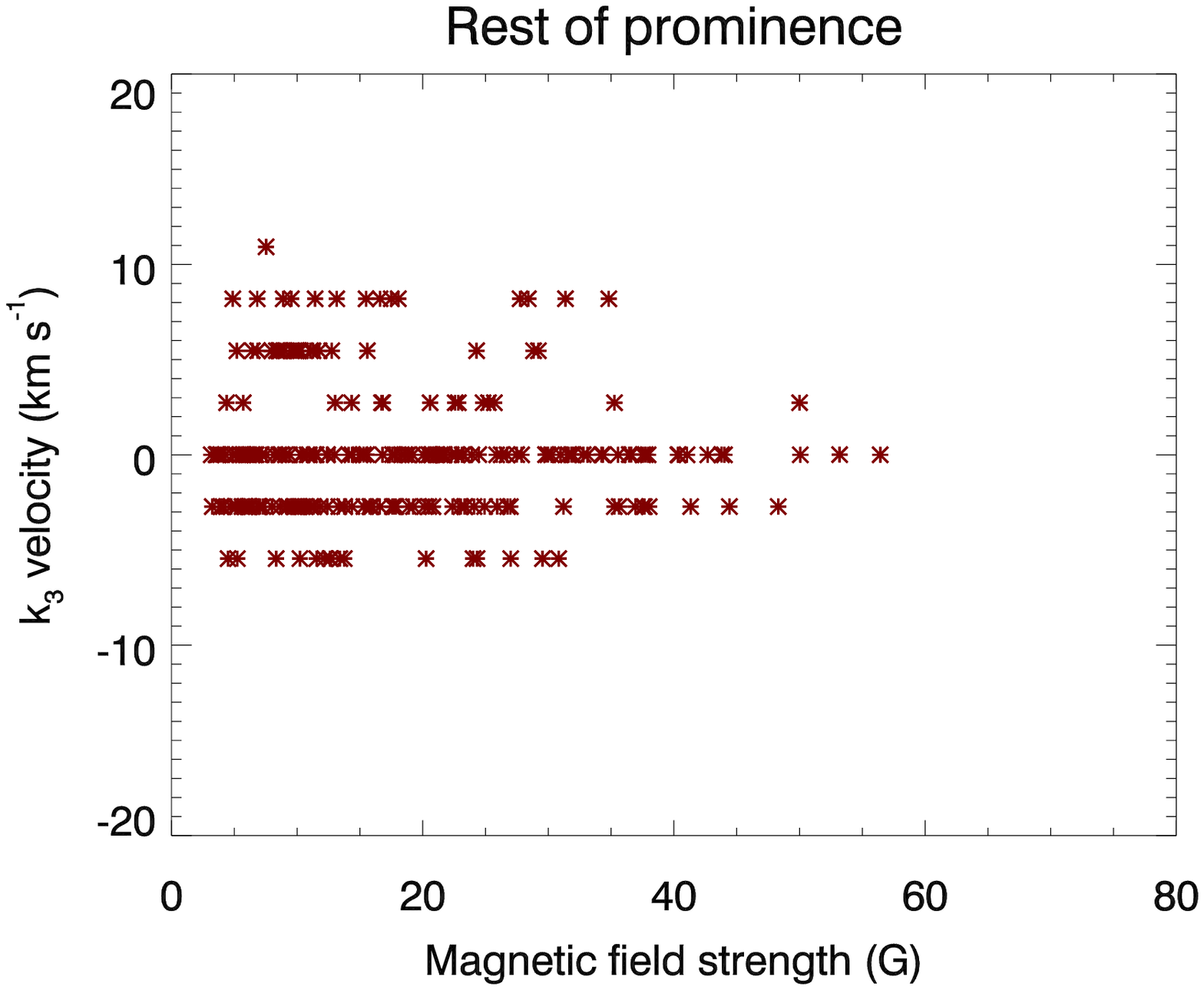}
\\
\includegraphics[width=0.45\hsize,clip=true,trim= 2.1cm 0 0.6cm 0]{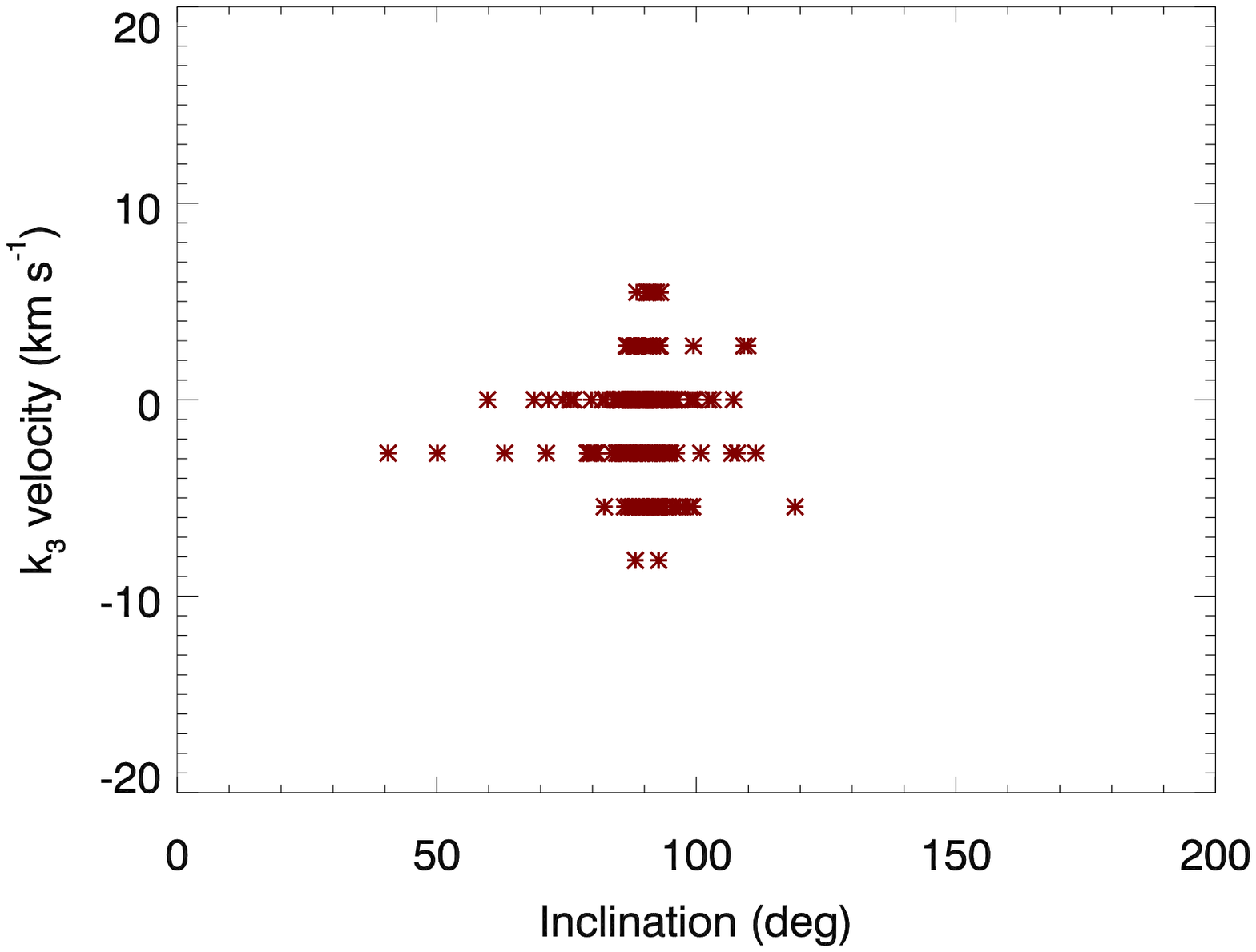}
\includegraphics[width=0.45\hsize,clip=true,trim= 2.1cm 0 0.6cm 0]{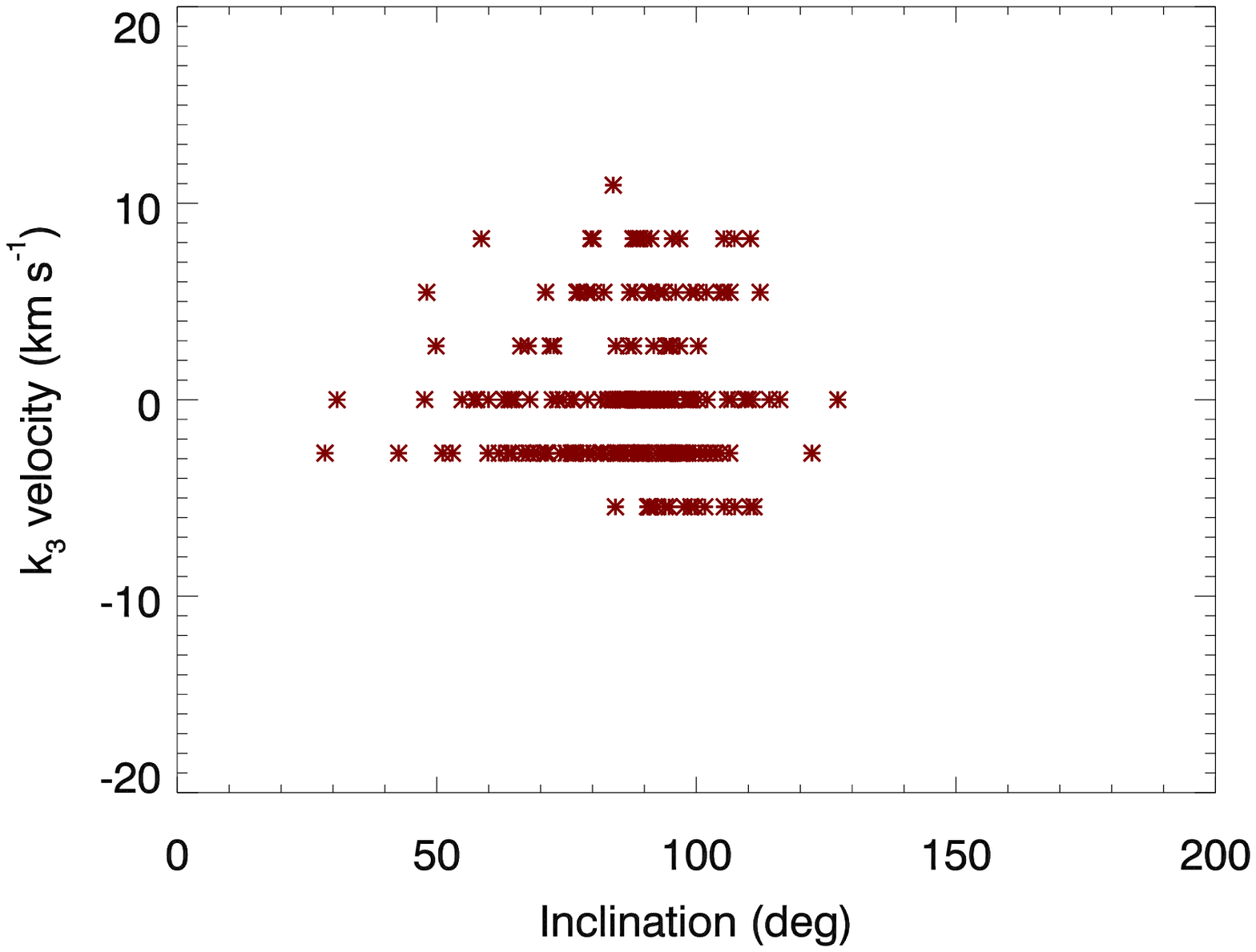}
\\
\includegraphics[width=0.45\hsize,clip=true,trim= 2.1cm 0 0.6cm 0]{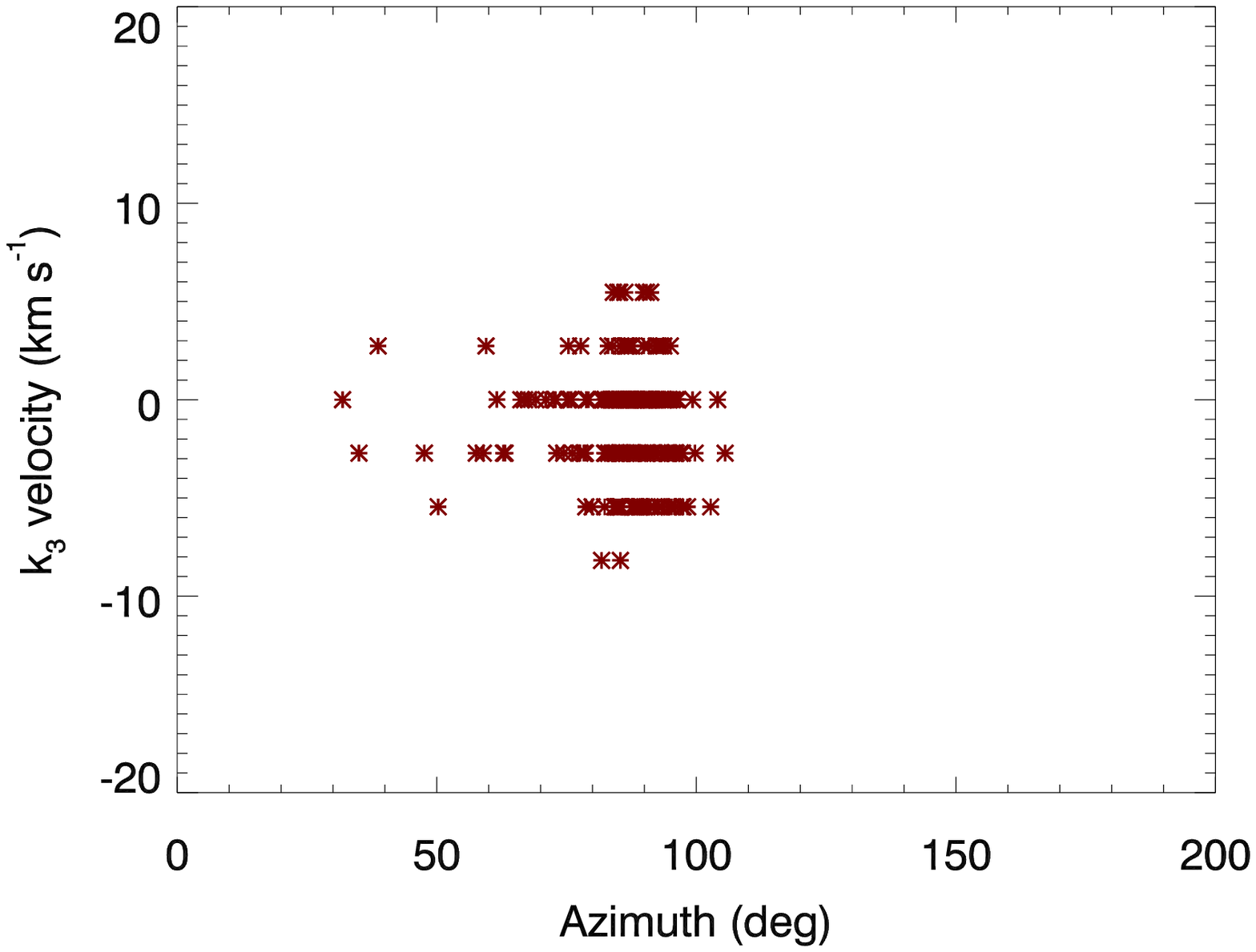}
\includegraphics[width=0.45\hsize,clip=true,trim= 2.1cm 0 0.6cm 0]{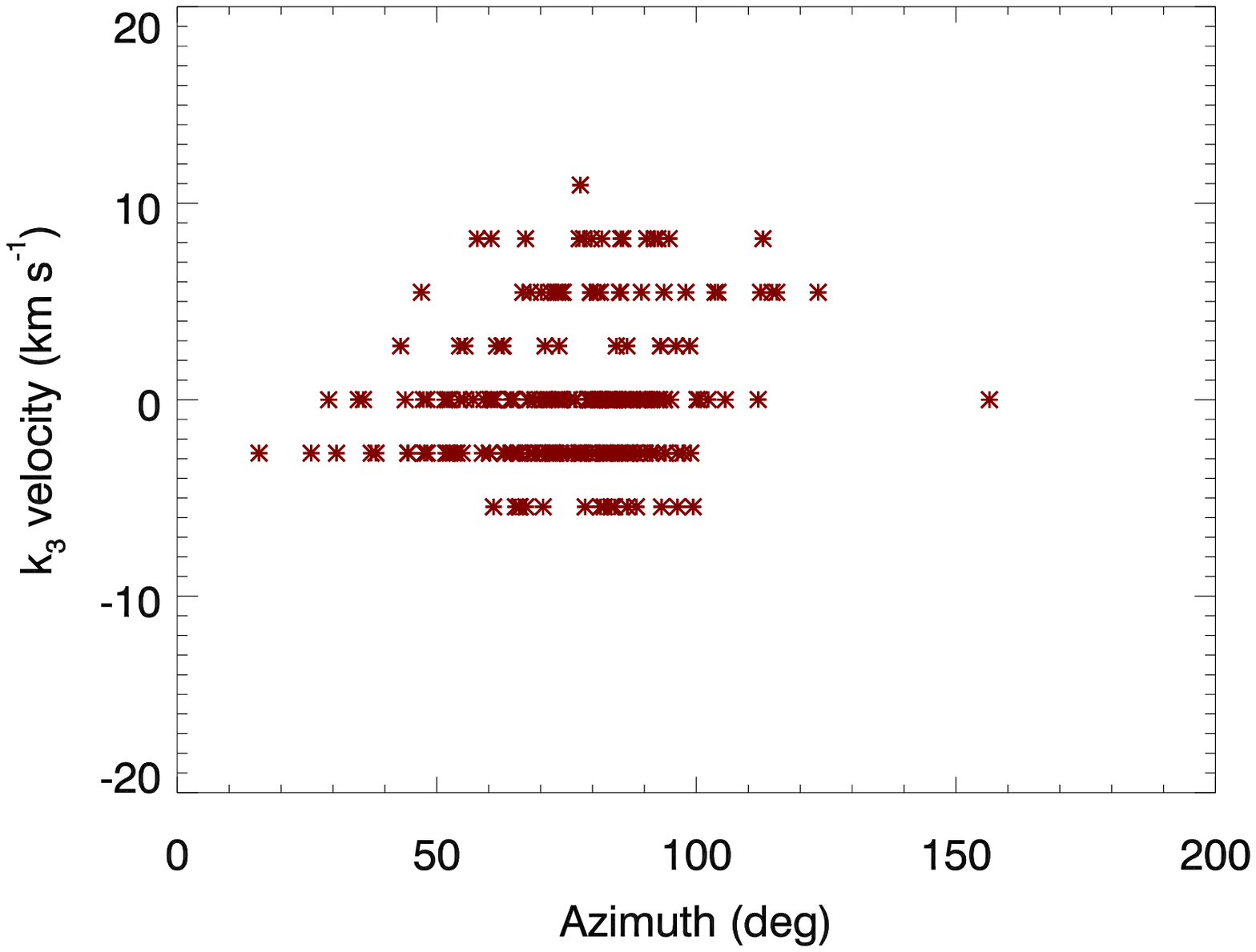}
\caption{Plots showing the \ion{Mg}{ii} k$_3$ Doppler shift, relative to the nominal rest wavelength of \ion{Mg}{ii},  against magnetic field parameters. Left column is points in the tornadoes, right column is points everywhere else in the prominence. Points are all taken from pixels where \ion{Mg}{ii} profiles are reversed. {  Inclination is with respect to the local vertical, and azimuth is with respect to the line of sight.}}
\label{fig:vk3}
\end{center}
\end{figure*}

Figure \ref{fig:vk3} shows plots of k$_3$ Doppler shift versus the magnetic field parameters, the same as seen in Figs.~\ref{fig:k2k3} and \ref{fig:khrat}. 
The points are at discrete values of velocity due to the spectral resolution of \textit{IRIS}. 
The position of the k$_3$ minimum cannot be measured more {accurately} without fitting the curve with a Gaussian, for example. 
%This is not the approach we take here, due to the fact that assuming a Gaussian profile is assuming a lot about the shape of the absorption profile itself. 
{  However, there is no reason to assume a Gaussian profile for an optically thick line. }
{We also note that the actual Doppler shifts encapsulated in each discrete point cover a range of v$_{k3}$ values. 
This is conveyed in the bottom panel of Fig.~\ref{fig:iris_hist}, where bin sizes reflect the spectral resolution of \textit{IRIS}.}

In the case of the k$_3$ Doppler shift versus magnetic field strength in the tornado (top left panel of Fig.~\ref{fig:vk3}), we see that higher l.o.s. velocities are found at higher magnetic field strengths, and at lower magnetic field strengths the k$_3$ Doppler shift is much closer to zero. 
However, in the rest of prominence case, the opposite appears to be true. %\textbf{Why. -- NL: May be related to differences in magnetic field structure in tornadoes and in the rest of the prominence (loop structure).}
This could be an indication that the plasma $\beta$ is lower in the tornado than in the rest of the prominence.
In all cases these velocities are within around $\pm$ 10 km s$^{-1}$, which is a similar value to that found previously in quiescent prominences with \textit{IRIS} (Paper I).
Comparing k$_3$ Doppler shift to inclination and azimuth of the magnetic field, we see again the inclination clustered around 90$^\circ$, and azimuth values of around 80--90$^\circ$ in the tornado points. 
%We also see a tail to lower inclinations and azimuths in the rest of prominence case, these points mostly having k$_3$ velocities close to zero. % similar pattern to those seen in the examples of Figures \ref{fig:k2k3} and \ref{fig:khrat}.

%__________________________________________________________________

\section{Correlation between THEMIS and EIS data}
\label{sec:eis_correlation}

In order to identify correlations between the profiles observed by EIS and the magnetic field parameters from THEMIS, we first need to characterise the spectral lines seen by EIS. 
To do this we follow the outline laid out by \citet{Young2007}, and fit Gaussian profiles to the spectral lines. 
As noted in that paper and elsewhere, complex blended lines are found in EIS data, many of which can be seen in prominence observations \citep[see][for more details]{Labrosse11,Levens2015}. 
We here follow a similar strategy to \citet{Levens2015} in the de-blending of these lines, where necessary. 

In the EIS study used here, spectral lines formed at a range of plasma temperatures are available -- from \ion{He}{ii} 256~\AA, formed at $\sim 30000$ K, up to \ion{Fe}{xv} at 283~\AA, formed at $\sim 2.5$ MK. 
Unfortunately the lower temperature lines available are mostly either part of a large blend with hotter components (such as \ion{He}{ii} and the \ion{O}{v} lines at 192~\AA), or suffer from poor signal-to-noise ratios (such as the \ion{O}{vi} lines around 184\AA\ or the \ion{Si}{vii} lines at 274 and 275~\AA). 
%As was mentioned in \ref{ssec:eisaia}, the \ion{He}{ii} 256\AA\ line could be de-blended well enough for co-alignment purposes, however we do not trust the assumptions made in the de-blending process enough to do any detailed spectral analysis on that line. 
%We therefore look to another line, or set of lines to compare to the magnetic field parameters from THEMIS. 

We ideally want to study two lines formed by the same ion whose intensity ratio shows some density sensitivity, a process outlined in \citet{Young2007}. 
For this analysis we chose the two \ion{Fe}{xii} lines at 195.119~\AA\ (the strongest line as observed by EIS) and 195.179~\AA\ which are de-blended by fitting two Gaussian profiles whose centroid positions are tied at a fixed distance apart. 
These are the same lines used by \citet{Levens2015} on another tornado-like prominence. 
Figure \ref{fig:eis_int} shows the two intensity maps for these lines. 
The prominence can be seen as two dark columns, with the northern column being much more visible than the southern one, as is the case in coronal AIA images (Figs.~\ref{fig:aiaboxed}, \ref{fig:eis_int}). 

{\ion{Fe}{xii} is formed at a plasma temperature of $1.5 \times 10^6$~K, much higher than the temperatures expected in a prominence. 
However, as previous analysis has shown \citep{Levens2015}, it appears that the tornado structure can be traced to temperatures as high as this through a PCTR, with hot plasma seemingly forming a `sheath' around the cool core. 
In fact, the core of the prominence is formed from many threads and sheaths may exist along each thread, as was suggest in the multi-threaded model of prominence formation \citep{Luna2012}.}
%It is also shown in \citet{Levens2015}, through differential emission measure (DEM) analysis, that there is more contribution at all temperatures when looking towards the tornado. 
%This suggests that there is hot plasma in the region of the tornado, which is consistent with the model of multi-thread prominence formation \citep{Luna2012}.}

\begin{figure}
\begin{center}
\includegraphics[width=0.9\hsize,clip=true,trim=6cm 0 3.8cm 1.9cm]{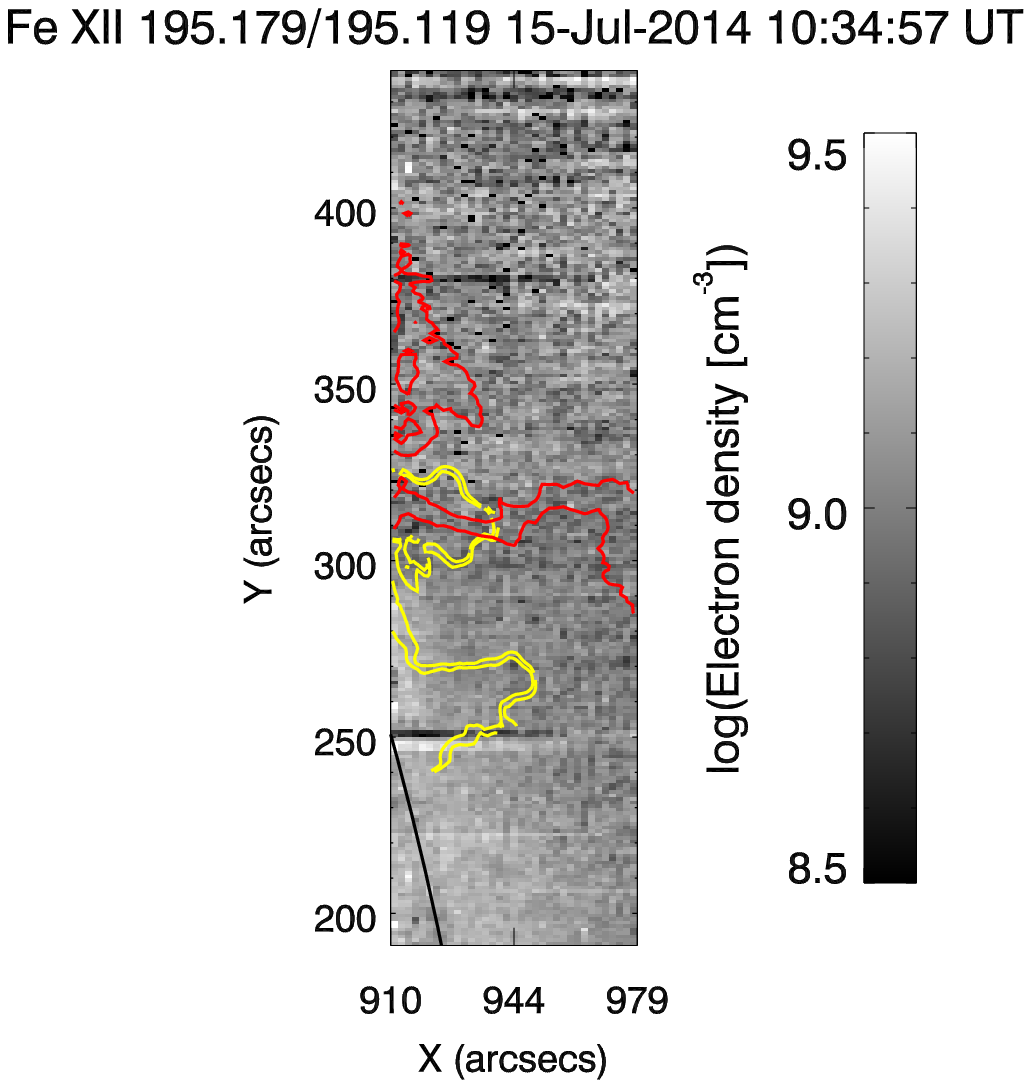}
\caption{Density map of the prominence region using the density sensitive \ion{Fe}{xii} lines at 195.119~\AA\ and 195.179~\AA. Yellow contours show outlines of the tornadoes as seen by THEMIS. {Red} contours are {22\% and 25\%} of the EIS 195.119~\AA\ {intensity}, {same as white contours} in Fig.~\ref{fig:eis_int}. Black line shows the solar limb position.}
\label{fig:eis_dens}
\end{center}
\end{figure}

These two iron lines show density sensitivity across a range of densities, so are suitable for this analysis. 
Using atomic data from CHIANTI v8.0 \citep{Dere1997,DelZanna2015} we can create a density map at the formation temperature of these two lines ($1.5 \times 10^6$~K). 
The result of this is shown in Fig.~\ref{fig:eis_dens}. 

\begin{figure}
\begin{center}
\includegraphics[width=0.9\hsize,trim=3cm 0 2cm 0,clip=true]{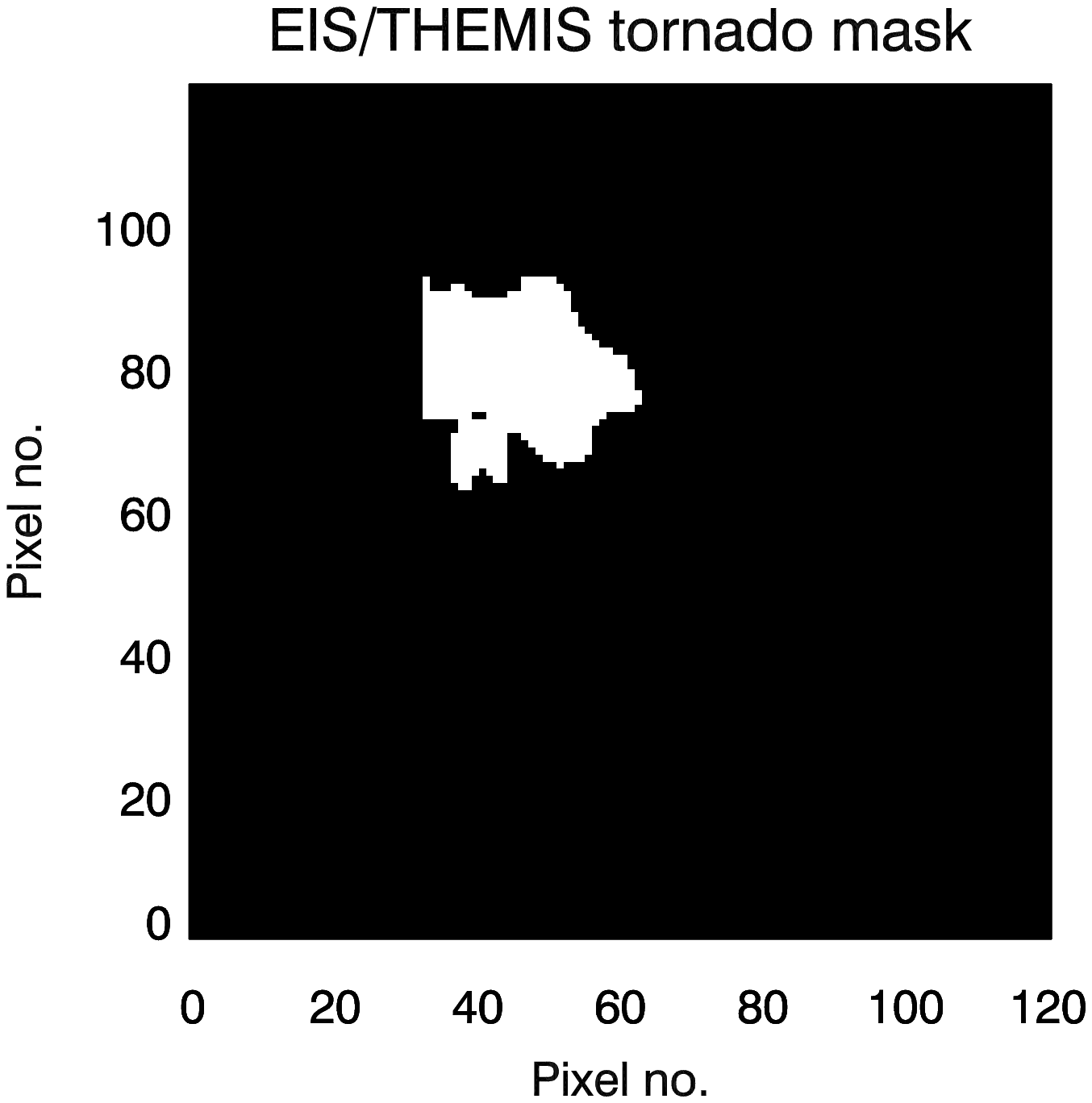}
\includegraphics[width=0.9\hsize,trim=3cm 0 2cm 0,clip=true]{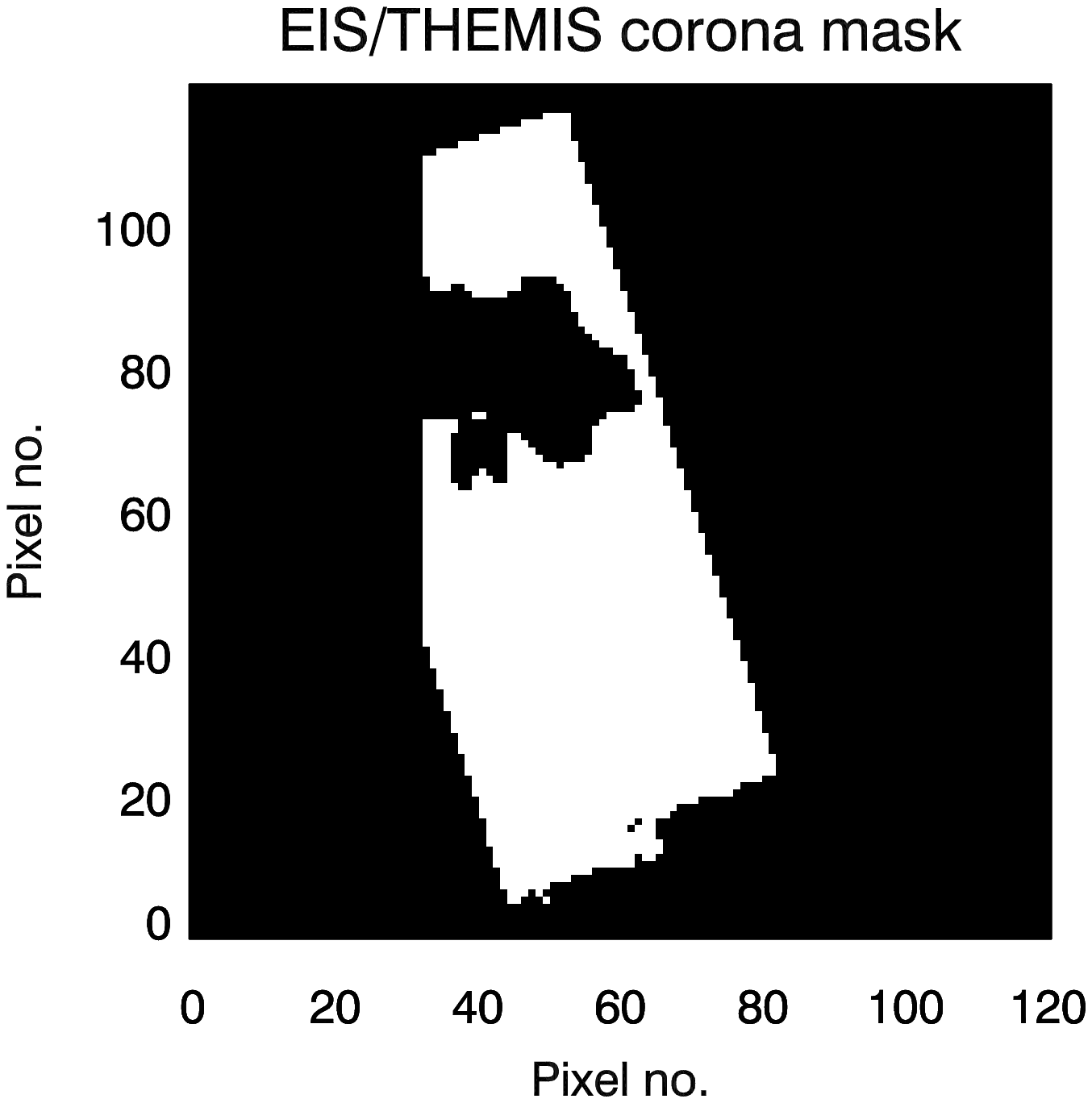}
\caption{Overlapped area between EIS and THEMIS maps for data sets on 15 July 2014. White area shows overlap. \textit{Top:} Area including the northern tornado, which is visible in the EIS 195 \AA\ line. \textit{Bottom:} Area including the surrounding corona.}
\label{fig:overlap_eis}
\end{center}
\end{figure}

\begin{figure*}
\begin{center}
\includegraphics[width=0.45\hsize,clip=true,trim= 1.8cm 0 0.6cm 0]{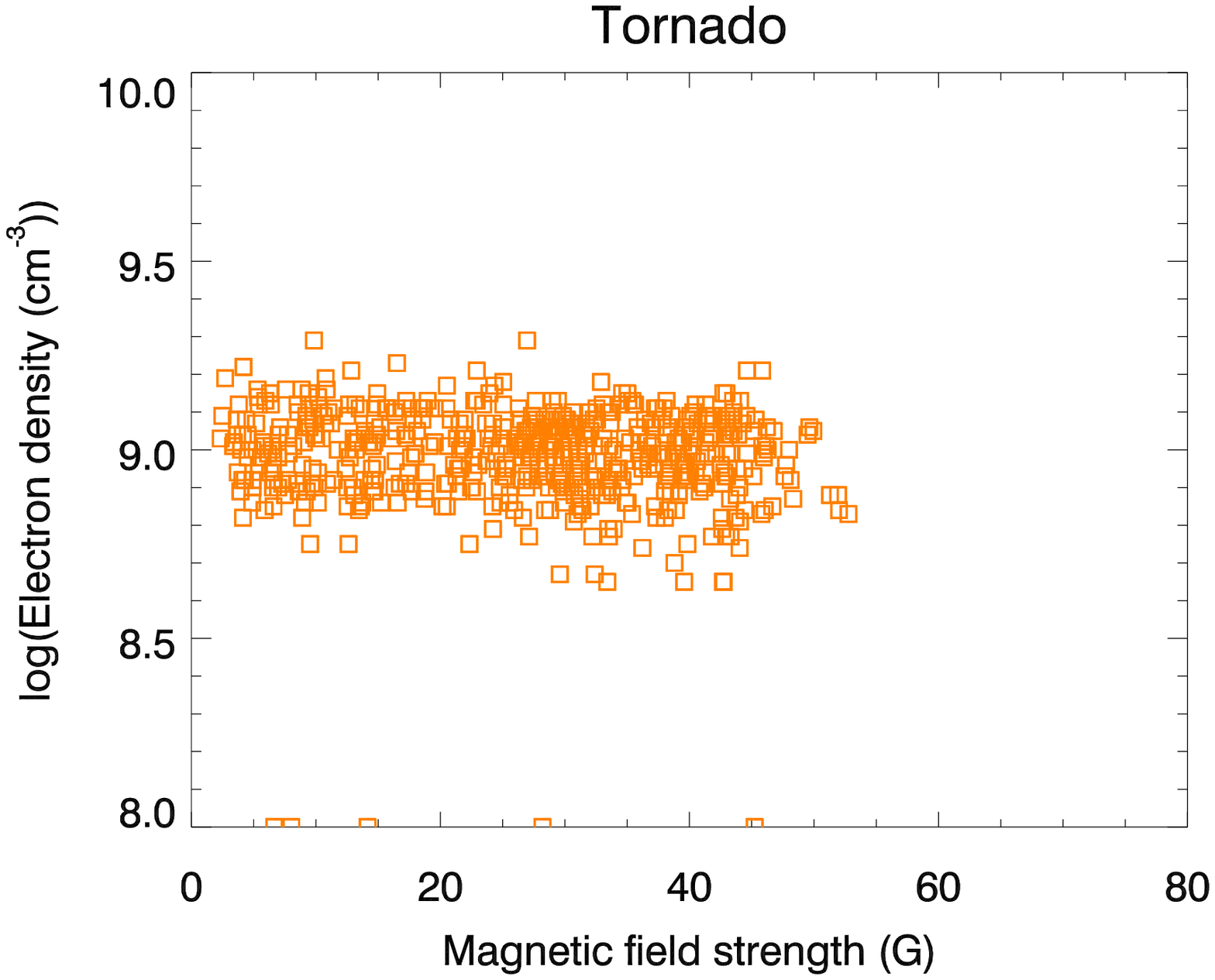}
\includegraphics[width=0.45\hsize,clip=true,trim= 1.8cm 0 0.6cm 0]{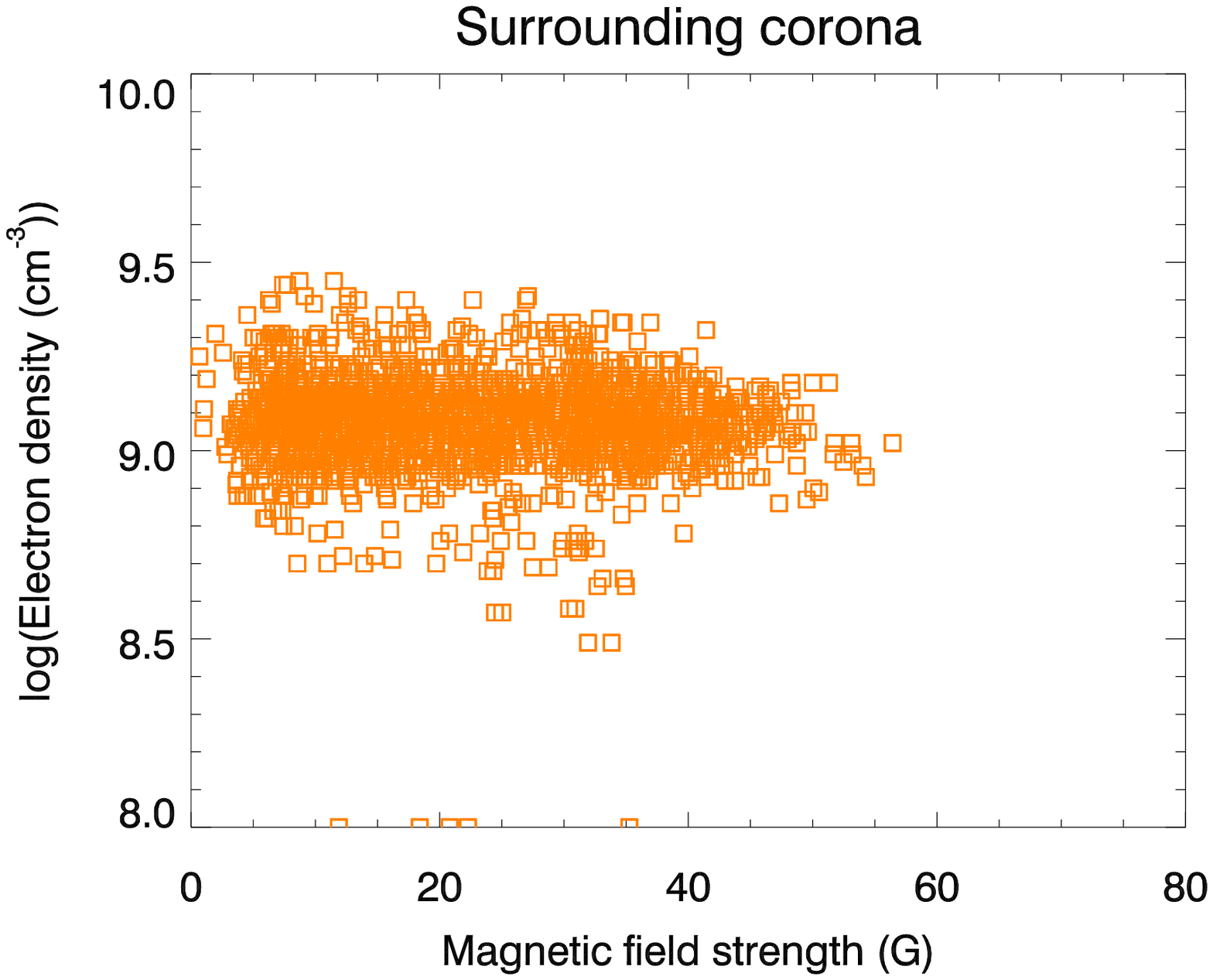}
\\
\includegraphics[width=0.45\hsize,clip=true,trim= 1.8cm 0 0.6cm 0]{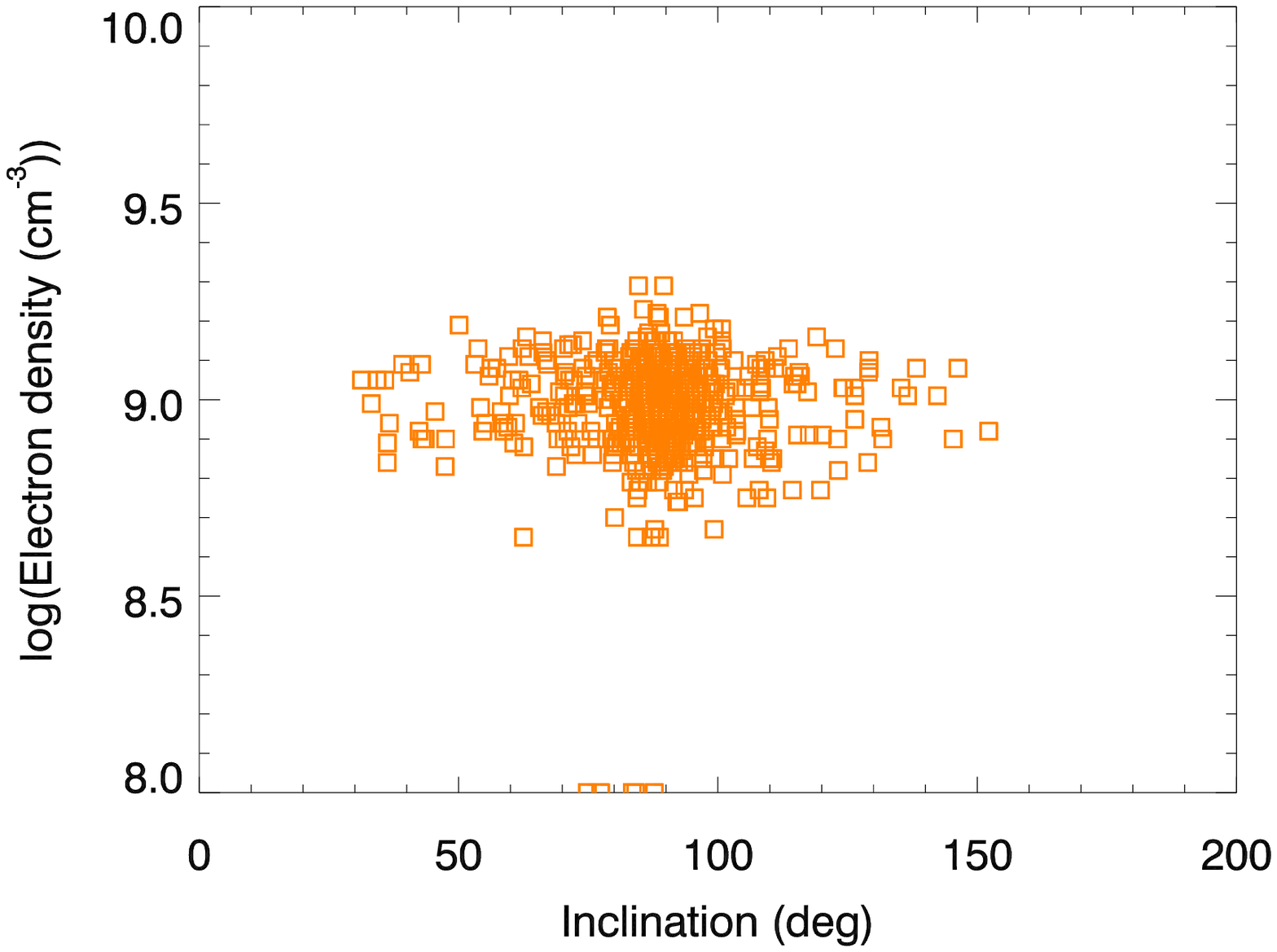}
\includegraphics[width=0.45\hsize,clip=true,trim= 1.8cm 0 0.6cm 0]{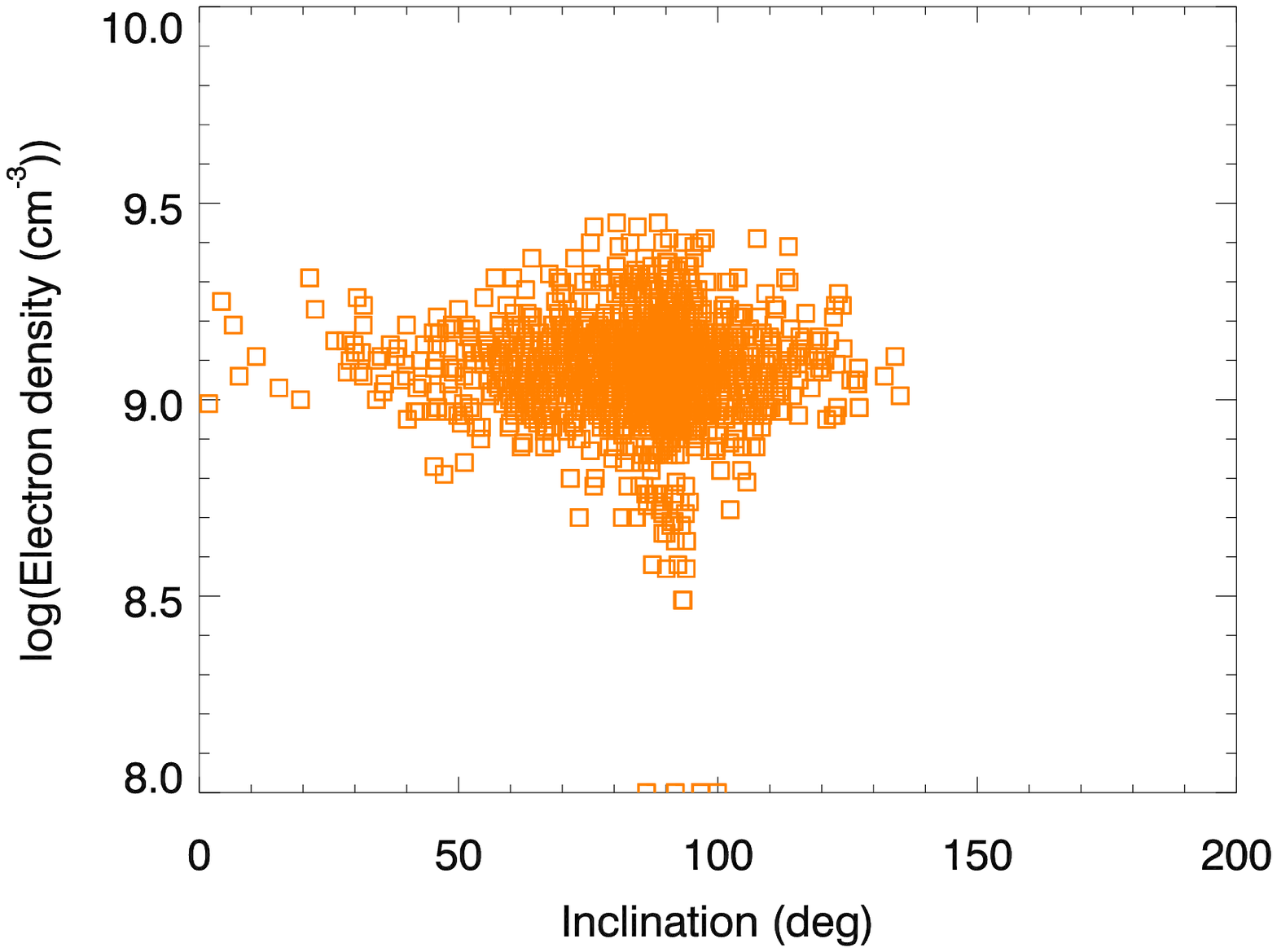}
\\
\includegraphics[width=0.45\hsize,clip=true,trim= 1.8cm 0 0.6cm 0]{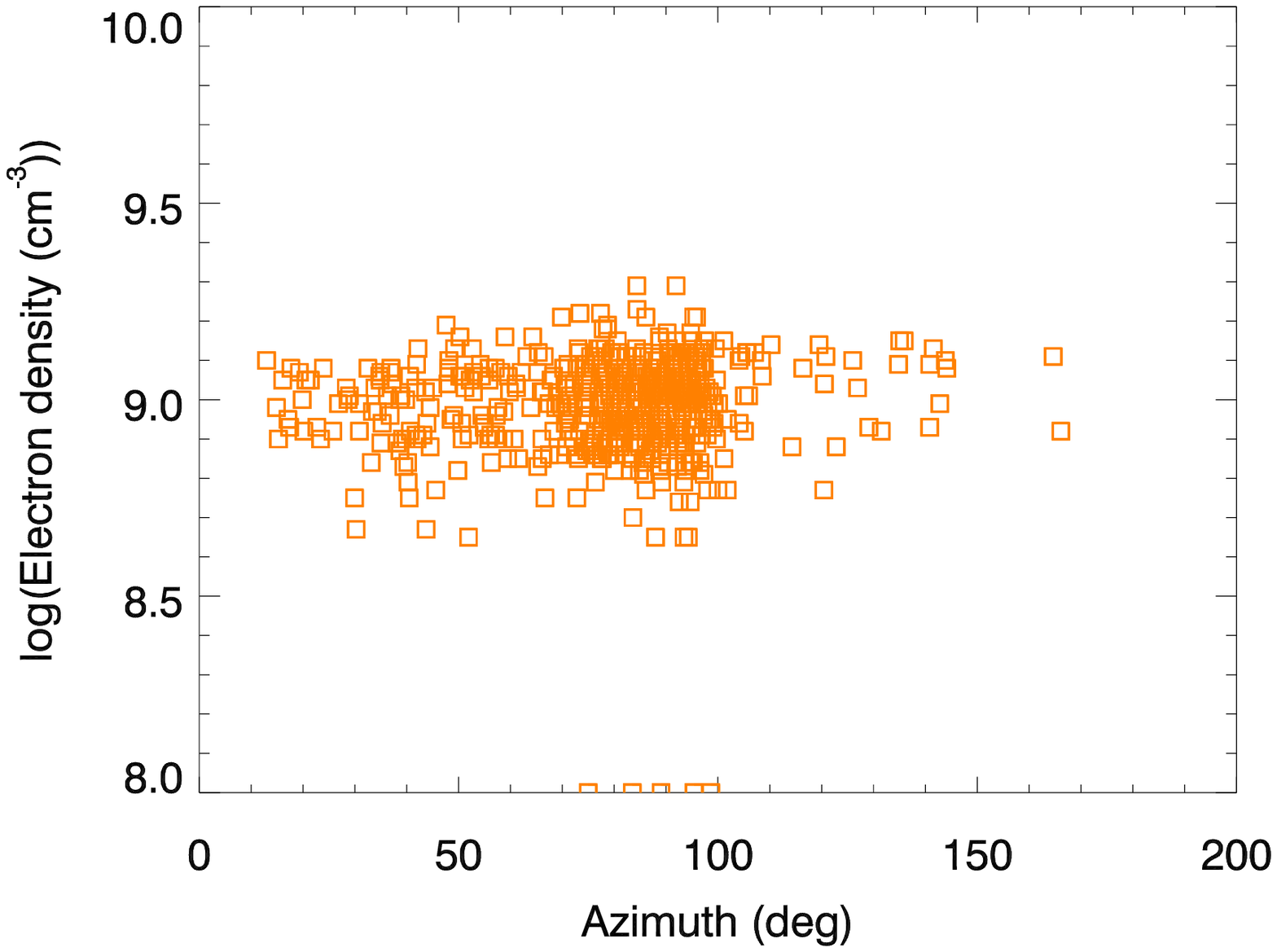}
\includegraphics[width=0.45\hsize,clip=true,trim= 1.8cm 0 0.6cm 0]{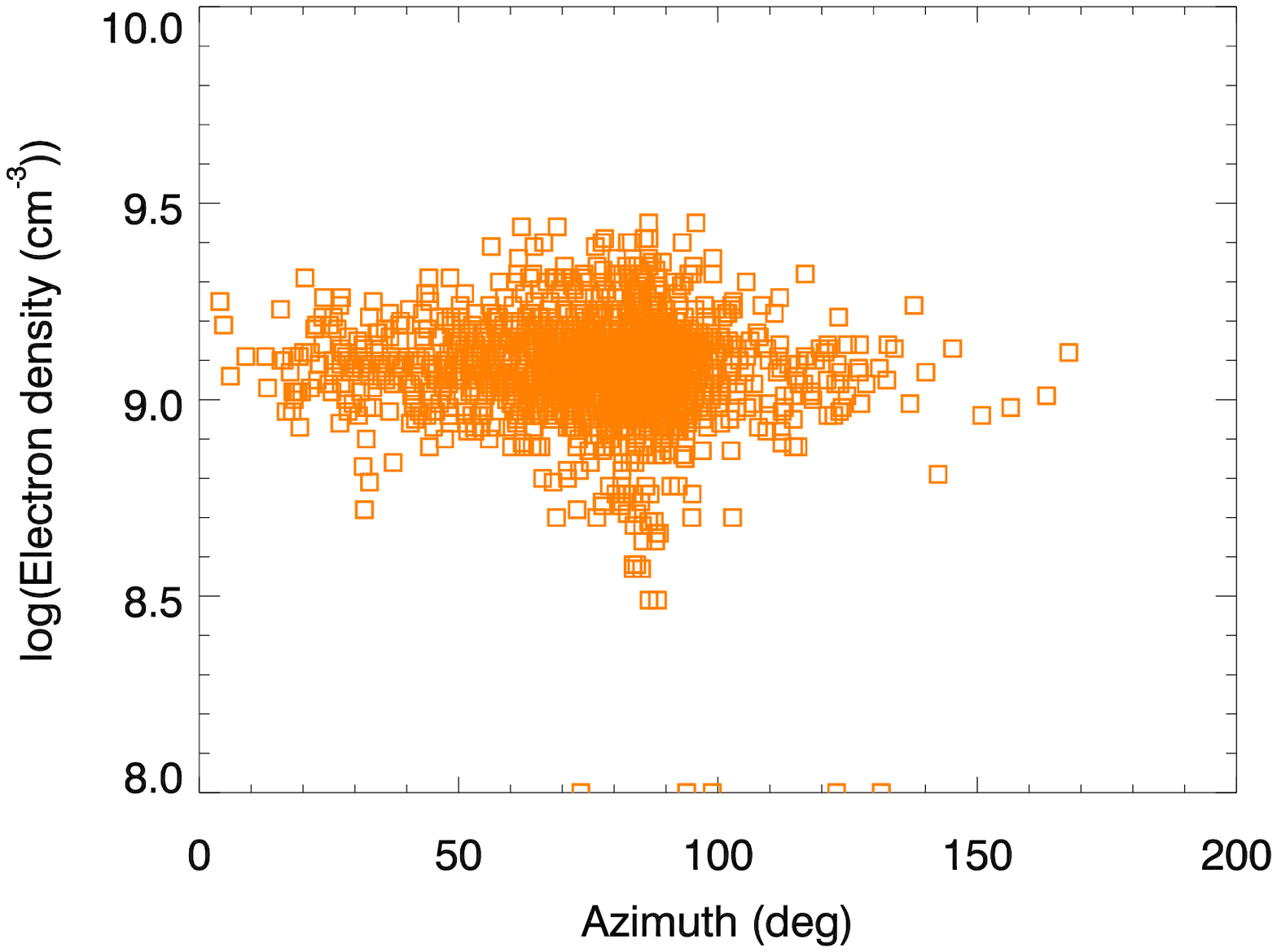}
\caption{Plots showing the electron density, calculated using the intensity ratio of the \ion{Fe}{xii} lines 195.119/195.179 observed by EIS, against magnetic field parameters. Left column is points in the tornadoes, right column is points in the surrounding corona.}
\label{fig:dens_scatter}
\end{center}
\end{figure*}

\begin{figure}
\begin{center}
\includegraphics[width=0.9\hsize,trim=1cm 0 0.7cm 1cm,clip=true]{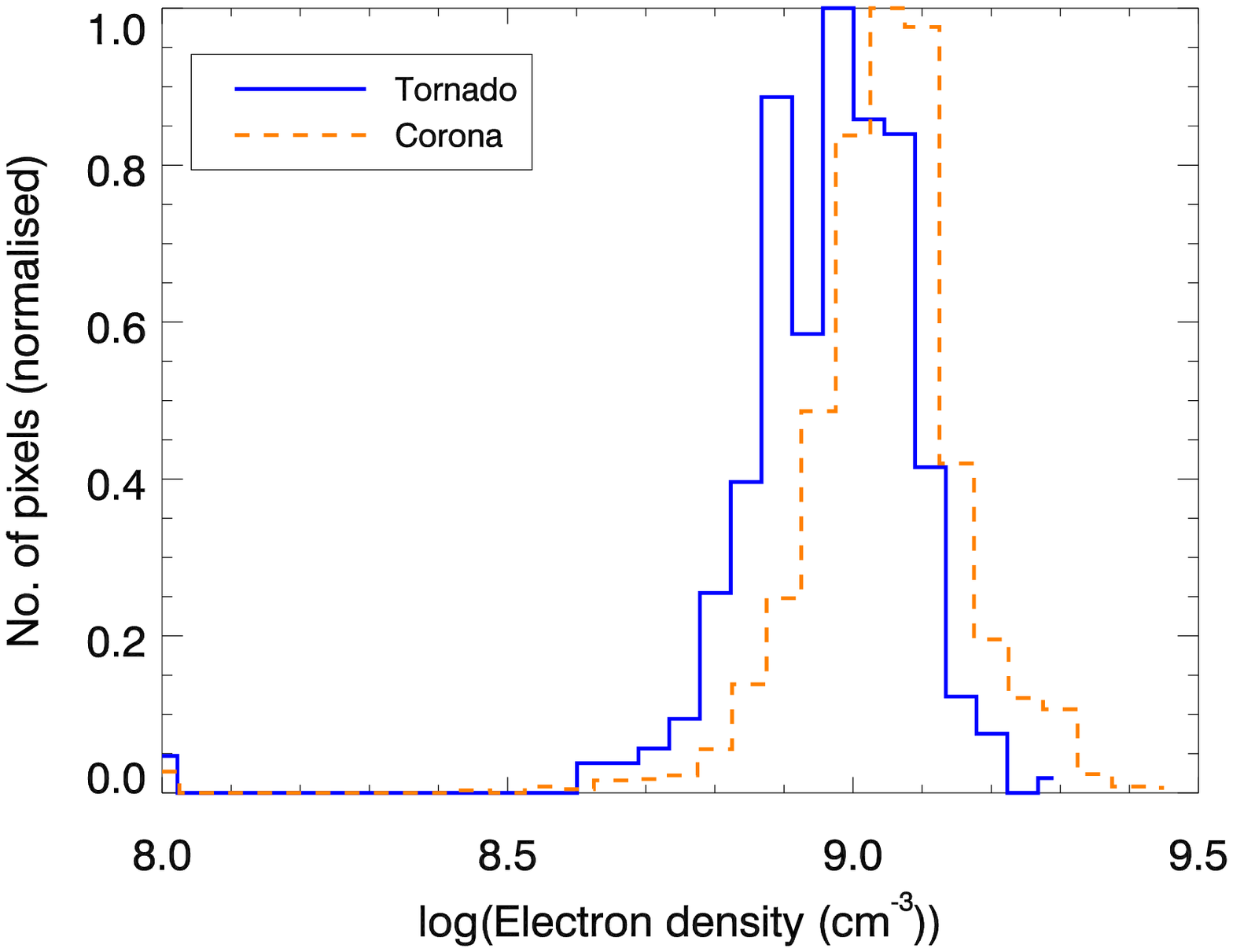}
\caption{Histograms of electron density showing the distribution of values in both the northern tornado ($\sim$ 600 pixels) and the surrounding corona ($\sim$ 3000 pixels).}
\label{fig:eis_hist}
\end{center}
\end{figure}

We then need to calculate the overlapping region between the EIS raster and the THEMIS raster, as was done with \textit{IRIS} in Sect.~\ref{sec:correlation}. 
The overlap masks for EIS and THEMIS are shown in Fig.~\ref{fig:overlap_eis}. 
Notably for this correlation we only consider `tornado' points to be those in the northern tornado. 
This is done because the southern tornado is barely visible in the \ion{Fe}{xii} lines considered here, mostly blocked by bright coronal emission in front of it. 
%It would therefore be false to claim that the emission in the region that is visible in the EIS data is coming from the tornado and not the surrounding corona, so we omit it from the `tornadoes' case. 
Following this, we can perform a statistical analysis between points in the EIS and THEMIS rasters. 
Figure \ref{fig:dens_scatter} shows the scatter plots of electron density versus magnetic field parameters from THEMIS. 
The panels in this plot are in the same order as those in Figures \ref{fig:k2k3}, \ref{fig:khrat} and \ref{fig:vk3}. 
We see in these plots a relatively small scatter in densities, with most points having {$\log{n_e}$ between 8.5 and 9.5}. %a density of between 10$^{8.5}$ and 10$^{9.5}$ cm$^{-3}$. 
%The scatter is much tighter in the rest-of-prominence case. %, however, with the points less scattered than in the tornadoes case. 
The mean electron density at $T\sim 1.5 \times 10^6$~K is lower in the tornado than in the surrounding corona, as is clear by looking at the histogram of Fig.~\ref{fig:eis_hist}. % (as the rest of the prominence is not visible at the wavelengths used here). 
%Taking moments of these distributions allow us to give a value to the mean electron density in each case. 
%In the tornado the mean density is $0.95 \times 10^{9}$ cm$^{-3}$, with standard deviation of 1.38 cm$^{-3}$, whereas in the surrounding corona the value is $1.15 \times 10^{9}$ cm$^{-3}$, also with standard deviation of 1.38 cm$^{-3}$.
In the tornado the mean density is $\log{n_e} = 8.98 \pm 0.14$, whereas in the surrounding corona the value is $\log{n_e} = 9.06 \pm 0.14$, {with $n_e$ in units of cm$^{-3}$}.
This is comparable with the results of \citet{Levens2015}.
The lower density in the tornado could be due to the volume blocking effect of cool material in the tornado region \citep{Heinzel2008} -- cool material in the tornado means that there is less hot \ion{Fe}{xii} plasma along that line of sight. 

%__________________________________________________________________

\section{Correlation between IRIS and EIS data}
\label{sec:iris_eis}

\begin{figure}
\begin{center}
\includegraphics[width=0.9\hsize,clip=true,trim= 1.8cm 0 0.6cm 1cm]{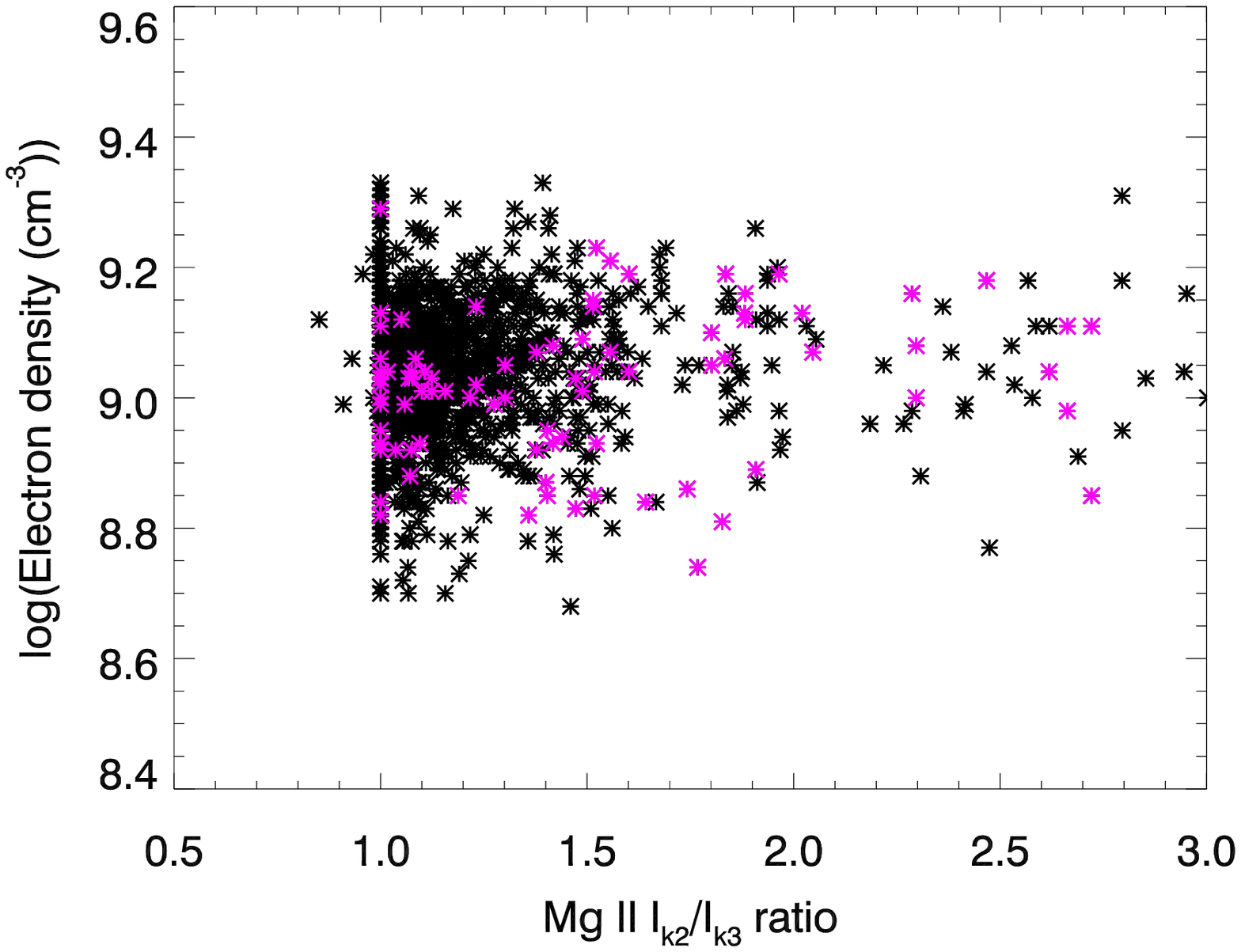}
\includegraphics[width=0.9\hsize,clip=true,trim= 1.8cm 0 0.6cm 1cm]{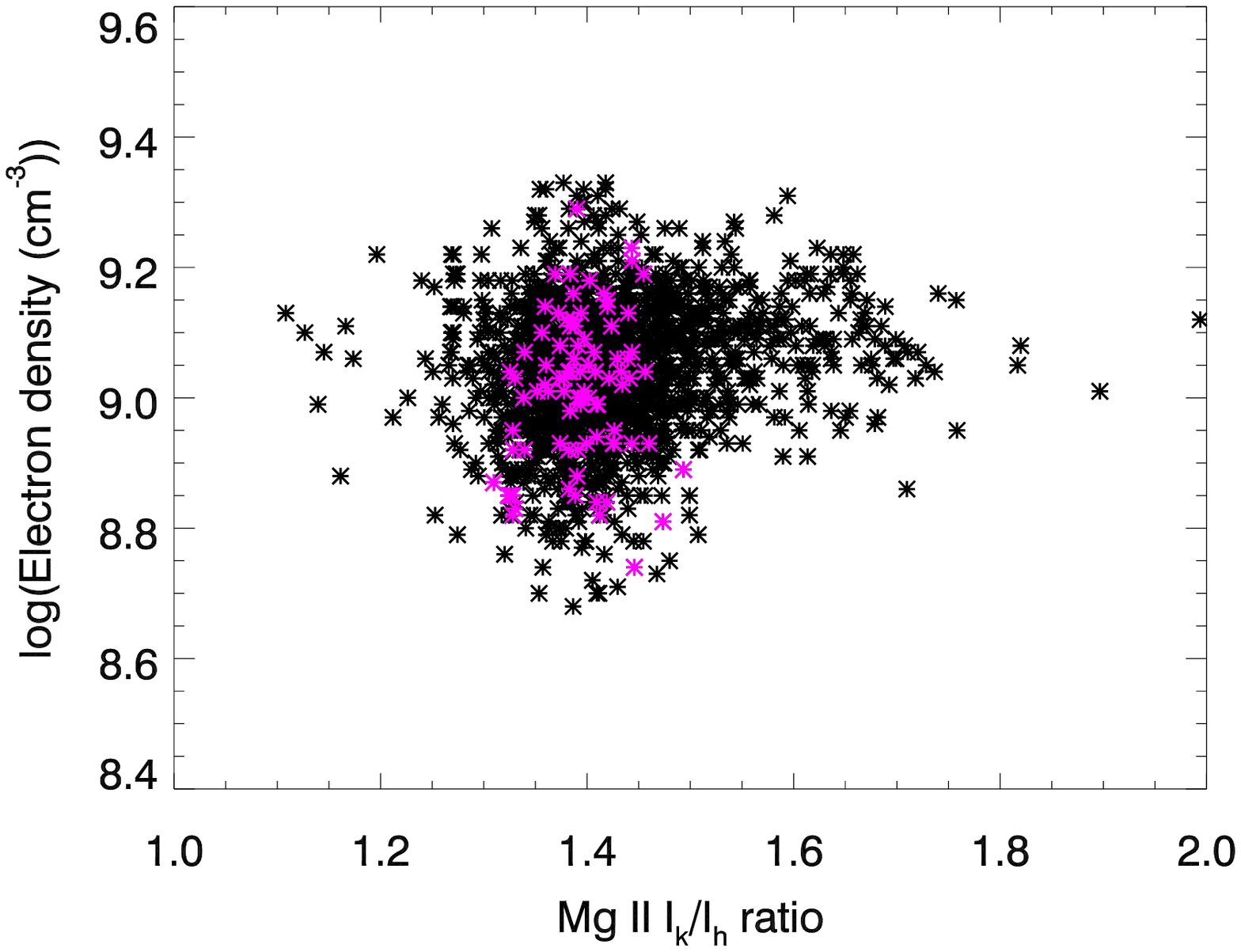}
\includegraphics[width=0.9\hsize,clip=true,trim= 1.8cm 0 0.6cm 1cm]{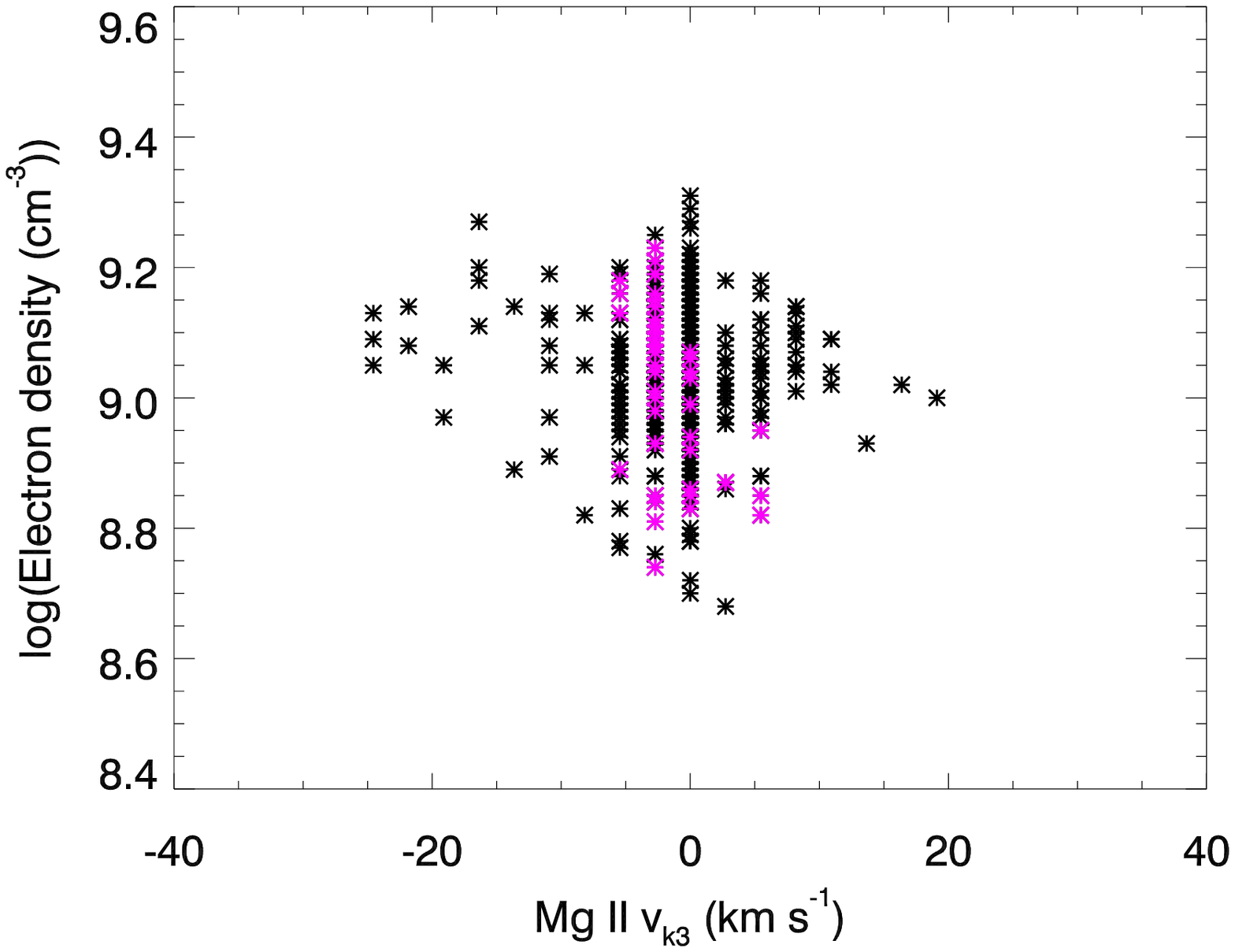}
\caption{Correlation plots showing electron density, calculated using the intensity ratio of the \ion{Fe}{xii} lines 195.119/195.179 observed by EIS, against: \textit{Top:} \ion{Mg}{ii} k$_2$/k$_3$ ratio. \textit{Middle:} \ion{Mg}{ii} k/h ratio. \textit{Bottom:} \ion{Mg}{ii} k$_3$ Doppler shift, from \textit{IRIS}. Magenta points show points in the northern tornado in both rasters. Black points are for the rest of the overlapped region.}
\label{fig:iris_eis}
\end{center}
\end{figure}

We also look for correlations between the \ion{Mg}{ii} line ratios and plasma parameters and the electron densities from EIS. 
Figure \ref{fig:iris_eis} shows plots of electron density versus (from top to bottom) \ion{Mg}{ii} k$_2$/k$_3$ ratio, k/h ratio and k$_3$ Doppler shift. 
As can be seen in Fig.~\ref{fig:aiaboxed}, the \textit{IRIS} raster lies entirely inside the EIS raster, so the overlap area is the \textit{IRIS} raster field of view. 
The EIS spatial resolution is lower than that of \textit{IRIS}, so the \textit{IRIS} data is binned spatially to match the EIS resolution, as described in Sect.~\ref{sec:correlation} for comparing \textit{IRIS} and THEMIS data.

The points are again divided into those in the northern tornado and those outside of it. 
This is done using the THEMIS brightness, although we note that this region misses out the highest altitude parts of the tornado. 
We only consider the northern tornado here, as in Sect.~\ref{sec:eis_correlation}. 

The plots in Fig.~\ref{fig:iris_eis} show that there are no meaningful correlations between the \textit{IRIS} line ratios or plasma parameters and the electron density. 
We note that the plasma emitting the lines observed by the two instruments is formed under very different circumstances: the \ion{Fe}{xii} emission seen by EIS is formed at coronal temperatures, and is optically thin, whereas the \ion{Mg}{ii} k line that \textit{IRIS} observes is formed at chromospheric temperatures, and is optically thick. 
{  This could explain why} these correlation plots do not show any patterns. 

%We note those points with a k$_2$/k$_3$ ratio of less than one in the top panel of Fig.~\ref{fig:iris_eis}. 
%These are points which are `complex', often flat topped, and the nature of the individual profiles mean that the identification of k$_2$ and k$_3$ peaks can be confused in the peak-finder algorithm. 
%As it is a small number of points we can conclude that it is the complexity of a few individual profiles that has caused this apparent `anti-reversal'. 
%We do not claim that these profiles are presenting any new physics. 

%\begin{figure}
%\begin{center}
%\caption{Correlation plot showing electron density, calculated using the intensity ratio of the \ion{Fe}{xii} lines 195.119/195.179 observed by EIS, against \ion{Mg}{ii} k/h ratio from \textit{IRIS}.}
%\label{fig:iris_eis_k_h}
%\end{center}
%\end{figure}
%
%\begin{figure}
%\begin{center}
%\caption{Correlation plot showing electron density, calculated using the intensity ratio of the \ion{Fe}{xii} lines 195.119/195.179 observed by EIS, against \ion{Mg}{ii} k$_3$ velocity from \textit{IRIS}.}
%\label{fig:iris_eis_vk3}
%\end{center}
%\end{figure}

%__________________________________________________________________

\section{Conclusion}
\label{sec:conclusion}

In this paper we try to ascertain whether there are correlations between magnetic field parameters, calculated from observations from a ground-based spectropolarimeter, and plasma parameters derived from UV and EUV spectroscopic observations from space, using data from the 15 July 2014 obtained during a joint observing campaign with the satellites \textit{IRIS} and \textit{Hinode}, and the telescope THEMIS in the Canary Islands.
This prominence has also been studied in \citet{Levens2016}. 
This data set was chosen as it shows fairly good coverage of the prominence of interest, with data available from each of our instruments with relatively good spatial overlap. 
THEMIS observations, however, were obtained a few hours after the satellite observations.

%{\bf The \textit{IRIS} spectra for this prominence provide \ion{Mg}{ii} profiles with a number of different characteristics. 
%As is discussed in Sect. \ref{ssec:iris_diagnostics}, we have a mixture of reversed profiles and single-peaked profiles, but also some that do not fit in either catagory. 
%We call these profiles `complex', and suggest three possible formation mechanisms for them -- 1) A blend of narrow, single-peaked profiles resulting from multiple threads along the line of sight with different l.o.s. velocities \citep[as explored in][]{Schmieder2014}, 2) The resultant profile of multiple optically thick threads with different l.o.s. velocities, which can act to remove emission in one of the peaks \citep[as in][for Lyman profiles]{Gunar2008}, 3) The reversal minimum position (most optically thick part of the line) is shifted relative to the rest of the profile due to movement of the frontmost part of the prominence, and so absorbs the emission of one of the peaks. 
%Of these formation mechanisms we prefer the third option for this prominence, however the other options have their own merits. 
%It may be worthwhile to investigate the true nature of these complex profiles in prominences, however that is beyond the scope of this paper.}

The first challenge is the co-alignment of data sets from different instruments. 
We use a 2D cross correlation method on similar images from each instrument to co-align the data, allowing us to compare the data on a pixel-by-pixel basis. 

From this study we {  conclude that there are no correlations} between the magnetic field parameters from THEMIS and the \ion{Mg}{ii} parameters from the \textit{IRIS} observation, nor between the magnetic field parameters and electron densities calculated from \textit{Hinode}/EIS. 

We find that the magnetic field is generally stronger within the tornadoes {  ($\sim$ 30~G)} than outside them {  ($\sim$ 20~G)}, and that the inclination and azimuth are the same as values found previously \citep{Levens2016b}. %However,  we should keep in mind that there is a lower signal in the \ion{He}{i} D$_3$ line outside the tornado, resulting in a lower SNR which could affect the inversion of the spectro-polarimetric data. 

We study the level of reversal in the \ion{Mg}{ii} h and k lines, and find that it varies from unreversed to a k$_2$/k$_3$ ratio of around 2.8, suggesting high optical thickness at these locations. 
The mean reversal level is found to be 1.23 in the tornadoes and 1.14 in the rest of the prominence, suggesting that there are relatively more reversed profiles in the tornadoes than elsewhere. 
%We also see that in the tornadoes there are a relatively greater number of reversed profiles than unreversed, when compared to the rest of the prominence (Fig.~\ref{fig:iris_hist}, top panel). 
%Figure \ref{fig:khrat} plots the k/h intensity ratio \emph{vs.} magnetic field parameters in areas of the prominence. 
We see a small spread of k/h ratios, {  ranging from around 1.3 to 1.5.} % in some cases, and up to values nearing 4. %, with average values of 2.38 in the tornadoes and 2.16 in the rest of the prominence. 
For optically thin, collisionally excited emission we would expect a k/h ratio of 2 \citep{Leenaarts2013}, with departures from this value indicating a departure from the optically thin regime. 
%Values of k/h of 4 would indicate that radiative excitation plays a larger role than collisional processes in the excitation of the \ion{Mg}{ii} ion \citep{Harra2014}. 
%The fact that we see ratios nearing this value indicates an increased importance of radiation in the emission of the k and h resonance lines -- this confirms that much of the prominence emission seen in these lines comes from the scattering of light from the solar surface. 
The mean k/h ratio takes a value of 1.41 in both the tornadoes and in the rest of the prominence. 
This is {  a similar value as} has been found previously \citep{Schmieder2014,Liu2015,Levens2016,Vial2016}.
The displacement of the k$_3$ feature from its stationary position at line centre gives an indication of how the frontmost layers of the prominence are moving along the line-of-sight. 
%Figure \ref{fig:vk3} shows the k$_3$ velocity versus magnetic field parameters for points inside and outside of the tornadoes. 
We measure Doppler shifts of the k$_3$ reversal between around $\pm$10 km s$^{-1}$ everywhere. 
This is comparable to the overall velocity distribution, outlined in Paper I. 
For a quiescent prominence, we do not expect the overall velocity to be high, and it follows in that case that the k$_3$ Doppler shift would also be low.

We also present a comparison of the electron density at a temperature of $1.5 \times 10^6$~K, as calculated from EIS \ion{Fe}{xii} observations, with the magnetic field parameters from THEMIS. 
The electron densities used here were estimated by calculating the intensity ratio of the \ion{Fe}{xii} 195.119~\AA\ and 195.179~\AA\ lines, and then comparing this to the density curve for that ratio as computed by CHIANTI.
%Figure \ref{fig:dens_scatter} shows the correlation plots. 
We find a small scatter of densities in the prominence, with electron densities generally between {$\log{n_e}$ between 8.5 and 9.5}, with a slightly larger scatter in the tornadoes pixels. 
The mean electron density is higher in the corona than in the tornado considered here, with values of $\log{n_e} = 9.06 \pm 0.14$ and $\log{n_e} = 8.98 \pm 0.14$ respectively. 
This is comparable to previous studies on the electron density in tornado-like prominences \citep{Levens2015}. 
The lower electron density along the l.o.s. of the tornado could be due to volume blocking by the cool material at the tornado location \citep{Heinzel2008}. 
%The mean electron density is comparable both in the tornado and in the surrounding corona.
%It therefore appears that points in the tornadoes are more likely to have larger electron densities than points in the rest of the prominence -- the opposite of what was found in \citet{Levens2015}. 
%We note, however, that at the temperature of these \ion{Fe}{xii} lines, we cannot see any prominence material in emission, but we observe the surrounding corona.%, so it may be more accurate to say that it is the `surrounding corona', rather than the `rest-of-prominence'. 
{  No correlation is found between} the electron density and the line and plasma parameters derived from the \ion{Mg}{ii} lines. 
%Figure~\ref{fig:iris_eis} shows the correlation plots, but no new information is found in them. 

{  While no clear correlation is found between line parameters and magnetic field parameters, we provide the first detailed maps of such parameters for a prominence including a tornado. }
{It is important to note that the observations are not strictly simultaneous. 
We also note that the overlap between the data sets does not fully cover the region of interest, so we do not have a complete picture of the correlation at all parts of the prominence, especially in the \textit{IRIS} \emph{vs.} THEMIS plots. 
More importantly, there are differences in the optical thickness of the lines used in this study, specifically between \ion{He}{i} D$_3$ and \ion{Mg}{ii} lines, which could be having an adverse effect on potential correlations as the magnetic field information retrieved from the inversion of the D$_3$ spectro-polarimetric data represents averaged quantities along the line of sight, whereas the \textit{IRIS} observations allow us to see mostly the frontmost part of the tornado and surrounding prominence.}
In this respect, similar studies in the future should make use of diagnostics based on spectral lines with comparable optical thicknesses. 
The recent observations by CLASP \citep{2012SPIE.8443E..4FK,2012ASPC..456..233K} and the available spectropolarimetric measurements in the optically thick hydrogen Lyman~$\alpha$ line may help to measure plasma and magnetic field parameters in similar regions of the prominence.

%__________________________________________________________________

\begin{acknowledgements}
The authors thank S. Gunar, B. Gelly, and the team at the THEMIS telescope for assisting with these observations. P.J.L. acknowledges support from an STFC Research Studentship ST/K502005/1. N.L. acknowledges support from STFC grant ST/L000741/1. We are grateful for the financial support from the International Space Science Institute where this work was presented, and thank our colleagues from International Team 374 for useful and insightful discussions. The AIA data are provided courtesy of NASA/\textit{SDO} and the AIA science team. \textit{IRIS} is a NASA small explorer mission developed and operated by LMSAL with mission operations executed at the NASA Ames Research Center and major contributions to downlink communications funded by the Norwegian Space Center (NSC, Norway) through an ESA PRODEX contract. \textit{Hinode} is a Japanese mission developed and launched by ISAS/JAXA, with NAOJ as domestic partner and NASA and STFC (UK) as international partners.
\end{acknowledgements}

%-------------------------------------------------------------------

%\bibliographystyle{aa}
%\bibliography{bibliography}

\end{document}